\documentclass[journal]{IEEEtran}
\usepackage{amsmath,amsfonts}
\usepackage{array}
\usepackage{textcomp}
\usepackage{stfloats}
\usepackage{url}
\usepackage{verbatim}
\usepackage{graphicx}
\usepackage{caption}
\usepackage{subcaption}
\usepackage{cite}
\usepackage{amsmath} 
\usepackage{amsthm}
\usepackage{graphicx}
\usepackage{pdfcomment}

\usepackage{graphicx}
\usepackage{algorithm}
\usepackage[noend]{algpseudocode}

\usepackage{breqn}
\usepackage{multirow}
\usepackage{psfrag}
\usepackage{url}
\usepackage{xcolor}
\usepackage{pdfcomment}
\usepackage{textcomp}
\usepackage{amsfonts}
\usepackage{soul}
\usepackage{minted}
\usepackage{paralist}

\newtheorem{theorem}{Theorem}[section]

\newtheorem{assumption}[theorem]{Assumption}

\usepackage{booktabs}
\usepackage{multirow}
\usepackage{caption}
\usepackage{color}
\begin{document}

\title{Quantum Vanguard: Server Optimized Privacy Fortified Federated Intelligence for Future Vehicles}

\author{Dev Gurung and Shiva Raj Pokhrel,~\IEEEmembership{Senior Member, IEEE}
       
\thanks{Authors are with the QUANTIMA Research Initiative, School of IT, Deakin University, Geelong, Australia.}
}

\markboth{Journal of \LaTeX\ Class Files,~Vol.~14, No.~8, August~2026}%
{Shell \MakeLowercase{\textit{et al.}}: A Sample Article Using IEEEtran.cls for IEEE Journals}

\maketitle
\begin{abstract}
This work presents vQFL (\textit{vehicular Quantum Federated Learning}), a new framework that leverages quantum machine learning techniques to tackle key privacy and security issues in autonomous vehicular networks. 
Furthermore, we propose a server-side adapted fine-tuning method, \textit{ft-VQFL}, to achieve enhanced and more resilient performance.
By integrating quantum federated learning with differential privacy and quantum key distribution (QKD), our \textit{quantum vanguard} approach creates a multi-layered defense against both classical and quantum threats while preserving model utility. 
Extensive experimentation with industry-standard datasets (KITTI, Waymo, and nuScenes) demonstrates that vQFL maintains accuracy comparable to standard QFL while significantly improving privacy guaranties and communication security. 
Our implementation using various quantum models (VQC, QCNN, and SamplerQNN) reveals minimal performance overhead despite the added security measures. 
This work establishes a crucial foundation for quantum-resistant autonomous vehicle systems that can operate securely in the post-quantum era while efficiently processing the massive data volumes (20-40TB/day per vehicle) generated by modern autonomous fleets. 
The modular design of the framework allows for seamless integration with existing vehicular networks, positioning vQFL as an essential component for future intelligent transportation infrastructure.
\end{abstract}

\begin{IEEEkeywords}
Quantum Federated Learning, Autonomous Vehicles, Differential Privacy, Quantum Key Distribution
\end{IEEEkeywords}

\section{Introduction}
\IEEEPARstart{A}{utonomous} vehicles generate unprecedented data volumes of 20 to 40 TB daily per vehicle \cite{chellapandiFederatedLearningConnected2024a} requiring sophisticated machine 
learning systems for perception, planning, and control under strict real-time constraints \cite{pokhrelFederatedLearningBlockchain2020}. Despite initial skepticism \cite{martinezChallengesAutonomousVehiclesTheoretical2018a}, commercial deployments by Waymo and Tesla demonstrate the technology's viability, yet three critical challenges persist: managing large vehicle fleets, overcoming latency in cloud-based learning, and mitigating safety risks from delayed decision-making.

Quantum federated learning (QFL) \cite{chen_introduction_2024} offers a promising solution by integrating quantum computing \cite{steane_quantum_1998}, quantum machine learning \cite{biamonte_quantum_2017}, and classical federated learning principles \cite{mcmahan_communication-efficient_2023, kairouz_advances_2021}. 
QFL leverages quantum mechanical phenomena superposition, entanglement, and interference to potentially accelerate computational tasks while preserving data locality. 
Recent advances in QFL architectures \cite{gurung_performance_2024, gurung_personalized_2024} demonstrate growing capabilities as quantum technologies mature beyond the limitations of classical computing miniaturization \cite{williamsExplorationsQuantumComputing2011}.

\begin{figure}
    \centering
    \includegraphics[width=0.75\linewidth]{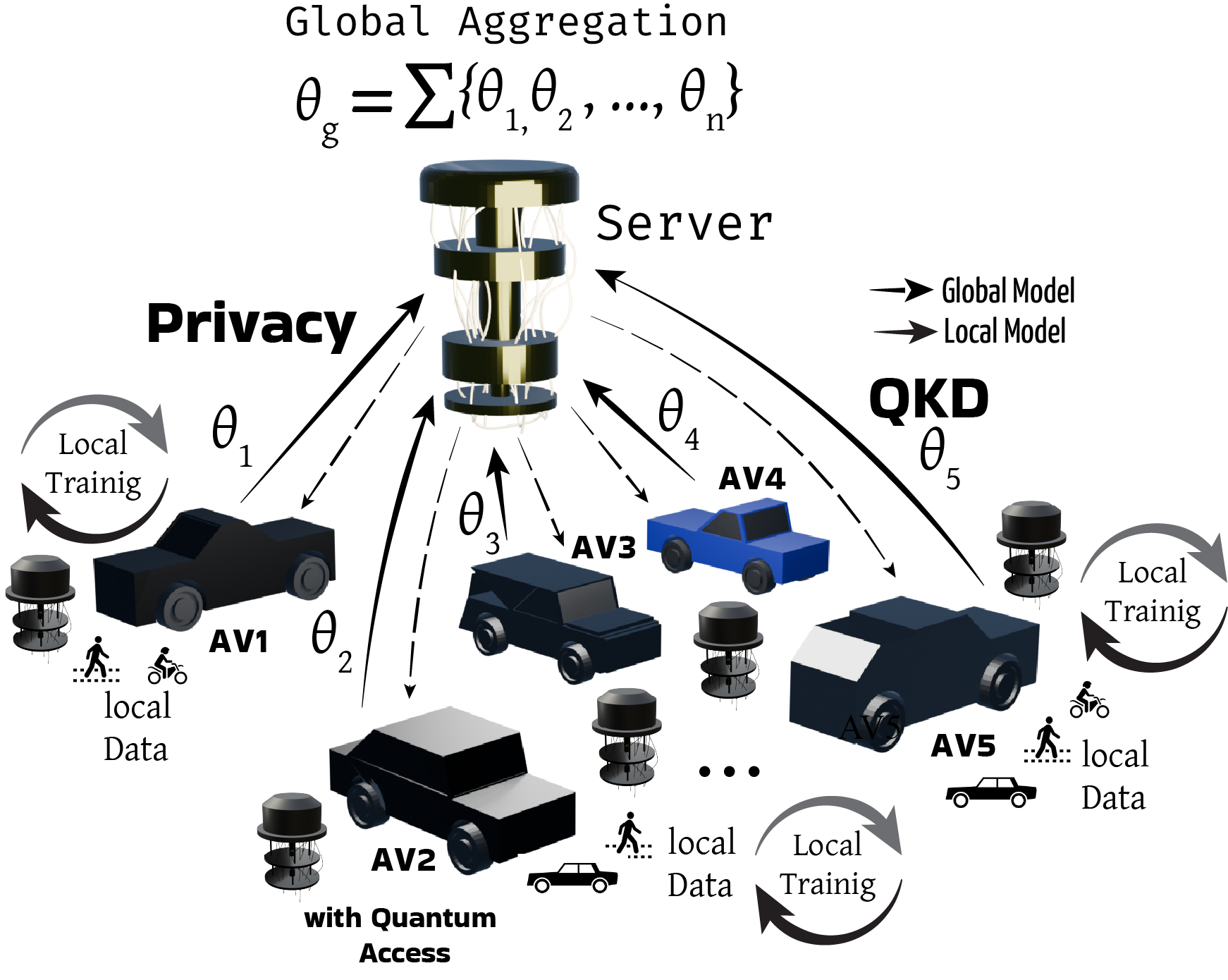}
    \caption{Comprehensive Architecture of vQFL Framework for Privacy-Preserving Secure Autonomous Driving Intelligence.} 
    \label{fig:vQFL}
\end{figure}

We introduce \textit{Quantum Vanguard}, a comprehensive framework for next-generation vehicular networks that strategically integrates quantum computing with distributed learning systems. This framework anticipates Figure \ref{fig:vQFL} as a layered security and performance architecture to create a quantum-resistant defense system while allowing quantum advantages for complex vehicular computational tasks. Note that developing \textit{Quantum Vanguard} is essential to address three critical challenges, namely, {computational demands, privacy requirements, and emerging quantum threats}, converging in future vehicular technology.

\noindent i) \textit{Computational Demands:} Autonomous vehicles generate 20-40TB of data daily per vehicle \cite{chellapandiFederatedLearningConnected2024a} requiring real-time processing with sub-millisecond latency. Quantum computing offers theoretical 
speedups from 
$O(N)$ to $O(\sqrt{N})$ for search problems and exponential advantages for specific optimization tasks 
\cite{biamonte_quantum_2017} critical for processing high-dimensional sensor fusion data from LiDAR, cameras, and radar simultaneously \cite{caesarNuScenesMultimodalDataset2020}.

\noindent ii) \textit{Privacy Vulnerabilities:} Conventional federated learning remains susceptible to model inversion attacks and gradient leakage, 
with recent research showing 87\% reconstruction accuracy from aggregated model updates \cite{li_privacy-preserving_2024}. For autonomous vehicles, 
such vulnerabilities could expose sensitive location data and behavioral patterns. 
Quantum approaches offer information-theoretic privacy guarantees rather than relying on computational hardness assumptions.

\noindent iii) \textit{Quantum Security Threats:} Large-scale quantum computers threaten classical cryptographic protocols securing vehicular communications. 
Shor's algorithm can break RSA and elliptic curve cryptography \cite{steane_quantum_1998}, while Grover's algorithm halves symmetric encryption security. With 10-15 year operational lifespans, vehicles deployed today will likely operate in a post-quantum environment, necessitating quantum-resistant security frameworks.

Recent research has explored various aspects of this domain. Li et al. \cite{li_privacy-preserving_2024} developed quantum protocols to prevent gradient inversion attacks through private inner-product estimation and quantum state concealment. 
In classical federated learning, Behera et al. \cite{beheraLargeModelAssistedFederated2025} proposed large model-assisted approaches for edge-based object detection, 
while Kishawy et al. \cite{eidkishawyFederatedLearningSystem2024a} focused on lane segmentation systems. 
However, these works do not address the comprehensive integration of quantum computing, differential privacy, and quantum-secure communications necessary for next-generation autonomous vehicle networks.

In this paper, we present vQFL as a \textit{proof of concept}, as shown in Figure \ref{fig:vQFL}, implementing the Quantum Vanguard concept through a novel framework that combines quantum computational advantages with multi-layered privacy and security guarantees. 
Our approach specifically addresses the computational, privacy, and security challenges of autonomous vehicle fleets while establishing a foundation for quantum-resistant intelligent transportation systems.

\subsection{Contribution}
The contributions of this work are listed below.
\begin{enumerate}
    \item We propose a novel vehicular quantum federated learning (vQFL) framework for autonomous vehicle driving networks for distributed quantum machine learning.
    Our proposed framework (vQFL) provides privacy on top of default data privacy provided by QFL framework and secure communication means through QKD.
    \item We further propose and investigate the concept of fine-tuning, or optimizing, the global model after the averaging phase through additional customizations. 
    This procedure aims to achieve improved performance at both the local and global model levels and is evaluated via experimental analysis, referred to as server-side optimized vehicular QFL (ft-vQFL).
    \item We present extensive theoretical and experimental analysis to demonstrate the practicality and adaptability of the proposed method in the field of autonomous vehicle technology with the Qiskit framework utilizing various models such as VQC, QCNN, and samplerQNN in various datasets including KITTI, nuScenes, and Waymo datasets.
\end{enumerate}

\subsection{Originality and novelty} Figure~\ref{fig:vQFL} presents an architectural overview of our proposed vQFL framework, illustrating the sophisticated integration of quantum computing paradigms with autonomous vehicle technology. The architecture depicts a network of connected autonomous vehicles (Tesla, Waymo, and other manufacturers) equipped with onboard quantum processors that collect and process multi-modal sensor data streams (high-definition LIDAR point clouds, multi-camera arrays, millimeter-wave radar, GPS, and inertial measurement units) locally.

Each vehicle in Figure \ref{fig:vQFL} independently trains quantum machine learning models, including variable Quantum Circuits (VQC), Quantum Convolutional Neural Networks (QCNN), and our samplerQNN (Quantum Neural Network) on its private driving data.
This localized training approach ensures that sensitive driving behaviors, route patterns, and environmental observations never leave the vehicle's secure computing environment.

The framework of Figure \ref{fig:vQFL} implements our novel three-tiered security approach: (1) Base-level QFL Privacy: Inherent data privacy from the federated architecture where only model 
parameters not raw data are shared; 
(2) Differential Privacy Layer: Application of calibrated Laplacian or Gaussian noise 
($\epsilon$) 
to model parameters before transmission, mathematically guaranteeing protection against model inversion and membership inference attacks; (3) Quantum Cryptographic Layer: Implementation of BB84 quantum key distribution protocol that leverages quantum superposition and measurement principles to establish information-theoretically secure communication channels. 
These encrypted model can be transmitted either to a central server in a hierarchical client-server topology  or directly between vehicles in a decentralized peer-to-peer network, depending on deployment requirements and available infrastructure.

The system illustrated in Figure \ref{fig:vQFL} performs intelligent federated aggregation of these quantum-secure updates
to create an improved global model. This model benefits from collective learning across diverse driving scenarios (urban environments, highways, adverse weather conditions) while maintaining strict privacy guarantees and quantum-resistant security. The bidirectional arrows in Figure \ref{fig:vQFL} represent the secure exchange of model parameters, with quantum-encrypted channels ensuring protection against both conventional cryptanalytic attacks and future quantum computing threats. The vQFL framework supports both synchronous and asynchronous communication patterns to accommodate varying connectivity conditions in real-world vehicular networks.

\section{Preliminaries and Background}
The convergence of quantum computing and autonomous vehicle technology creates new opportunities to address critical challenges in vehicular networks. Autonomous vehicles generate massive data volumes (20-40 TB daily per vehicle) that must be processed with ultra-low latency for safe operation \cite{caesarNuScenesMultimodalDataset2020}. These vehicles rely on sophisticated perception systems for object detection, lane segmentation, and trajectory prediction under various environmental conditions \cite{mundhedaTeacherguidedOffroadAutonomous2025WorkshopAAAI, zhangCopilot4DLearningUnsupervised2023}.

QFL offers a promising approach by extending classical federated learning with quantum computational capabilities \cite{gurungQuantumFederatedLearning2023}. QFL employs parameterized quantum circuits, quantum data encoding, and qubit-based processing within a hybrid quantum-classical framework where quantum circuits perform computations while classical optimizers update the circuit parameters \cite{chen_introduction_2024}. This paradigm leverages quantum mechanical phenomena superposition, entanglement, and interference to 
potentially accelerate specific computational tasks relevant to autonomous driving. 

However, collaborative learning in vehicular networks introduces significant privacy and security concerns. Differential privacy (DP) addresses these challenges by providing mathematically rigorous privacy guarantees, ensuring that model outputs differ by at most a factor of $e^\epsilon$ (with exception probability $\delta$) when trained on datasets differing by a single record \cite{holohanDiffprivlibIBMDifferential2019, jiLessMoreRevisiting}. For communication security, QKD enables information-theoretically secure key exchange based on quantum mechanical principles, making it resistant to both classical and quantum attacks \cite{liaoSatellitetogroundQuantumKey2017}.

The integration of these technologies, QFL, differential privacy, and QKD creates
a comprehensive framework to address the unique challenges of autonomous vehicles: computational efficiency for real-time decision making, preservation of privacy for sensitive driving data, and quantum-resistant security for vehicle-to-vehicle and vehicle-to-infrastructure communications \cite{martinezChallengesAutonomousVehiclesTheoretical2018a, eidkishawyFederatedLearningSystem2024a}. 
We aim to explore and exploit such a joint framework to establish a foundation for next-generation autonomous vehicle systems that can operate securely and efficiently in the post-quantum era.

\begin{table*}[!htbp]
    \centering
    \caption{Selected related work in FL for AVs}
    \label{tab:related_work}
    \resizebox{0.8\textwidth}{!}{
    \begin{tabular}{|p{1.5cm}|p{2cm}|p{5.5cm}|p{1cm}|p{1.2cm}|p{2cm}|p{2cm}|}
        \toprule
        \textbf{Work} & \textbf{Technique} & \textbf{Features} & \textbf{Model} & \textbf{Simulation} & \textbf{Datasets} & \textbf{Type} \\
        \midrule
        Semi-SynFed \cite{liangSemiSynchronousFederatedLearning2022} & Semi-synchronous FL & Select appropriate nodes, dynamic waiting time, dynamic aggregation scheme & Classical & Open datastreet & FMNIST, CIFAR10, TSRD & CNN \\
        \midrule
        LMFL \cite{beheraLargeModelAssistedFederated2025} & Large Model assisted FL & Object detection, synchronous/asynchronous approach, hierarchical structure & Classical & TensorFlow & KITTI & LENET \\
\midrule
          FedLane \cite{eidkishawyFederatedLearningSystem2024a} & Secure and Efficient FL & Lane Segmentation, real-time inference, complicated road scenarios & Classical & TensorFlow & TuSimple, CUlane & U-Net, ResUNet etc. \\
          \midrule
          Copilot4D \cite{zhangCopilot4DLearningUnsupervised2023} & World Models for AV via Discrete Diffusion & Unsupervised world models, Use discrete Diffusion, tokenize sensor observations, prediction via discrete diffusion & Classical & - & nuScenes, KITTI, Argoverse 2 & Discrete Fusion MOdel, VQVAE-like \\
          \midrule
          BDFL \cite{chenBDFLByzantineFaultToleranceDecentralized2021} & BFT decentralized FL & Privacy Preserving, Peer-to-peer, HydRand protocol, PVSS Scheme, Multi-Object recognition, Privacy Preserving & Classical & PyTorch & MNIST, KITTI & MultiLayer Perceptron Model, CNN \\
          \midrule
          FADNet \cite{nguyenDeepFederatedLearning2022} & Deep FL for AD & Peer to peer deep FL, decentralized, Model Stability, Convergence, Handle Imbalance data & Classical & Carla, Gazebo & Udacity data & silo Model \\
        \midrule
        \textbf{This work} & Vehicular QFL & Quantum private and secure QFL & Quantum & Qiskit & nuScenes, Waymo, KITTI & VQC \\
        \bottomrule
    \end{tabular}
    }
\end{table*}

\subsection{Literature}
The technological progress of autonomous vehicles has surpassed human expectations, with notable
implementations and services already being offered by major tech companies such as Waymo Taxi and Tesla
Robotaxi.
One of the main reasons for this is the considerable developments and advances in the field of artificial intelligence, particularly deep learning \cite{autonomousVehiclesAndSystems2023}.
In this section, we conclude various works done in the field of machine learning for autonomous vehicles specifically focusing primarily on federated distributed learning.

Some recent work, also shown in Table \ref{tab:related_work}, in the field of classical FL includes the semi-synchronous FL protocol with dynamic
aggregation on the Internet of Vehicles \cite{liangSemiSynchronousFederatedLearning2022}.
The key idea behind the proposed solution is to improve machine learning performance and efficiency through steps like first selecting
appropriate nodes for dynamic asynchronous aggregation based on computational capacity, network capacity, etc.
Similarly, in terms of object detection, Behera et al. \cite{beheraLargeModelAssistedFederated2025}
proposed a large model-assisted FL for object detection for the autonomous vehicle in edge.
The work focuses on combining the strength of both synchronous and asynchronous methods and further
organizing devices into a hierarchical structure to optimize model convergence, as well as to
mitigate any stragglers and dropout problems.
In terms of lane segmentation, Kishawy et al. \cite{eidkishawyFederatedLearningSystem2024a} proposed the secure and efficient FL system, FedLane, for AV for complicated and dynamic road scenarios.
Zhang et al. \cite{zhangCopilot4DLearningUnsupervised2023} proposed unsupervised world models, Copilot4D, for AV using discrete diffusion.
It is a world-modeling approach in which sensor observations are first tokenized with VQVAE and prediction is then performed through discrete diffusion.
Other works include BDFL, a Byzantine fault-tolerant decentralized FL for AVs by Chen et al. \cite{chenBDFLByzantineFaultToleranceDecentralized2021}.
It is a decentralized FL approach for privacy-preserving addressing multi object recognition task etc. 

Nguyen et al. \cite{nguyenDeepFederatedLearning2022} proposed the peer-to-peer deep FL approach to train deep architectures in a decentralized manner and design a new federate AV network (FADNet) to improve model stability, convergence, etc.
Xu et al. \cite{xuSecureFederatedLearning2023} proposed secure FL for quantum autonomous vehicle networks based on the QKD protocol, the SAGIN-enabled QAVN model, and the FL framework.

From the start of machine learning approaches such as the transportable neural network approach to AVs \cite{kehtarnavazTransportableNeuralnetworkApproach1998}, the integrated framework for decision making and motion planning for AVs \cite{hangIntegratedFrameworkDecision2020} to different decentralized approaches in AVs including FL, 
the key consideration and network are in a classical form similar to conventional computing. 
Considering recent advances in the field of quantum computing and additional threat to quantum attacks on classical computers and classical problems, it would be an overstatement that a novel quantum proof and a quantum way of decentralized approaches is integral to the survival and adaptation of the AV industry in the quantum era.
Thus, our work fundamentally is different from all the related works and as far as we know is the first work towards quantum FL for AVs with SOTA real-world datasets such as Waymo, KITTI, nuScenes etc. 

\section{Proposed vechicular QFL}
We introduce a secure and privacy-preserving framework for vehicular quantum federated learning (vQFL). This framework is comprised of three primary components: \begin{enumerate} 
    \item A QFL framework tailored for self-driving vehicles.
    \item A privacy-preserving strategy applied to QFL models.
    \item A secure QKD communication protocol for QFL.
\end{enumerate}

We propose a QFL framework for a vehicular scenario where autonomous vehicles can locally train on their data
facilitating distributed machine learning with access to quantum computational capability.
The Algorithm \ref{alg:qfl_algorithm} shows the main steps involved in the vehicular QFL framework in terms of
coordination between vehicles and the server, along with some other details such as topology, data distribution, etc.

\begin{figure*}
    \centering
    \includegraphics[width=0.9\linewidth]{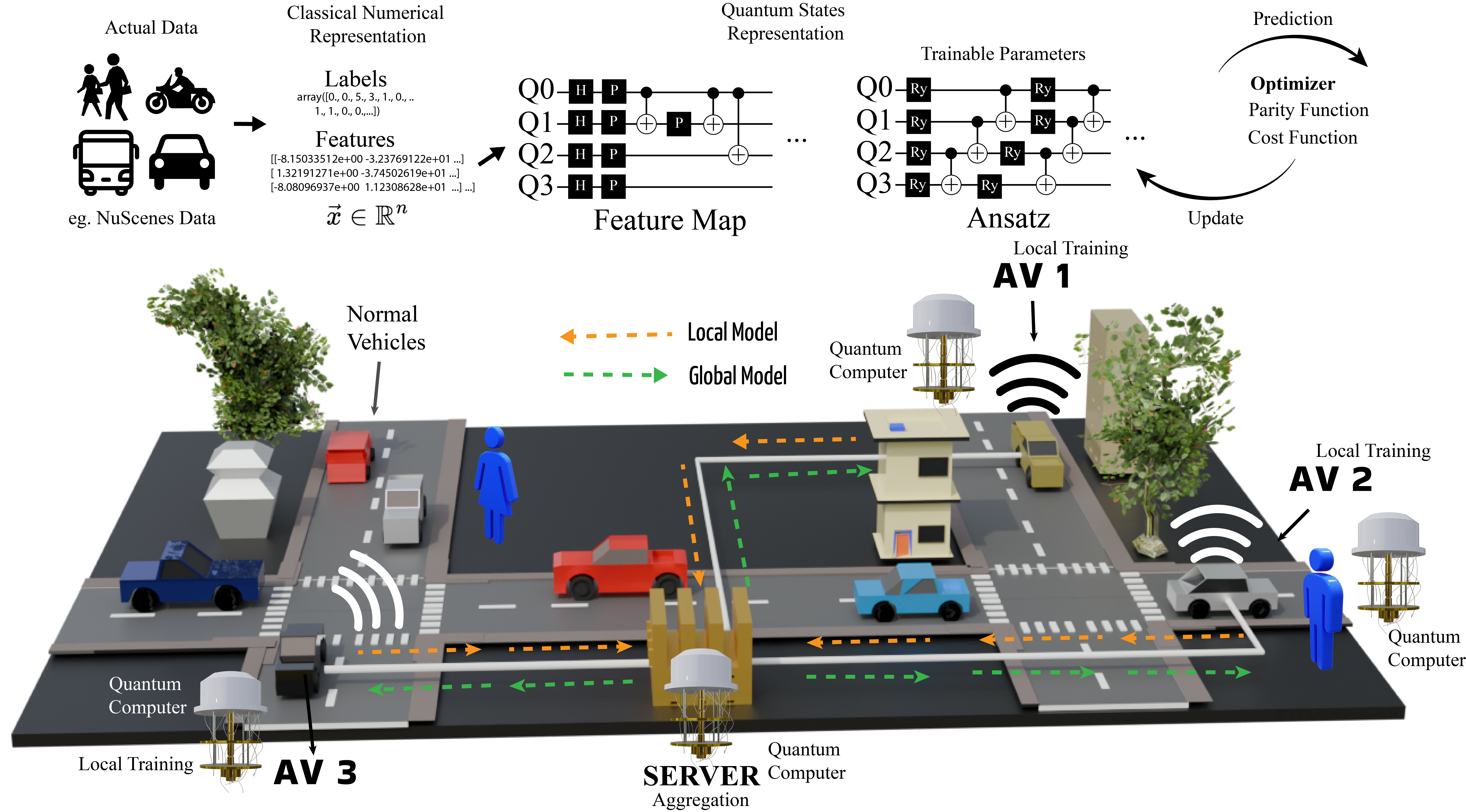}
    \caption{\textit{Proposed privacy-preserving secure vehicular QFL Framework: Autonomous vehicles collaboratively train global model; vQFL framework integrates quantum computing with autonomous vehicle technology. The architecture involves connected vehicles with onboard quantum processors training models locally on private driving data. 
    Our three-tiered privacy and security approach combines: (1) Base-level QFL privacy where only model parameters are shared; (2) Differential privacy layer applying calibrated noise to parameters; and (3) Quantum cryptographic layer implementing BB84 protocol for secure communications. The system can be extended to both client-server and peer-to-peer topologies, enabling collaborative intelligence while maintaining quantum-resistant security.}}
    \label{fig:enter-label}
\end{figure*}

In our proposed framework, we implement privacy measures for model where we add noise to the model parameters.
This step helps prevent any adversary or malicious server from inferring information from the model about the data that were trained, which is presented in Algorithm \ref{alg:dp_integration}.
On top of default QFL and privacy protocol, we add additional layer providing security with QKD implementation mechanism that works with QFL framework.
Our design involves creating QKD secret keys, encryption, and decryption, etc. which are detailed in Algorithm \ref{alg:qkd_encryption_decryption}.

\begin{figure}
    \centering
    \includegraphics[width=0.6\linewidth]{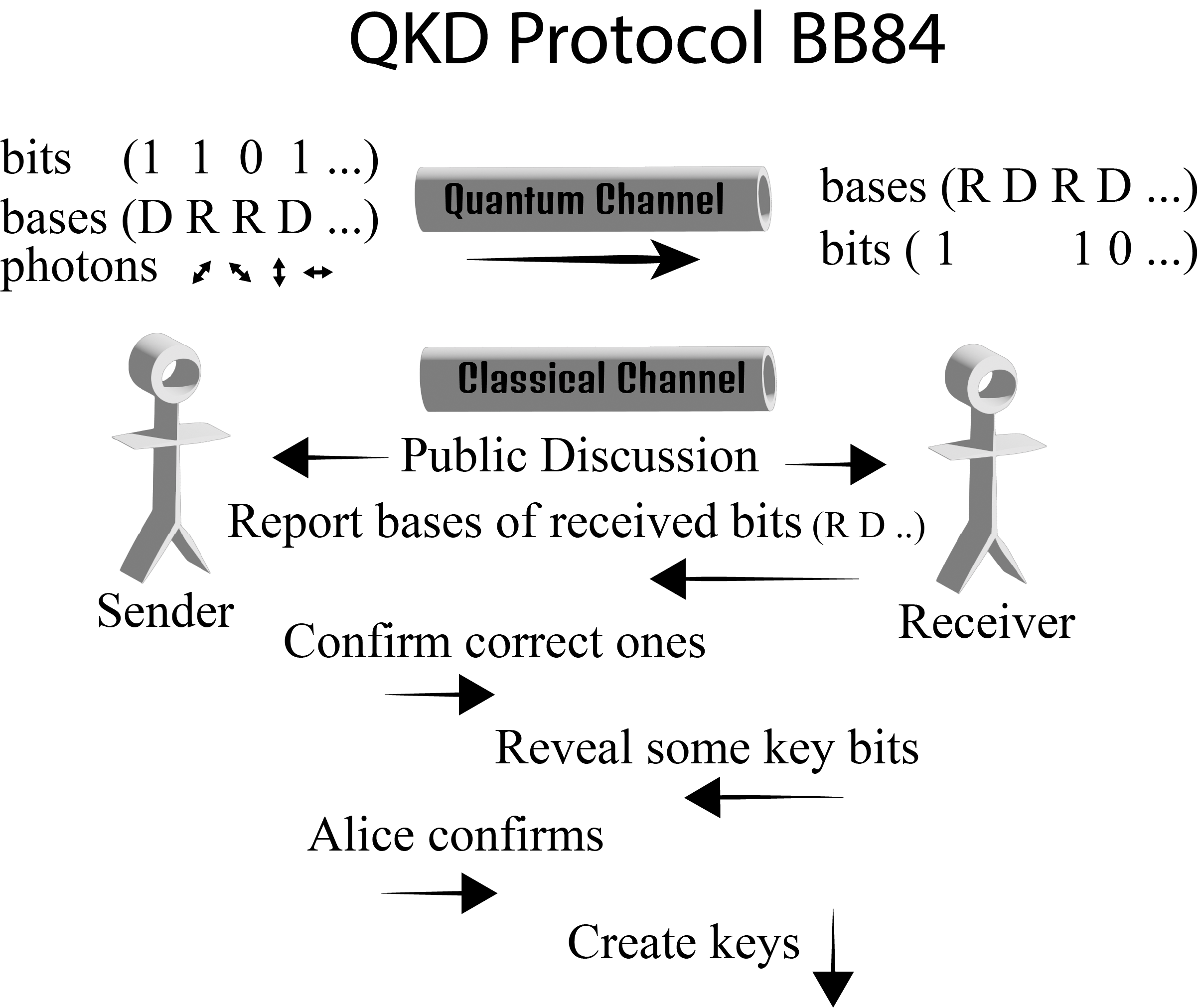}
    \caption{QKD Protocol: Standard BB84 protocol for Quantum Key Distribution \cite{bennettQuantumCryptographyPublic2014}.}
    \label{fig:qkd_protocol}
\end{figure}

\begin{figure}
    \centering
    \includegraphics[width=0.7\linewidth]{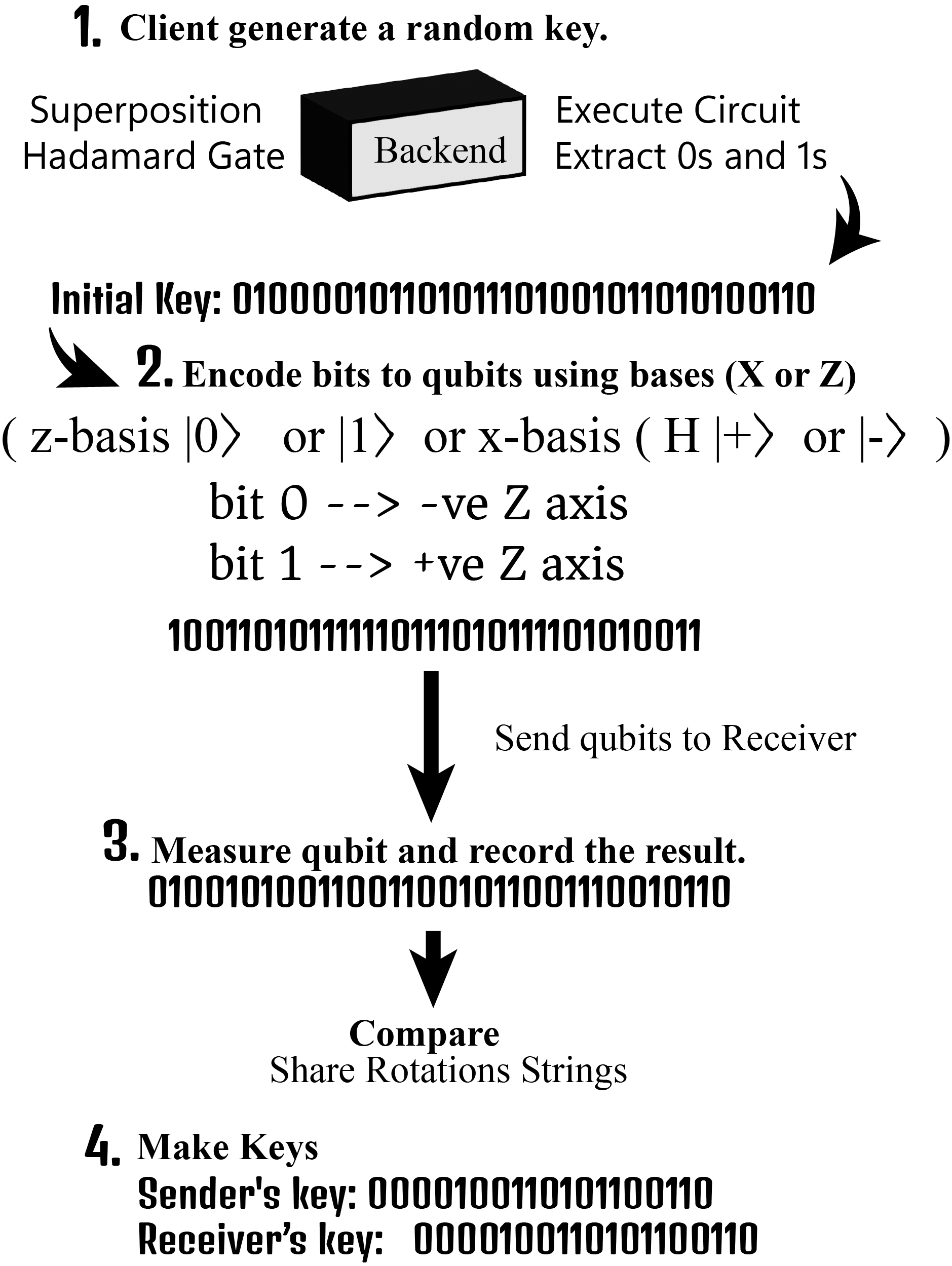}
    \caption{QKD Generation: Key Generation implemented in this work using backend and following BB84 Protocol.}
    \label{fig:qkd_generate_key}
\end{figure}

\begin{figure}
    \centering
    \includegraphics[width=0.5\linewidth]{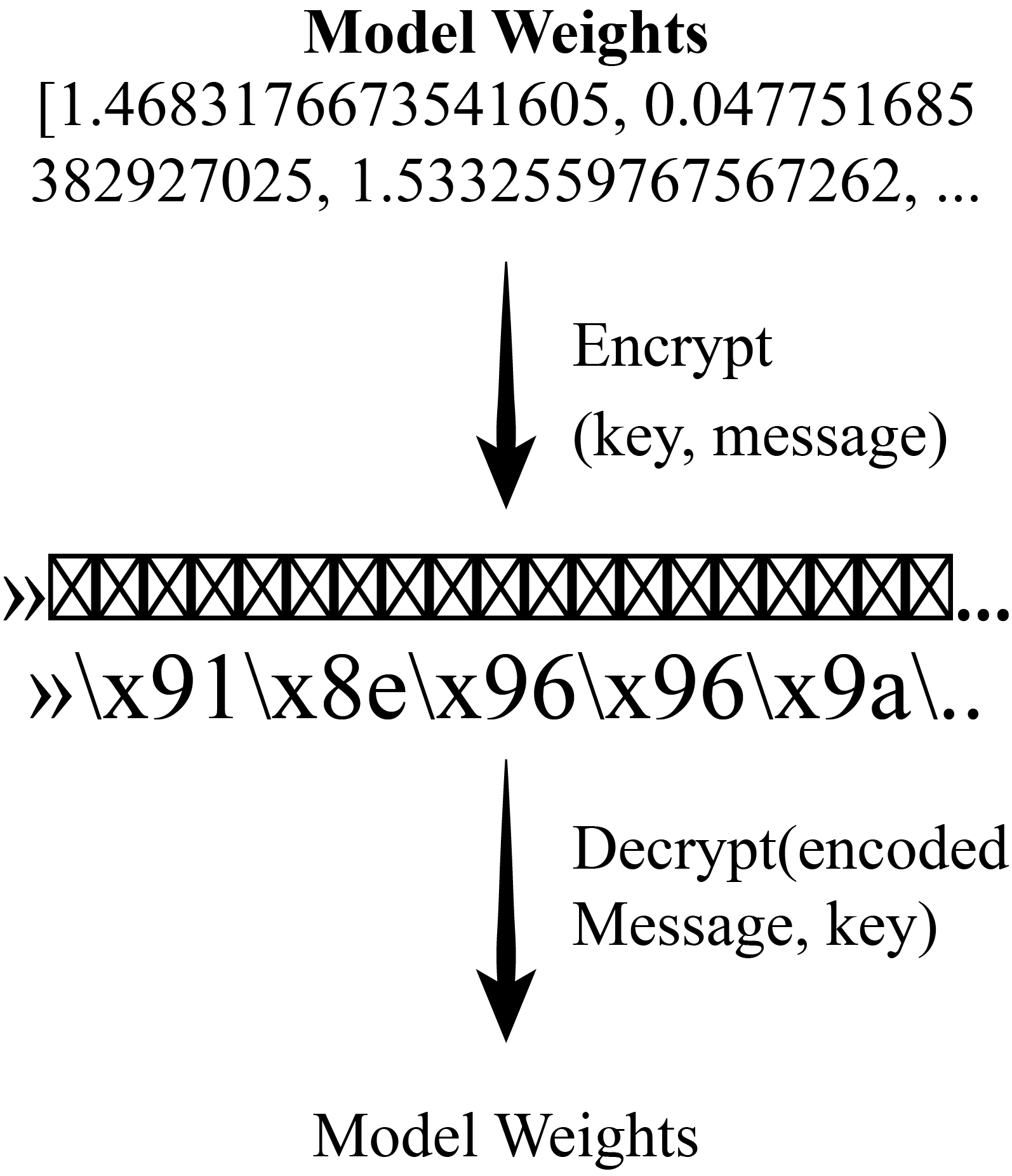}
    \caption{QKD Encryption: Encryption and Decryption of model weights as performed in the work (sample preview)}
    \label{fig:qkd_encryption}
\end{figure}

\begin{algorithm}
\caption{Privacy-Preserving Quantum-Secure vQFL}
\label{alg:qfl_algorithm}
\begin{algorithmic}[1]
\State \textbf{Notations:}
\State $ K $: Number of clients, indexed $ k = 1, \dots, K $; $ S $: Central server.
\State $ \mathcal{D}_k = \{ (x_i^{(k)}, y_i^{(k)}) \}_{i=1}^{n_k} $: Local dataset of client $ C_k $, size $ n_k $.
\State Model types: $ m \in \{ \text{VQC}, \text{QCNN}, \text{SamplerQNN} \} $.
\State Global parameters at round $ t $: $ \theta^{(t)} $; Local loss: $ \mathcal{L}_k(\theta) = \frac{1}{n_k} \sum_{i=1}^{n_k} \ell(\theta; x_i^{(k)}, y_i^{(k)}) $.
\State Optimizers: Gradient-based (e.g., Quantum Natural Gradient) or gradient-free (e.g., COBYLA).
\State DP parameters: $ (\epsilon, \delta) $, noise $ \mathbf{z}_k \sim \mathcal{N}(0, \sigma^2 \mathbf{I}) $.
\State QKD for secure key generation; $ T $: Number of rounds.
\State \textbf{Initialize:} Server sets $ \theta^{(0)} $, broadcasts to clients via QKD channels.
\For{$ t = 1 $ to $ T $}
\For{each client $ C_k $ in parallel}
\State Receive $ \theta^{(t-1)} $.
\State Local training: Update $ \theta_k^{(t)} $ to minimize $ \mathcal{L}_k(\theta) $.
\State \quad - Gradient-based: $ \theta_k^{(t)} \leftarrow \theta^{(t-1)} - \eta \nabla_\theta \mathcal{L}_k(\theta^{(t-1)}) $.
\State \quad - Gradient-free: $ \theta_k^{(t)} = \arg\min_\theta \mathcal{L}_k(\theta) $.
\State Add DP noise: $ \tilde{\theta}_k^{(t)} \leftarrow \theta_k^{(t)} + \mathbf{z}_k $, $ \sigma \propto \frac{\sqrt{\ln(1.25/\delta)} \cdot C}{\epsilon n_k} $.
\State Send $ \tilde{\theta}_k^{(t)} $ to server via QKD channel.
\EndFor
\State \textbf{Server:} Compute $ \theta^{(t)} \leftarrow \frac{1}{\sum_k n_k} \sum_{k=1}^K n_k \tilde{\theta}_k^{(t)} $.
\State Optimize global model (if desired).
\State Broadcast $ \theta^{(t)} $ via QKD channels.
\EndFor
\State \textbf{Focus:} DP noise ensures privacy against model inversion; QKD provides quantum-secure parameter exchange.
\end{algorithmic}
\end{algorithm}

\begin{algorithm}
\caption{Privacy Mechanism}
\label{alg:dp_integration}
\begin{algorithmic}[1]
    \State \textbf{Input:} Model parameters $\theta = [\theta_1, \theta_2, \dots, \theta_n]$, privacy budget $\epsilon > 0$, sensitivity $\Delta > 0$
    \State \textbf{Output:} Noisy parameters $\tilde{\theta} \in \mathbb{R}^n$
    \Procedure{AddDPNoise}{$\theta, \epsilon, \Delta$}
        \State Train local model to obtain parameters $\theta \in \mathbb{R}^n$.
        \State Determine parameter shape $n = |\theta|$.
        \State Compute noise scale $\sigma = \frac{\Delta}{\epsilon}$.
        \State Generate noise $\eta \sim \text{Laplace}(0, \sigma)^n$.
        \State Compute noisy parameters $\tilde{\theta} = \theta + \eta$.
        \State \Return $\tilde{\theta}$
    \EndProcedure
\end{algorithmic}
\end{algorithm}

\begin{algorithm}
\caption{Security Protocol}
\label{alg:qkd_encryption_decryption}
\begin{algorithmic}[1]
    \State \textbf{Input:} Message $m \in \{0,1\}^n$, desired key length $n$
    \State \textbf{Output:} Ciphertext $c \in \{0,1\}^n$, decrypted $m' \in \{0,1\}^n$
    \Procedure{RandomStringGen}{$n$}
        \State Initialize quantum backend $\mathcal{B}$.
        \State Define registers $Q = \{q_i\}_{i=1}^n$, $C = \{c_i\}_{i=1}^n$.
        \State Construct quantum circuit $\mathcal{C}$, applying $H^{\otimes n}$ to $Q$.
        \State Execute $\mathcal{C}$ on $\mathcal{B}$ , get bit string $s \in \{0,1\}^n$.
        \State Generate bases $b_A, b_B \in \{\text{X}, \text{Z}\}^n$ for both.
        \State Encode qubits with $s$, $b_A$: apply $X$ if $s_i = 1$, $b_{A,i} = \text{X}$.
        \State Transmit encoded qubits $|{\psi}\rangle = \bigotimes_{i=1}^n |{\psi_i}\rangle$ to receiver.
        \State Measure qubits in basis $b_B$ to get $s' \in \{0,1\}^n$.
        \State Shift: keep bits where $b_{A,i} = b_{B,i}$, get key $k \in \{0,1\}^m$, $m \leq n$.
        \State \Return $k$
    \EndProcedure
    \Procedure{EncryptionDecryption}{$m, k$}
        \State Ensure $|k| = |m| = n$.
        \State Compute ciphertext $c = m \oplus k$
        \State Transmit $c$ to receiver.
        \State At receiver,  decrypt $m' = c \oplus k$.
        \State \Return $c, m'$
    \EndProcedure
\end{algorithmic}
\end{algorithm}


With QKD, we encrypt the model weights before sending them to the server and vise versa. 
The model weights are differentially private before encryption and thus maintaining multi layer privacy and security to the communication framework.
In this work, we have followed the BB84 protocol for QKD \cite{bennettQuantumCryptographyPublic2014}.
Clients and server are connected through the quantum channel, where they exchange qubits which are processed to generate encryption and decryption key (symmetric key) to send private message using the classical channel by using methods like one-time padding.

The implemented protocols are as follows: 
For message $m$, random binary strings $\in \{0,1\}^N$ of length $s$ are generated simulating a quantum circuit where 
the H gate is applied to each of the $s$ qubits initialized to $|0\rangle$, and then measured in the Z base to obtain random bits.
Mathematically, for length $s$, 
initialize qubits  $|\psi_0 \rangle = |0\rangle^{\otimes s}$ and then apply the Hadarmard gate as 
\[
 |\psi\rangle = H^{\otimes s} |\psi_0\rangle = \bigotimes_{j=1}^s \frac{1}{\sqrt{2}} (|0\rangle + |1\rangle)_j.
\]
The measurement in Z-basis results in outcome sequence $R=(r_j)^s_j$, where $r_j = 0$ or $1$ with the probability of $\frac{1}{2}$
independent (uniform over $ \{0,1\}^s $).
Now, a random initial key is generated $D^A$ along with the sender's basis choices ($B^A$) and the receiver's basis choices ($B^B$).

Each client device prepares $N$ qubits based on their bits $D^A = (d^A_j)^N_{j=1}$ and bases $B^A = (b^A_j)^N_{j=1}$.
However, server measures using bases $B^B = (b^B_j)^N_{j=1}$.
Mathematically, 
For each position $j = 1$ to $N$, 
the Device prepares the $j^{th}$ qubit as, 
\begin{align}
    |\psi \rangle = H^{b_j^A} X^{d_j^A} |0 \rangle
\end{align}
such that 
\begin{itemize}
    \item If $ b^A_j = 0 $ (rectilinear basis): $ |\psi_j\rangle = |d^A_j\rangle $ ($ |0\rangle $ or $ |1\rangle $).
    \item If $ b^A_j = 1 $ (diagonal basis): $ |\psi_j\rangle = H |d^A_j\rangle = 
    \begin{cases} 
    |+\rangle & \text{if } d^A_j = 0, \\
    |-\rangle & \text{if } d^A_j = 1.
    \end{cases} 
    $
\end{itemize}
The qubit is transmitted to the server. 
After receiving it, the server applies its basis transformation
\[
| \phi_j \rangle = H^{b^B_j} \, |\psi_j\rangle
\]
and measures $|\phi_j\rangle$ in z-basis and get outcome $ r_j \in \{0,1\} $, with probability $ |\langle r_j | \phi_j \rangle|^2 $.
The server result is equal to $R^B = (r_j)^N_{j=1}$.
The bases $B^A$ and $B^B$ are shared publicly whereby bits that matches bases match are kept.

Let $ S = \{ j \in \{1, \dots, N\} \mid b^B_j = b^A_j \} $ (expected $ |S| \approx N/2 $).
Then, keys are generated on the basis of the bases and results of measurements.

For our model weights $\theta_i$ in the form $[m_1, m_2, m_3...m_n]$ for the client $i$, the secret key generated using the QKD protocol $K = k_1k_2..k_s$ for length $s$ where $n=s$ after shortening the secret key to match the message for encryption, we obtain encrypted weights $E=e_1e_2..e_n$ by
\[
E = (e_i)_{i=1}^n
\]
where, 
$e_i = f(m_i, k_i) =  chr(ord(m_i) + 2 * ord(k_i) \text{ mod } 256$, $ord(x)$ is the unicode point of character $x$ (0-127 for ASCII characters) using the encoding function $f$ and the operation of the mod ensures the range $[0, 255]$.
It is experimentally depicted in Figure \ref{fig:qkd_encryption}.
To decrypt the message, we reverse the encryption procedure using $f(e_i, k_i) =  chr(ord(e_i) - 2 * ord(k_i) \text{ mod } 256$ to encode the message character $e$.

Let $\vec{\theta} \in \mathbb{R}^d$ be the input parameter vector with $d$ elements, 
$\epsilon > 0$ be the privacy budget, $S > 0$ be the sensitivity, then noise scale. \[
b = \frac{S}{\epsilon}.
\]
The noise vector $N \in R^d$ is generated element-wise as follows:
\begin{itemize}
    \item For the Laplace mechanism, $N_i = Laplace(0,b, size=\theta.shape)$.
    \item For the Gaussian mechanism, $N_i \sim \mathcal{N}(0, \sigma^2), \quad \forall i = 1, \dots, d.$ where $\sigma = \sqrt{2 \ln\left( \frac{1.25}{\delta} \right)} \cdot b.$
\end{itemize}
Finally noisey params, 
\[
 \theta' = max(\theta + N, 0)
\]

QFL is in some way similar to classical FL in that clients and devices train locally with their dataset collaboratively and contribute to the global model while keeping the data private at the local device.
However, there are multiple aspects of the QFL setup.
QFL has some peculiar features different from the classical approach of learning.
First, we need to encode the classical dataset into a quantum representation.
The main goal with the federated averaging version of QFL is such that 
\[
\min_{\theta} \; \sum_{k=1}^{K} \frac{n_k}{n} \, \mathcal{L}_k(\theta)
\]
where, $\mathcal{L}_k(\theta)$ is the local loss in client $k$ using its parameterized quantum circuit, $n_k$ is the number of samples in client $k$ as $n = \sum_{k=1}^K n_k$.
Whereas, at each local client evaluates a quantum loss function using a parameterized quantum circuit $U(\theta)$ as, 
\[
\mathcal{L}_k(\theta) = \mathbb{E}_{x \sim \mathcal{D}_k} \left[ \ell\left( f_\theta(x), y \right) \right]
\]
where, $f_\theta(x)$ is the output of the quantum model (eg. expectation value from quantum measurement) and $l$ is a classical loss function (like MSE or cross-entropy).
For parameter aggregation, we have the following. 
\[
\theta^{(t+1)} = \sum_{k=1}^{K} \frac{n_k}{n} \, \theta_k^{(t)}
\]
where, $\theta_k^{(t)}$ is the locally trained parameter at round $t$ on client $k$.

The following steps are carried out in the proposed framework.
Suppose that we have model weights as a vector $\theta \in \mathbb{R}^n$ where 
$\theta = (a_1, a_2, ... , a_n)$.
For DP, we produce a perturbed version $\theta'$ by adding a noise vector $\eta \in \mathbb{R}^n$ as, 
\[
\theta' = \theta + \eta
\]
Two popular approaches used for the choice of the noise distribution for $\eta$ are the Laplace mechanism (for $\epsilon-DP$) and the Gaussian mechanism for $(\epsilon, \delta)-DP$.
With the Laplace mechanism,  we have 
\[
\Delta_1 = max ||\theta - \theta'||_1
\]
where, $\Delta_1$ is the $l_1$ sensitivity of the weights, i.e., the maximum $l_1$ norm bound on how much $\theta$ can vary under perturbations.
Each component $n_i$ (for i=1 to n) is drawn independently from a Laplace distribution as, 
\[
\eta_i \sim Lap(0, \frac{\Delta_1}{\epsilon})
\]
such that 
\[
p(x | 0, b) = \frac{1}{2b} exp(-\frac{|x|}{b}), 
\]
with scale $b = \frac{\Delta_1}{\epsilon}$

With Gaussian mechanism, suppose that $\Delta_2$ is the $l_2$ sensitivity of the weights, i.e. the maximum $l_2$
norm bound on variables in $\theta$ as, 
\[
\Delta_2 = max ||\theta - \theta'||_2.
\]
Then, the noise vector is multivariate Gaussian as 
\[
\eta \sim \mathcal{N}(0, \sigma^2I_n)
\]
with variance calibrated as, 
\[
\sigma \geq \frac{\Delta_2\sqrt{2ln(1.25/\delta)}}{\epsilon}
\]

Let $\theta \in R^d_{\geq 0}$ denote the vector of model parameters, 
where $d$ is the dimension corresponding to the flattened shape of the input array $params$.
Let $\Delta$ be the sensitivity parameter, $\epsilon > 0$ be the privacy budget, $m \in \{laplace, gaussin\}$ be the noise 
mechanism, and $\delta$ be the $decimals$ be the number of decimals for rounding.
Then, operation on the model parameters can be summed as 
\[
\tilde{\theta} = R_\delta \circ C_0 \left( \theta + \eta \right),
\]
where, 
 $\eta \in \mathbb{R}^d$ is the noise vector, 
$C_0 : \mathbb{R}^d \rightarrow \mathbb{R}^d_{\geq 0}$ is the component wise clipping operator such that 
$[C_0(x)]_i= max(x_i,0)$ for $i=1,...,d$.
$R_\delta: \mathbb{R} \rightarrow \mathbb{R}^d$ is the component wise rounding operator to $decimal$ places.

Assuming $\Delta$ is chosen appropriately, as the sensitivity of the parameters $\theta$, the mechanisms provide differential privacy.
Both the mechanism with or without clipping and rounding preserves the privacy of DP.
Since clipping is a deterministic, data-independent function that maps to the non-negative, and rounding introduces quantization which is also similar but with finite-precision arithmetic, rounding can slightly amplify the effective privacy loss



QKD enables two parties to share cryptographic keys with provable security, even when the quantum communication channel is potentially insecure \cite{korzhProvablySecurePractical2015}. 
Building on this, we extend the unconditional security proof of the BB84 QKD protocol to cover any arbitrary attack permitted by quantum mechanics, under the assumption that the eavesdropper possesses unbounded computational resources and full quantum capabilities, as formulated in \cite{bihamProofSecurityQuantum2000}.
From \cite{bihamProofSecurityQuantum2000}, assumptions for the security proof of QKD include:
\begin{assumption}[Secure Classical Channel]
    The client and server share a classical communication channel that is immune to jamming or is authenticated using a short, pre-shared secret key.
\end{assumption}
The classical channel cannot be jammed or disrupted by an adversary, ensures messages sent over the classical channel are genuine and not tampered with.
\begin{assumption}[Adversary Limitations]
The adversary can attack the quantum channel and eavesdrop on all classical channel communications but 
cannot compromise the security of the client's or server's laboratories.
\end{assumption}
This means that the adversary can attack the quantum communication channel and passively listen to all communication on the classical channel, but the physical setup and devices (where quantum states are prepared) of the client and the server are secure.
\begin{assumption}[Qubit-Based Communication]
The sender transmits quantum information using qubits, which are two-state quantum systems.
\end{assumption}
Transmitting information using qubits which are quantum systems with two possible states.

\begin{assumption}[Unlimited Adversary Technology]
    The adversary has access to unlimited quantum technology, including quantum memory and quantum computing capabilities.
\end{assumption}
This assumption states that the adversary has access to unlimited quantum technological resources, including memory and quantum computers.

In the BB84 protocol, the sender and recipient use four possible quantum states, known as BB84 states, distributed over three distinct bases. This includes employing the ``spin" notation and its association with the ``computational basis" notation as $|0_z\rangle = |0\rangle$, $|1_z\rangle = |1\rangle$, $|0_x\rangle = \frac{1}{\sqrt{2}}(|0\rangle + |1\rangle)$
and $|1_x\rangle = \frac{1}{\sqrt{2}}(|0\rangle - |1\rangle)$.

In this context, after the sender has transmitted states and compared bases, 
a shared key is produced if both the sender and receiver utilize the same basis. 
Following this, a shifted key is established which is subsequently used to formulate the final key.

To model the potential threat, we can have general attack, sequential attack, etc.
With general attack, an adversary may attack qubits by initially executing a unitary
transformation on both the sender's qubit and its probe. 
This probe is retained in memory and utilized only once all classical data, including bases of all bits, 
choices of test bits, and test bit values, are obtained from both the sender and receiver.
Using this information, an optimal measurement is carried out on the probe to extract as much information as possible about the resulting secret key.
For $Q_i$ quantum systems that send the sender to the receiver, 
Adversary may coherently intercept all $n$ systems and apply an arbitrary quantum channel $\mathcal{A}: Q^n \rightarrow EQ^n$ (General Attack), 
subsequently forwarding the $Q^n$ output systems ($n$ quantum systems) to the legitimate receiver \cite{metgerSecurityQuantumKey2023}.
With the following assumption, 
\begin{assumption}[Single Quantum System Constraint]
    The adversary can possess only one quantum system $Q_i$ at a time.
\end{assumption}
This implies that the adversary cannot simultaneously hold or manipulate multiple quantum systems during the attack on the quantum communication channel.

Regarding a sequential attack strategy, the sender dispatches the systems $ Q_1, Q_2, 
\ldots, Q_n $ in a specific sequence, necessitating that adversary's attack also follow the same sequence. 
Consequently, the attack is characterized by a series of mappings $ A_i: E'_{i-1} Q_i \to E'_i Q_i $, where $ E'_{i-1} $ represents the adversary's knowledge prior to interacting with $ Q_i $, $ E'_i $ denotes her updated information post-processing of $ Q_i $, and $ Q_i $ is forwarded to the receiver after the adversary's intervention.

For vQFL setting with $N$ clients, $T$ communication rounds, the full attack can be given by the tensor product map, 
\[
\mathcal{A}^{\otimes n}: Q^n \rightarrow EQ^n  \quad \forall N, T
\]

The security and robustness of QKD can be examined in more detail as follows.  
Conditioned on passing the test, adversary’s information about sender’s key becomes exponentially small, such as
\[
I(A; E | T = \text{pass}) < A_{\text{info}} e^{-\beta_{\text{info}} n}
\]
where, $I(A; E \mid T = \text{pass})$ denotes the mutual information between the sender's key $A$ and the adversary's knowledge $E$, conditioned on the event that the test is passed, 
and $n$ represents the number of qubits used in the protocol.
However, this condition is not satisfied in the presence of attacks such as the SWAP attack. Similarly
, it also fails under a half-SWAP attack, 
in which the adversary, with probability $\frac{1}{2}$, leaves the sender's qubit unchanged so that it reaches the receiver without disturbance, causing the test at the receiver's end to pass with high probability.
With the remaining half of the probability, adversary executes a SWAP attack: 
it intercepts the qubits sent by sender, transmits randomly prepared states to receiver instead, and only later performs measurements on sender’s original qubits.
So, on average, $I(A;\varepsilon) = \frac{m}{2}$ and
$P(T=pass) \geq \frac{1}{2}$.

Accordingly, the appropriate security requirement should bound the joint probability that both undesirable events occur simultaneously: namely, 
that adversary obtains non-negligible information ($ I_{\text{adversary}} \geq A_{\text{info}} e^{-\beta_{\text{info}} n} $) and that the protocol still passes the test (T = pass).
Thus, the security criterion is \cite{bihamProofSecurityQuantum2000}:
\[
P \left[ (T = \text{pass}) \land (I_{\text{adversary}} \geq A_{\text{info}} e^{-\beta_{\text{info}} n}) \right] < A_{\text{luck}} e^{-\beta_{\text{luck}} n}
\]
where,  
\( T \) denotes the outcome of the test and  
\( I_{\text{adversary}} = I(A; E \mid i_T, C_T, b, s) \) represents 
the amount of information adversary possesses about the key, given the protocol parameters \( (i_T, j_T, b, s) \) that are revealed by sender and receiver.
The test is considered passed when \( C_T = i_T \sim j_T \) satisfies \( |C_T| \leq n_{\text{allowed}} \). 
Sender and receiver can enhance security by choosing a larger number of bits \( n \).
In that case, the probability that the protocol both passes the verification test and leaves adversary with substantial information decreases exponentially.
Moreover, regarding reliability, the likelihood that the final keys held by each client and the Server disagree is exponentially small, namely
\[
P(A \neq B) \leq A_{rel}e^{-B_{rel}n}
\]

The most general strategy available to adversary is to couple an ancillary quantum probe to the transmitted qubits, apply a global unitary operation $U$ on the combined system of all qubits and her probe, and then forward the resulting, potentially disturbed, qubits on to the receiver.
Subsequently, once the sender and receiver have disclosed all classical information, it performs the optimal measurement on her probe.
The unitary operator \(U\) is expressed as follows.
\[
U(|0\rangle|i\rangle) = \sum_j |E'_{i,j}\rangle|j\rangle
\]
where, $|E'_{i,j}\rangle$ denote the states of adversary’s probes when the sender transmits $|i\rangle$ and the receiver obtains $|j\rangle$, 
$i$ is a bit string that the sender encodes using a chosen base set $b$, and the receiver measures a string $j$ in the same bases; $|0\rangle$ represents a probe initialized by an adversary in some unknown state.

For QKD to be secure, it satisfies the following:
\begin{enumerate}
    \item \textbf{Correctness}: If the protocol does not abort, the sender’s key $K$ and the receiver’s key $K'$ are equal with high probability, such that
    \[
    \Pr[K \neq K' \mid \text{no abort}] \leq \varepsilon_{\text{cor}}.
    \]
    The probability of the keys being different, since the protocol does not abort, is bounded by a small error term $\varepsilon_{cor}$. This accounts for the robustness against adversarial interference that might cause discrepancies between keys.
    \item \textbf{Secrecy}: The adversary’s knowledge of the final key is negligible, with the trace norm distance between the actual state $\rho_{KE|\Omega}$ (conditioned on no abort) and the ideal state $\tau_K \otimes \rho_{E|\Omega}$, where $\tau_K$ is the maximum mixed state in the system $K$ and $\Omega$ is the event without abort, bounded by a small error term $\varepsilon_{sec}$
    \[
    \|\rho_{KE|\Omega} - \tau_K \otimes \rho_{E|\Omega}\|_1 \leq \varepsilon_{\text{sec}}.
    \]
    This property guarantees that the adversary gains negligible information about the final key.
    Secrecy is quantified by comparing the actual quantum state of the key and adversary’s system ($ \rho_{KE|\Omega} $, conditioned on the protocol not aborting) to an ideal state where the key is maximally mixed ($ \tau_K \otimes \rho_{E|\Omega} $).
    \item \textbf{Completeness}: For a given noise model, there exists an honest adversary behavior such that the protocol aborts with a probability that is small bounded by $\varepsilon_{comp}$ as
    \[
    \Pr[\text{abort}] \leq \varepsilon_{\text{comp}}.
    \]
    This property ensures that the protocol is robust against noise in the communication channel.
    This also ensures that the protocol is practical and is likely to succeed under realistic conditions.
\end{enumerate}

Thus, the QKD protocol is $(\varepsilon_{\text{cor}} + \varepsilon_{\text{sec}}/2)$-secure.

\begin{figure*}
    \centering
    \includegraphics[width=0.7\linewidth]{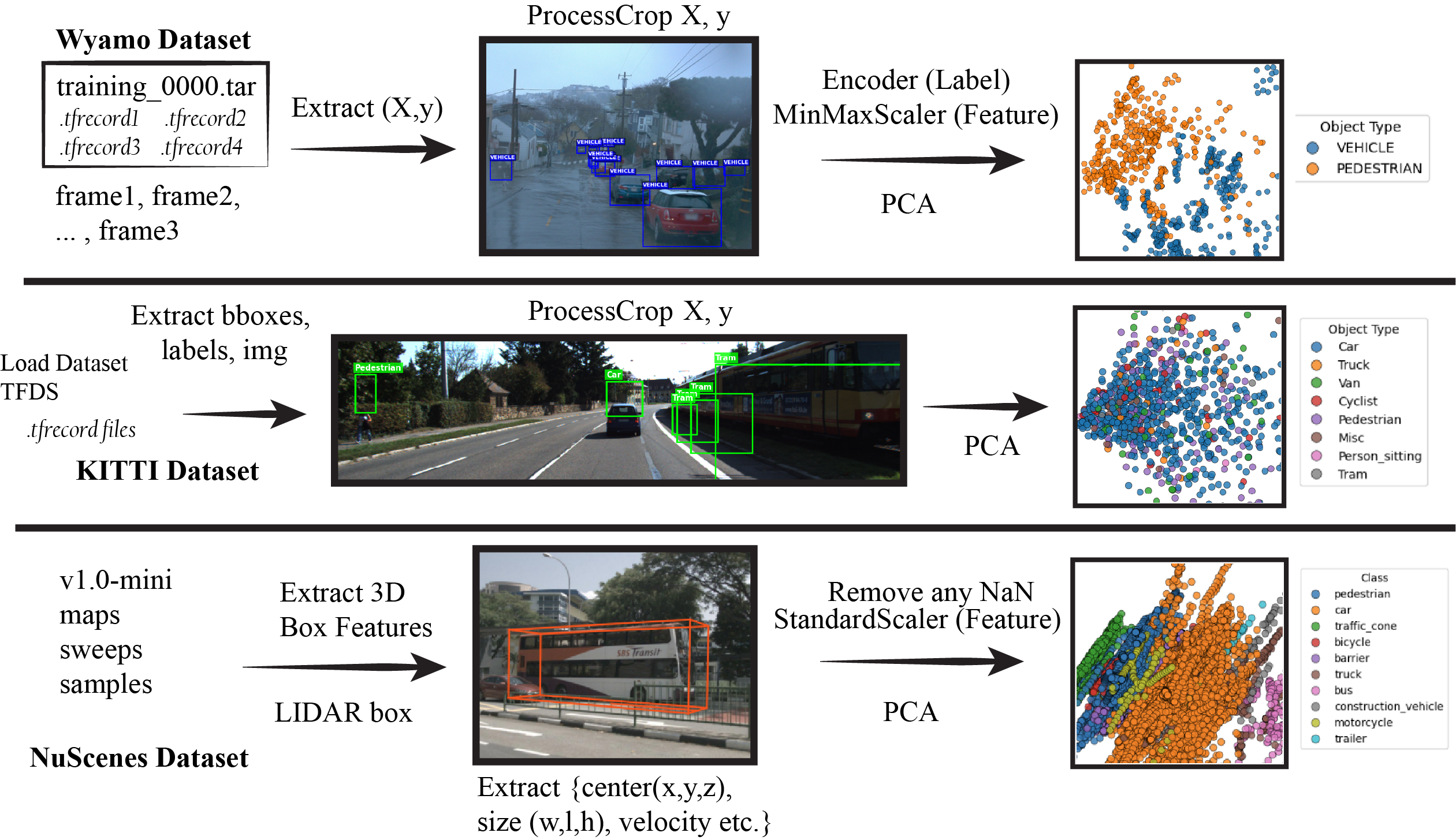}
    \caption{These SOTA datasets are huge and challenging to process. A brief process use to perform experimental analysis is shown for all datasets - Waymo, KITTI and nuScenes.}
    \label{fig:data_processing}
\end{figure*}

\section{Experimental Analysis}
This section highlights the method we used to prepare the datasets, define our analysis metrics, and various results, including ablation studies.
\subsection{Datasets.}
In terms of dataset, we utilize three state-of-the-art datasets: Waymo dataset\footnote{https://github.com/waymo-research/waymo-open-dataset}, KITTI dataset\footnote{https://www.cvlibs.net/datasets/kitti/} and nuScenes\footnote{https://github.com/nutonomy/nuscenes-devkit} dataset.

\begin{figure}[!htbp]
    \centering
    \begin{subfigure}{0.45\columnwidth}
        \centering
        \includegraphics[width=\linewidth]{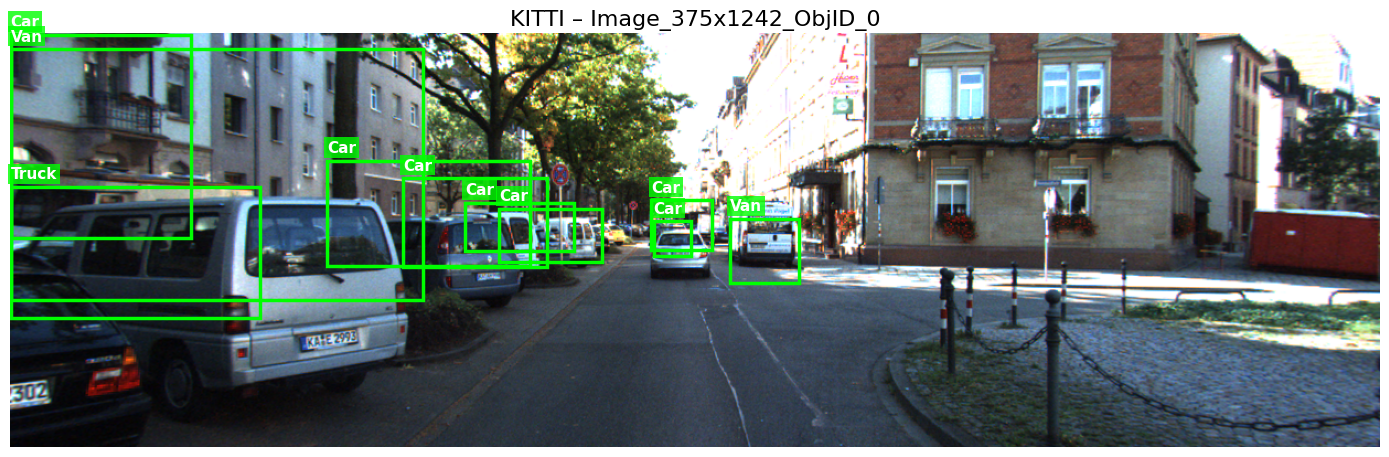}
        \caption{Random Sample 1: 10 Objects Detected  - Objects Car, Van Truck etc.}
        \label{fig:kitti_sample_image}
    \end{subfigure}
    \begin{subfigure}{0.45\columnwidth}
        \centering
        \includegraphics[width=\linewidth]{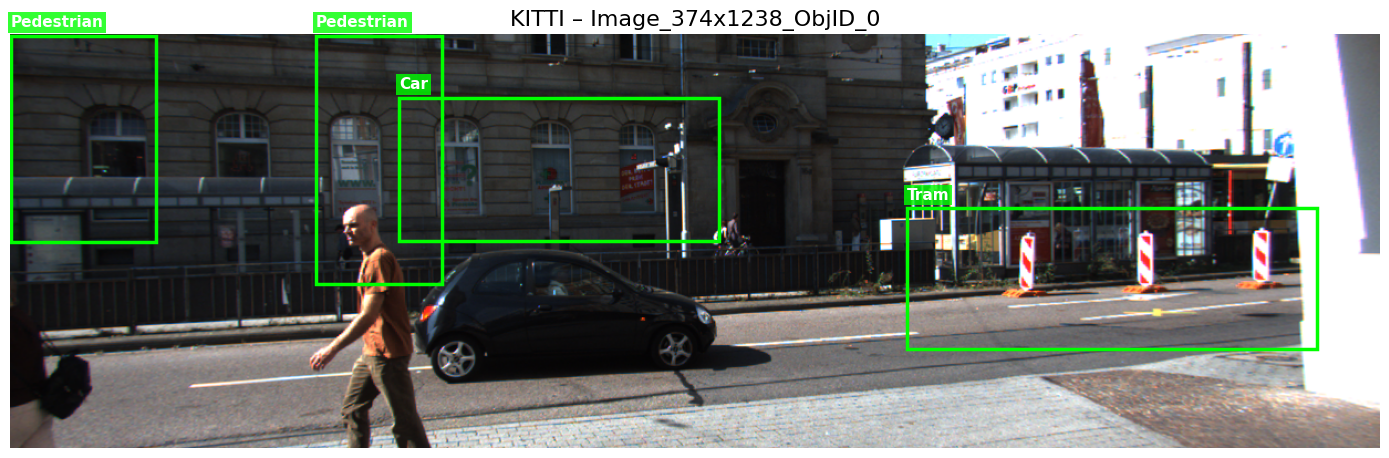}
        \caption{Random Sample 2: 4 Objects Detected - Tram, Car, Pedestrian etc.}
        \label{fig:kitti_sample_image_id0}
    \end{subfigure}
    \begin{subfigure}{0.4\columnwidth}
        \centering
        \includegraphics[width=\linewidth]{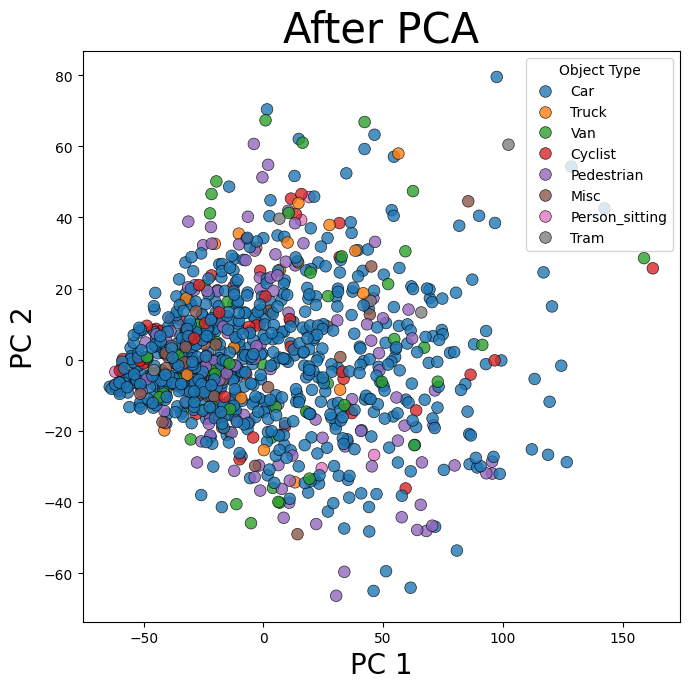}
        \caption{PCA; 1000 Samples}
        \label{fig:kitti_pca_100_samples}
    \end{subfigure}
    \begin{subfigure}{0.4\columnwidth}
        \centering
        \includegraphics[width=\linewidth]{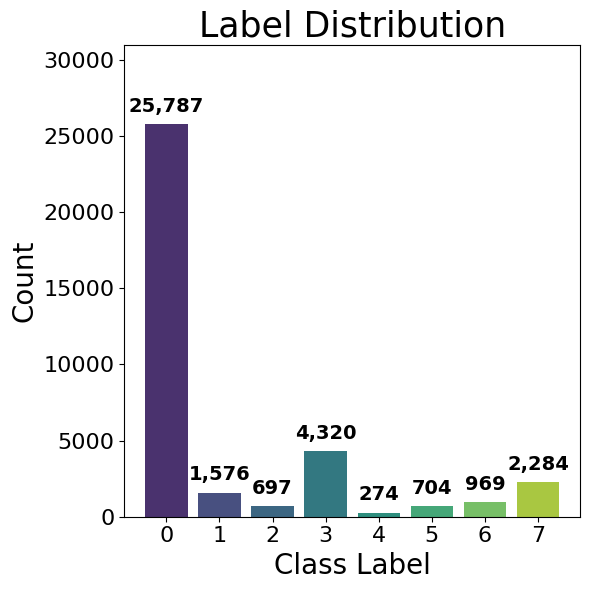}
        \caption{Label Distribution}
        \label{fig:kitti_label_distribution}
    \end{subfigure}
    \caption{KITTI Dataset: (a, b) Sample images with bounding boxes, (c) PCA figure and (d) Label Distribution.}
    \label{fig:kitti_dataset}
\end{figure}

\textbf{KITTI.} 
This dataset comprises a collection of vision tasks developed using an autonomous driving platform. 
It includes an object detection dataset featuring monocular images and bounding boxes, with 7,481 training images annotated with 3D bounding boxes.
It is an autonomous driving dataset developed by the Karlsruhe Institute of Technology, Germany, and Toyota Technological Institute at Chicago, USA \cite{geigerVisionMeetsRobotics2013}.
We utilized TensorFlow Datasets to extract/download the dataset.
Object type labels are Car, Van, Truck, Pedestrian, Person\_sitting, Cyclist, Tram, and Misc which are represented from numbers 0 to 7.
Bounded boxes and labels are extracted with a specific batch size. 
The complete data set contains both both train set (6347), the test set (711), and the validation set (423).
We extract small 64 x 64 grayscale image patches (cropped) from every annotated object detected in every image.
The other steps involved normalizing the bounding box into real pixel coordinates, cropping the object tightly from left camera image, resizing to 64 x 64 pixels using smooth bilinear resizing which are converted to grayscale, flattened into 4096 length vector and the patches saved with their class names.
The train dataset is distributed among the devices while the test dataset is used for the server device.
We follow the StandardScaler or MinMaxScaler and PCA transformation with label encoding applied in some cases, as shown in Figure \ref{fig:data_processing}.
After crop patching, we have 33414 patches for training and 3197 patches for validation.
However, we only take a small sample of randomized dataset samples for our experimental analysis, such as 2000 samples distributed among training devices, while 100 samples for server test.
In terms of label distribution, we can observe in Figure \ref{fig:kitti_label_distribution} that there is a class imbalance between label (0) which is ``car" and others. 
Figures \ref{fig:kitti_sample_image} and \ref{fig:kitti_sample_image_id0} show sample images with bounding boxes and Figure \ref{fig:kitti_pca_100_samples} shows 100 samples with the first components after PCA implementation.

\begin{figure}[!htbp]
    \centering
    \begin{subfigure}{0.4\columnwidth}
        \centering
        \includegraphics[width=\linewidth]{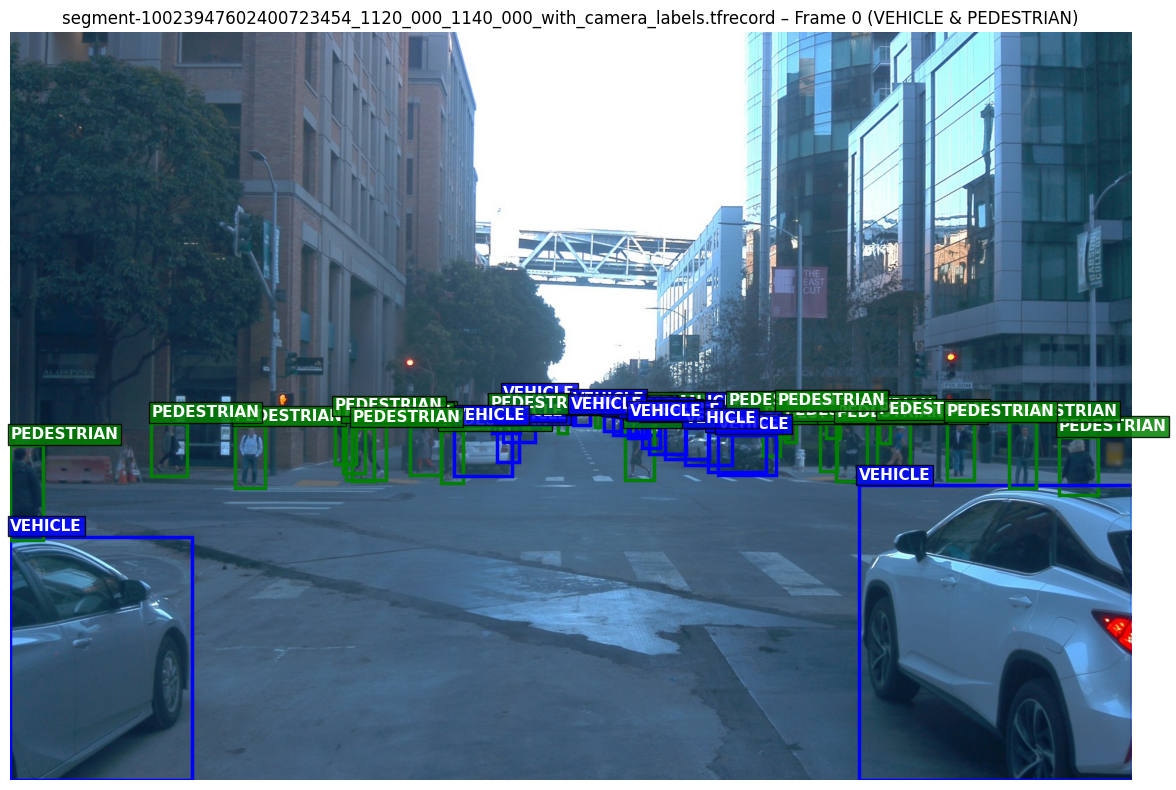}
        \caption{Frame 0}
        \label{fig:waymo_image_sample}
    \end{subfigure}
    \begin{subfigure}{0.4\columnwidth}
        \centering
        \includegraphics[width=\linewidth]{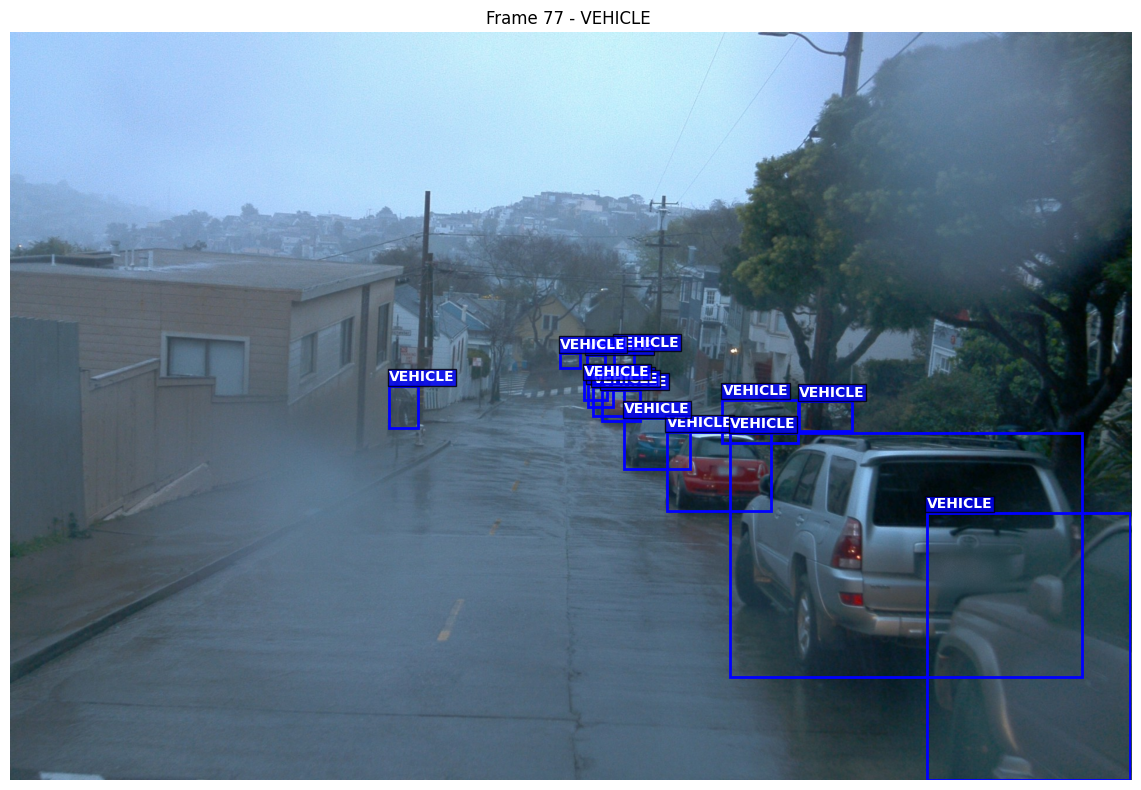}
        \caption{Frame 77}
        \label{fig:waymo77_frame}
    \end{subfigure}
    \begin{subfigure}{0.4\columnwidth}
        \centering
        \includegraphics[width=\linewidth]{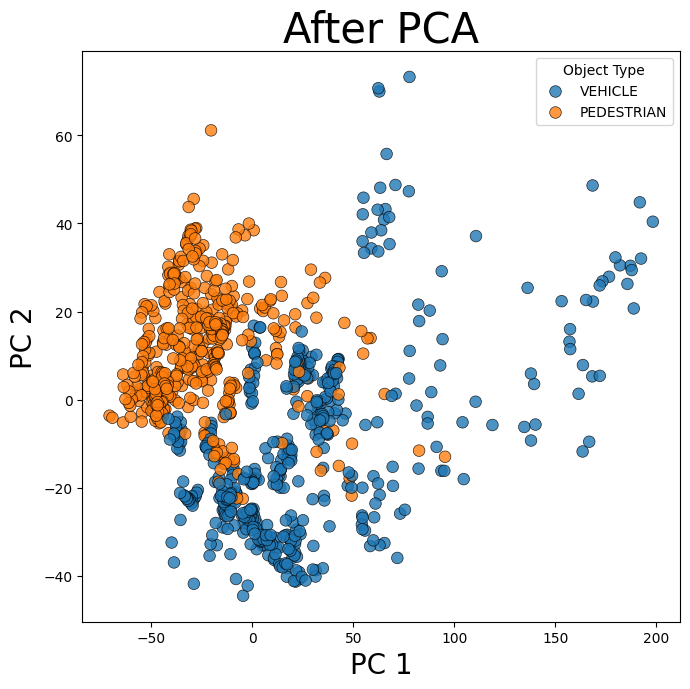}
        \caption{PCA; 100 Patches}
        \label{fig:waymo_pca_1000_samples}
    \end{subfigure}
    \begin{subfigure}{0.4\columnwidth}
        \centering
        \includegraphics[width=\linewidth]{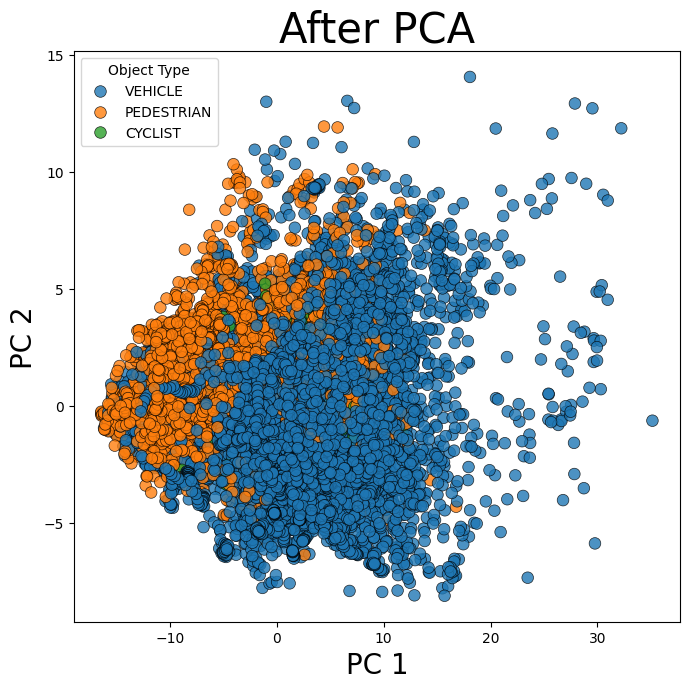}
        \caption{PCA; All}
        \label{fig:waymo_full_onthat_tar_pca}
    \end{subfigure}
     \begin{subfigure}{0.4\columnwidth}
        \centering
        \includegraphics[width=\linewidth]{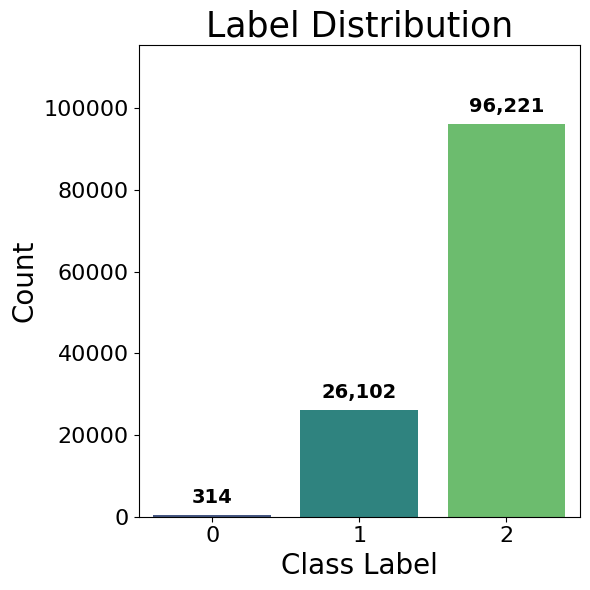}
        \caption{Label Distribution}
        \label{fig:waymo_label_distribution1}
    \end{subfigure}
    \caption{Waymo Dataset: (a, b) Sample images, (c, d) PCA 100/full, (e) Label distribution in that sample of the dataset.}
    \label{fig:waymo_dataset}
\end{figure}

\textbf{Waymo.}
Waymo Open Dataset is a large-scale, high quality diverse dataset consisting of 1150 scenes that each span 20 seconds 
\cite{SunWaymo2020CVPR}.
The data captured across a range of urban and suburban geographies are well synchronized and high quality calibrated LiDAR and camera data exhaustively annotated with 2D (camera image) and  3D (LiDAR) bounding boxes providing strong baseline for 2D and 3D detection and tracking masks.
Figures \ref{fig:waymo_image_sample} and Figure \ref{fig:waymo77_frame} show visualization for objects in frames 0 and  77, respectively, which is in the scene of Waymo dataset TFRecord file from training\_000.tar file, specially focusing on front camera images and drawing 2D bounding boxes around the detected objects.
The object types selected are vehicle, pedestrian and cyclist for visualization, which is present in a single TFRecord file extracted from a Waymo training dataset.
To process the dataset, we first load one TFRecord archived in .tar files, which reads one segment file containing multiple frames.
Each sample is an entire front image that is cropped for objects that are flattened, and each label is one type of object.
We used only small samples for our experimental analysis.
Waymo and KITTI data experiments were performed in the wsl windows environment using a local jupyter notebook, while others (also KITTI and nuScenes) were performed in the Google colab environment.
Figures \ref{fig:waymo_full_onthat_tar_pca}, \ref{fig:waymo_pca_1000_samples} and \ref{fig:waymo_label_distribution1} show the plot of PCA components 100, all sample and label distribution, respectively.
As we can observe, from that particular .tar file, we have more label 2 samples while almost very low 0 label classes, as seen in Figure \ref{fig:waymo_label_distribution1}.

\begin{figure}[!htbp]
    \centering
    \begin{subfigure}{0.4\columnwidth}
        \centering
        \includegraphics[width=\linewidth]{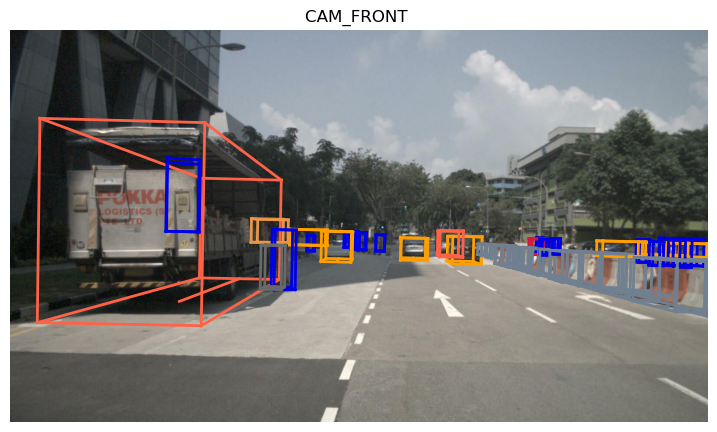}
        \caption{Sample 1}
        \label{fig:nuscenes_data_cam_front}
    \end{subfigure}
    \begin{subfigure}{0.4\columnwidth}
        \centering
        \includegraphics[width=\linewidth]{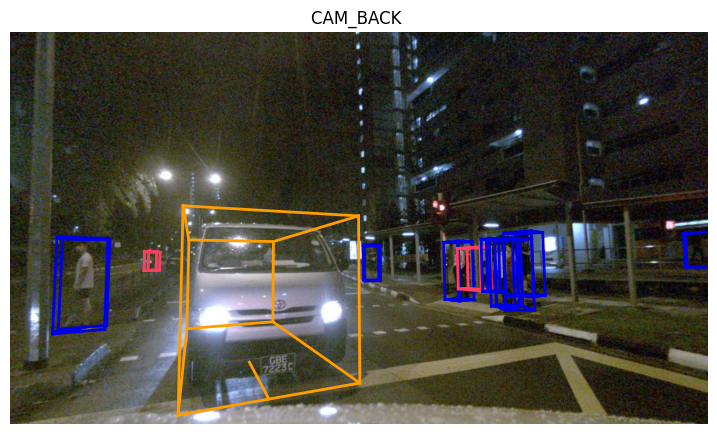}
        \caption{Sample 2}
        \label{fig:nuscenes_sample_400}
    \end{subfigure}
    
    \begin{subfigure}{0.3\columnwidth}
        \centering
        \includegraphics[width=\linewidth]{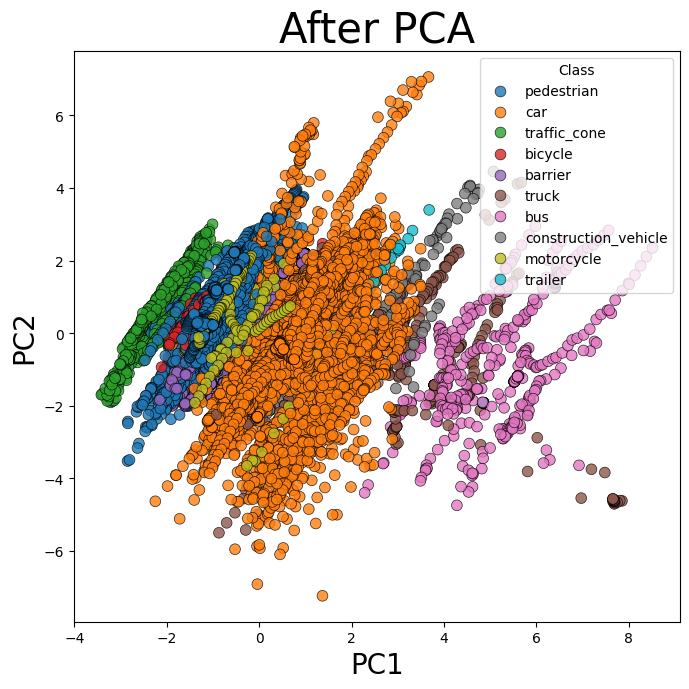}
        \caption{PCA, Standard}
        \label{fig:nuscenes_pca_full}
    \end{subfigure}
     \begin{subfigure}{0.3\columnwidth}
        \centering
        \includegraphics[width=\linewidth]{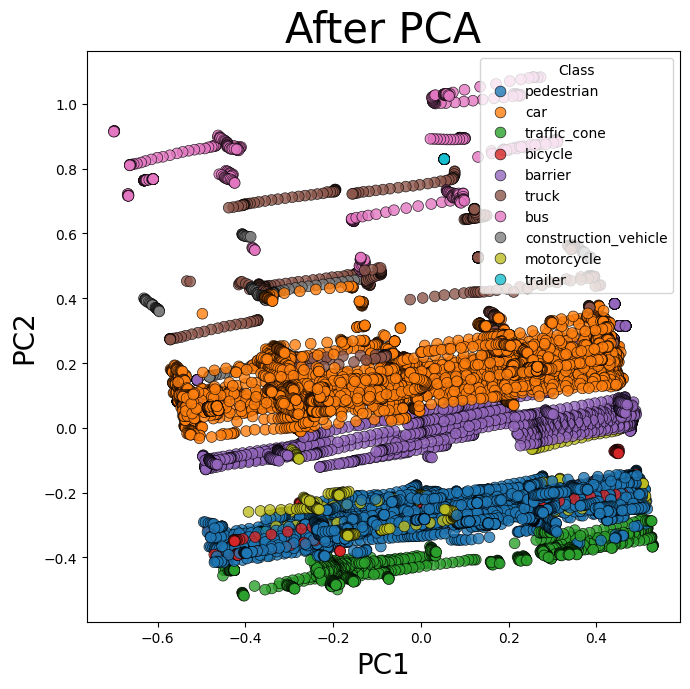}
        \caption{PCA; MinMax}
        \label{fig:nuscenes_minmaxscaler_andPCA}
    \end{subfigure}
     \begin{subfigure}{0.3\columnwidth}
        \centering
        \includegraphics[width=\linewidth]{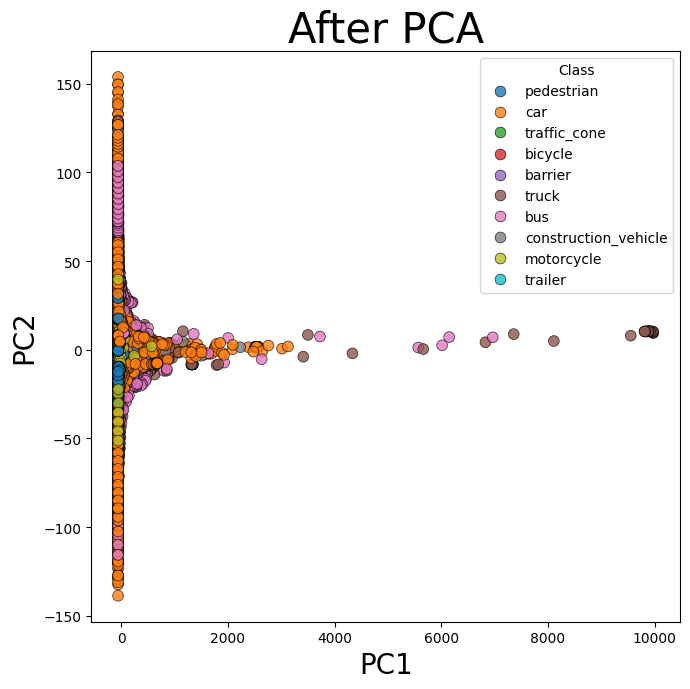}
        \caption{Only PCA}
        \label{fig:nuscenes_pca_without_scalerapplied}
    \end{subfigure}
     \begin{subfigure}[b]{0.45\columnwidth}
        \centering
       \includegraphics[width=\columnwidth]{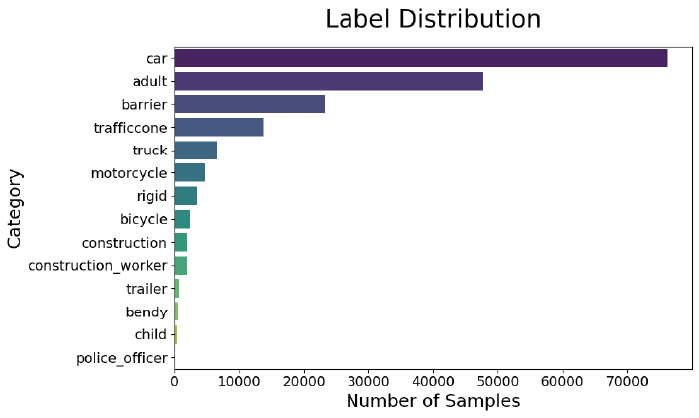}
    \caption{All Labels}
    \label{fig:nuscenes_label_distribution}
    \end{subfigure}
    \begin{subfigure}[b]{0.45\columnwidth}
        \centering
       \includegraphics[width=\columnwidth]{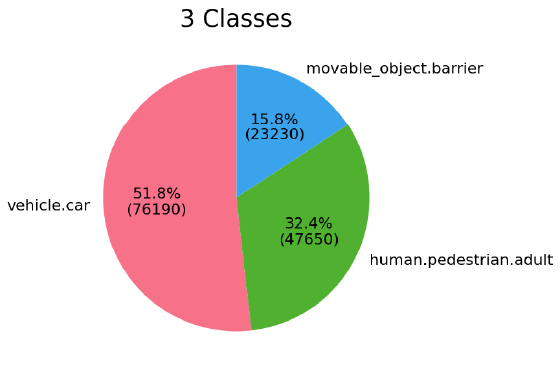}
    \caption{3 Labels}
    \label{fig:nuscenes_3labels}
    \end{subfigure}
     \caption{nuScenes Dataset v1.0-mini CAM FRONT: (a, b) samples of a scene which multiple objects and their bounding boxes with annotations; v1.0-mini, (c, d, e) Effect of PCA with standarScaler, MinMaxScaler or with no Scaler; (f) Label Distribution, (g) Only some selected label distribution.}
    \label{fig:nuscenes_data_eda}
\end{figure}

\textbf{nuScenes.}
The nuScenes dataset consists of various scenes with 23 categories, 8 attributes, 12 sensors, etc. \cite{caesarNuScenesMultimodalDataset2020}.
We utilized the v1.0-mini version of the dataset, which has 10 scenes within which there are 404 samples containing
additional sample data and annotations. 
There are various scenes within the dataset such as scene-0061 that contains parked truck, construction, etc. while scene-0103 contains many vehicles right, waiting for the car, etc.
Each sample data set contains data from RADAR\_FRONT, RADAR\_FRONT\_LEFT, LIDAR\_TOP, etc.
A sample CAM\_FRONT scene is shown in Figure \ref{fig:nuscenes_data_cam_front} which shows various vehicle types with their bounding boxes.
To extract features and labels, we first localize nuScenes data v1.0-mini.
Then, class mapping is either done to convert to numerical labels or keep the categorical labels as it is.
Then, we extract 3D box features for every annotation in every sample.
The LIDAR box is retrieved via get\_sample\_data().
The extracted features are center (x,y,z), size (w,l,h), yaw, velocity(vx, vy, vz), num\_lidar\_points.
Also, we remove any NaN rows, which is followed by StandardScaler/MinMaxScaler, PCA and train test split methods.
For this dataset, we extract input features from boxes for each annotation, which consists of `xyz: [x, y, z]' location, dimensions `wlh: [w, l, h]', orientation, velocity, and number of lidar points. 
For each annotation, we assign a category such as `human.pedestrian.adult', `vehicle.bicycle' etc. as in Table \ref{tab:class_mapping} which we converted into their numerical representation from 0 to 9 for barrier:0 to truck:9 (in some categorical labels were not changed).   
Once we extract all the annotation features and remove any `nan' values, we have altogether 185,080 samples of input feature matrix with their corresponding labels.
We also apply a test with PCA to reduce from 11 features to 4 feature dimensions and in some cases discard certain features and keep only 7 features and perform no PCA.
And also, experiment with all samples along with small sample of 1000 training samples distributed among 10 devices with 100 samples for server device for testing.

Figures \ref{fig:nuscenes_data_cam_front} and \ref{fig:nuscenes_sample_400} show the sample images while Figures \ref{fig:nuscenes_pca_full}, \ref{fig:nuscenes_minmaxscaler_andPCA} and \ref{fig:nuscenes_pca_without_scalerapplied} show PCA component plots with scalers used StandardScaler, MinMaxScaler or no Scaler, respectively.
In Figure \ref{fig:nuscenes_label_distribution}  we show the number of labels with all labels and in Figure \ref{fig:nuscenes_3labels}, we select only a few labels and show in the pie chart. 
The label class imbalance seems to be the issue while training due to which there were performance issues with the full dataset for nuScenes.

\begin{table}[h]
\centering
\caption{nuScenes class mapping; Only few classes used for some experiments}
\label{tab:class_mapping}
\resizebox{0.6\columnwidth}{!}{
\begin{tabular}{>{\ttfamily}l >{\ttfamily}l}
\toprule
\textbf{Original Class} & \textbf{Mapped Class} \\
\midrule
movable\_object.barrier             & barrier \\
vehicle.bicycle                     & bicycle \\
vehicle.bus.bendy                   & bus \\
vehicle.bus.rigid                   & bus \\
vehicle.car                         & car \\
vehicle.construction                & construction\_vehicle \\
vehicle.motorcycle                  & motorcycle \\
human.pedestrian.adult              & pedestrian \\
human.pedestrian.child              & pedestrian \\
human.pedestrian.construction\_worker & pedestrian \\
human.pedestrian.police\_officer    & pedestrian \\
movable\_object.trafficcone         & traffic\_cone \\
vehicle.trailer                     & trailer \\
vehicle.truck                       & truck \\
\bottomrule
\end{tabular}
}
\end{table}

\subsection{Tools \& Metrics}
\begin{figure}
    \centering
    \includegraphics[width=0.8\linewidth]{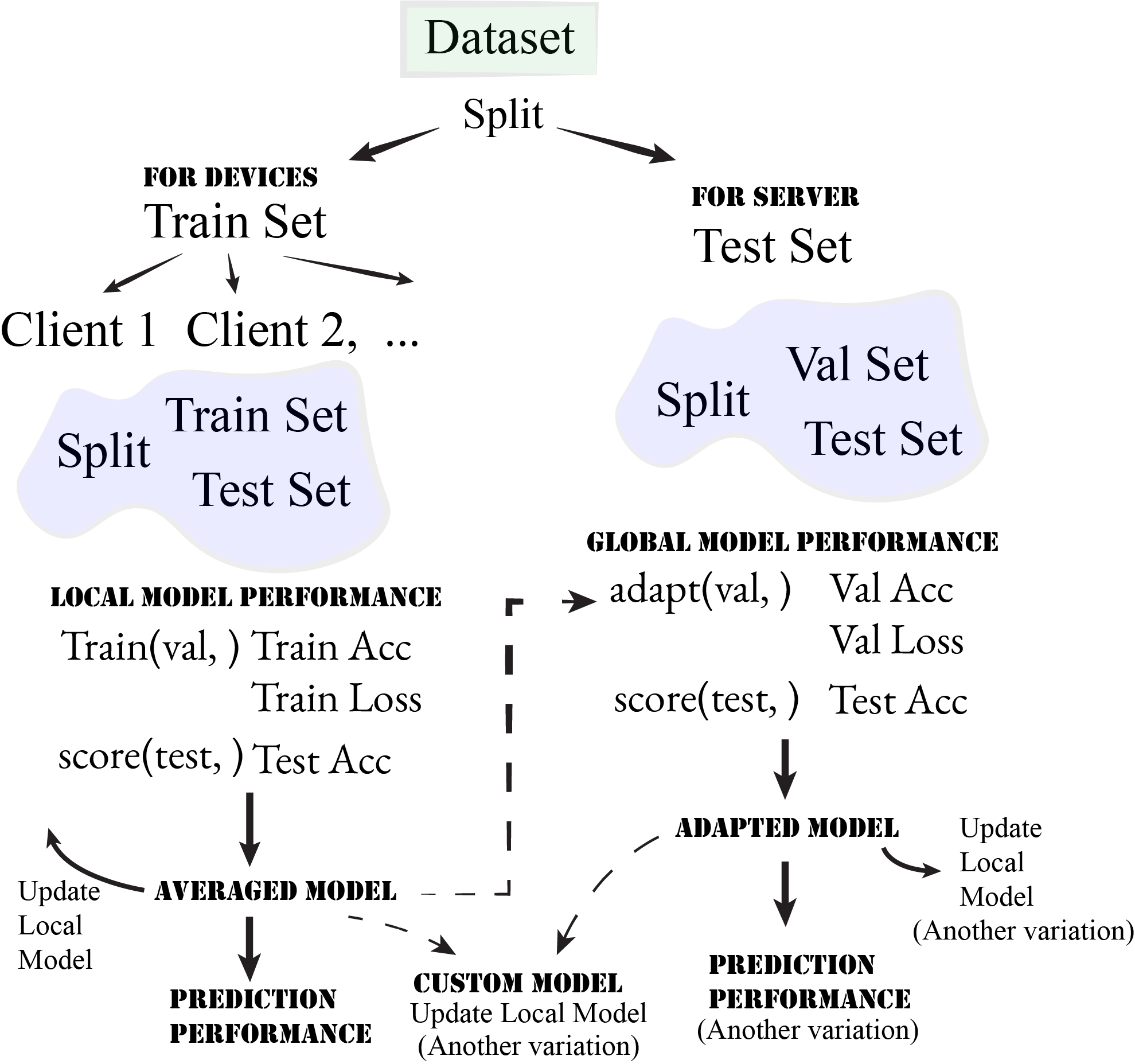}
    \caption{Data Split, Metrics Definition etc - The methods used to prepare datasets, define metrics and results in our work.}
    \label{fig:metrics}
\end{figure}

In terms of a quantum machine learning tool, we utilize the Qiskit library.
Within Qiskit library, we utilize Quantum Convolutional Neural Network (QCNN), Variational Quantum Classifier (VQC) and SamplerQNN with Network Classifier. 
The metrics for analysis are accuracy at the server and device level, the objective values of the local optimizer functions, and the total communication time required to complete each communication round, as also shown in Figure \ref{fig:metrics}.

\begin{itemize}
    \item Global FineTuned (FT) Model Performance: Global Model FT Test Accuracy, Validation Loss, and Validation Accuracy - These metrics are computed on the server device, which has a set of validation and test set dataset. 
    Once we obtain the global average model, we validate it against the server dataset.
    Here, we learn how capable the average global model is of optimizing or learning on the validation set and performing on the test set again. This is still fitting the average model on the validation set of the server which follows same optimization steps as in training in local model devices. 
    Depending upon the ablation study, this fine tuned model is only used for performance analysis and not to update devices local model unless mentioned otherwise in this work.
    \item Local Model Performance: Test Accuracy, Train Accuracy, Train Loss, Train Time - Each device has their own train set and test set dataset as well. Training accuracy tests how well the device model is able to learn with each new updated average model parameter, which also computes the train loss. Whereas, test accuracy is performance of newly trained local model before sending to server on test set at local device.
    \item Global Prediction Model Performance: Prediction test accuracy, validation accuracy, and prediction loss - This metric is prediction performance of a newly formed average global model or a customized/adapted global model depending upon the experiment and ablation study.
 The prediction validation accuracy and the loss result is in the validation set of the server data set and the prediction test result with the test set of the server dataset.
    We include these metrics to measure the performance of the global model in performing prediction accuracy on the dataset of the server test set.
    This model is different and separate from the optimized FL global model performance, and thus we term it as global prediction model.
    \item Communication Time: Communication time involves the total time required to complete one communication round.
\end{itemize}

\subsection{Models - VQC, SamplerQNN, QCNN}
\begin{figure}[!htbp]
    \centering
    \begin{subfigure}[b]{0.24\columnwidth}
        \centering
       \includegraphics[width=\columnwidth]{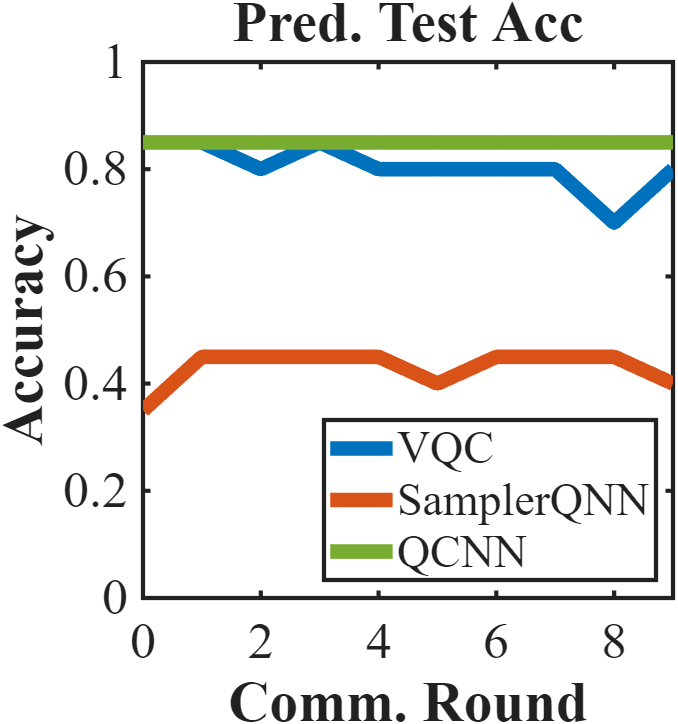}
    \caption{Pred. Test}
    \label{fig:prediction_test_acc_models_kitti}
    \end{subfigure}
    \begin{subfigure}[b]{0.24\columnwidth}
        \centering
       \includegraphics[width=\columnwidth]{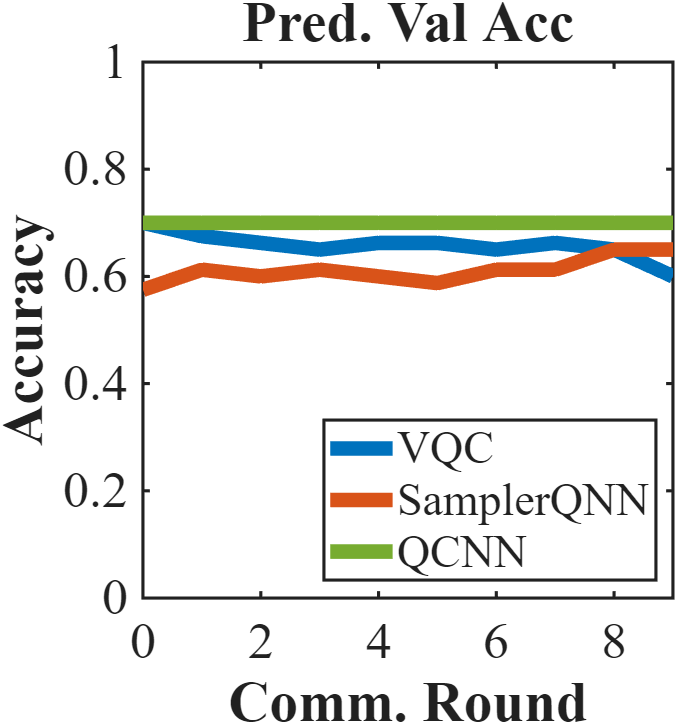}
    \caption{Pred. Val}
    \label{fig:prediction_val_acc_models_kitti}
    \end{subfigure}
    \begin{subfigure}[b]{0.24\columnwidth}
        \centering
       \includegraphics[width=\columnwidth]{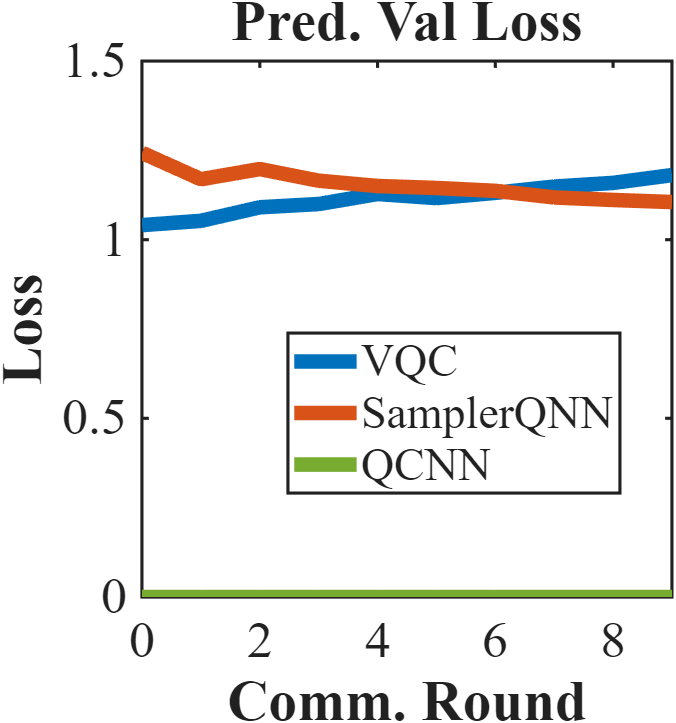}
    \caption{Pred. Loss}
    \label{fig:prediction_val_loss_models_kitti}
    \end{subfigure}
       \begin{subfigure}[b]{0.24\columnwidth}
        \centering
       \includegraphics[width=\columnwidth]{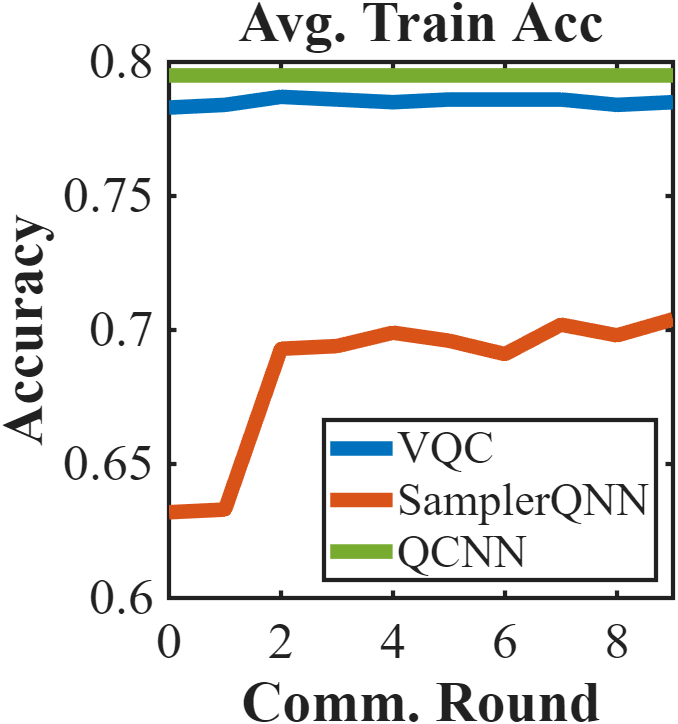}
    \caption{Avg. Train}
    \label{fig:avg_devices_train_acc_model_kitti}
    \end{subfigure}
      \begin{subfigure}[b]{0.24\columnwidth}
        \centering
       \includegraphics[width=\columnwidth]{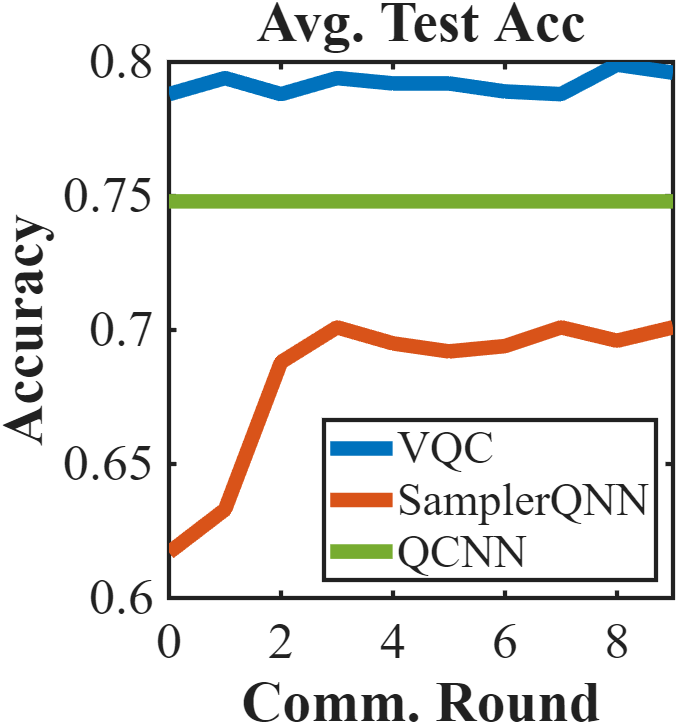}
    \caption{Avg. Test}
    \label{fig:avg_devices_test_acc_model_kitti}
    \end{subfigure}
  \begin{subfigure}[b]{0.24\columnwidth}
        \centering
      \includegraphics[width=\columnwidth]{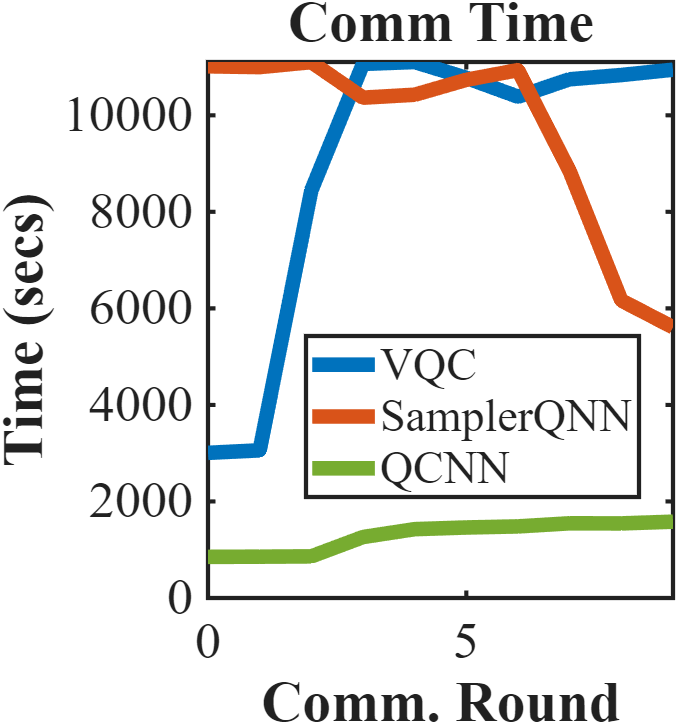}
    \caption{Time}
    \label{fig:comm_time_models_comparison_kitti}
    \end{subfigure}
    \begin{subfigure}[b]{0.24\columnwidth}
        \centering
       \includegraphics[width=\columnwidth]{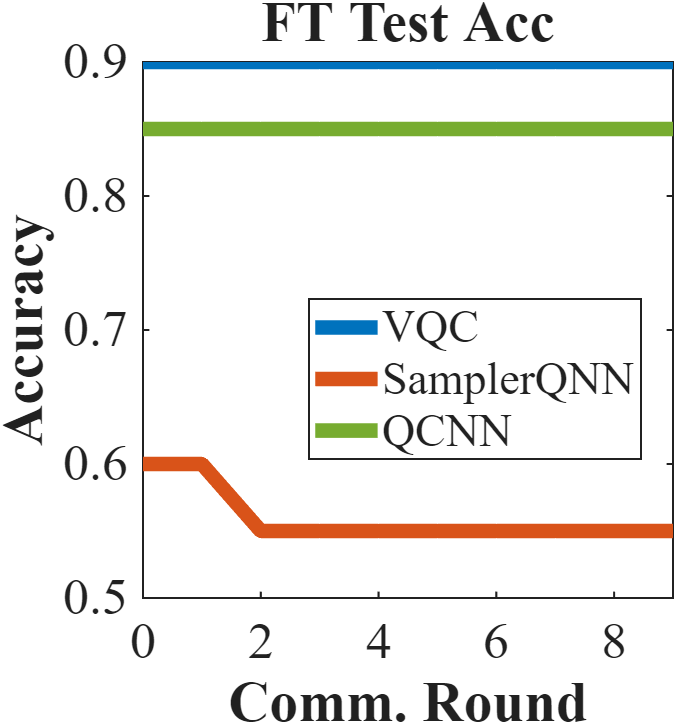}
    \caption{FT Test}
    \label{fig:global_ft_test_acc_models_kitti}
    \end{subfigure}
      \begin{subfigure}[b]{0.24\columnwidth}
        \centering
       \includegraphics[width=\columnwidth]{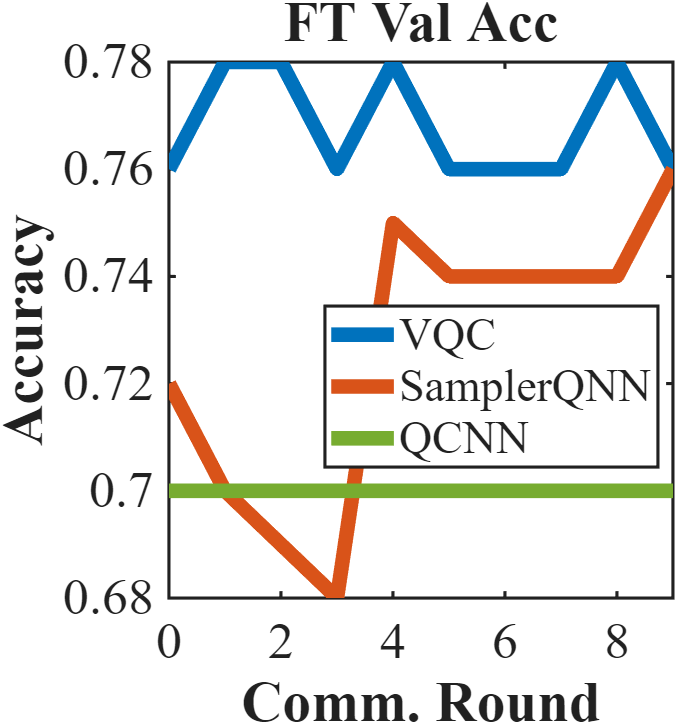}
    \caption{FT Val}
    \label{fig:global_ft_val_acc_models_kitti}
    \end{subfigure}
    \caption{Models Comparison on KITTI Dataset; Prediction Model, Fine-Tuned Model, Average Devices Performance, Communication Time}
    \label{fig:model_performance_kitti}
\end{figure}

We perform series of experiments to compare between various models like VQC, QCNN and NeuralNetworkClassifier with SamplerQNN with all datasets: KITTI, nuScenes, and Waymo.
With the KITTI dataset, we see that the prediction model performance is better with QCNN as shown in Figures \ref{fig:prediction_test_acc_models_kitti}, \ref{fig:prediction_val_acc_models_kitti} and \ref{fig:prediction_val_loss_models_kitti} for the prediction test accuracy, validation accuracy, and loss, respectively. While SamplerQNN performs worst in terms of prediction test accuracy, val accuracy, and in average device performance, as shown in Figures \ref{fig:avg_devices_test_acc_model_kitti} and \ref{fig:avg_devices_train_acc_model_kitti}.
With communication time, QCNN seems to be faster than in Figure \ref{fig:comm_time_models_comparison_kitti}.
However, in terms of fine tuning process and performance, we see better results with VQC as shown in Figures \ref{fig:global_ft_test_acc_models_kitti} and \ref{fig:global_ft_val_acc_models_kitti} for test accuracy and validation accuracy, respectively.

\begin{figure}[!htbp]
    \centering
    \begin{subfigure}[b]{0.24\columnwidth}
        \centering
       \includegraphics[width=\columnwidth]{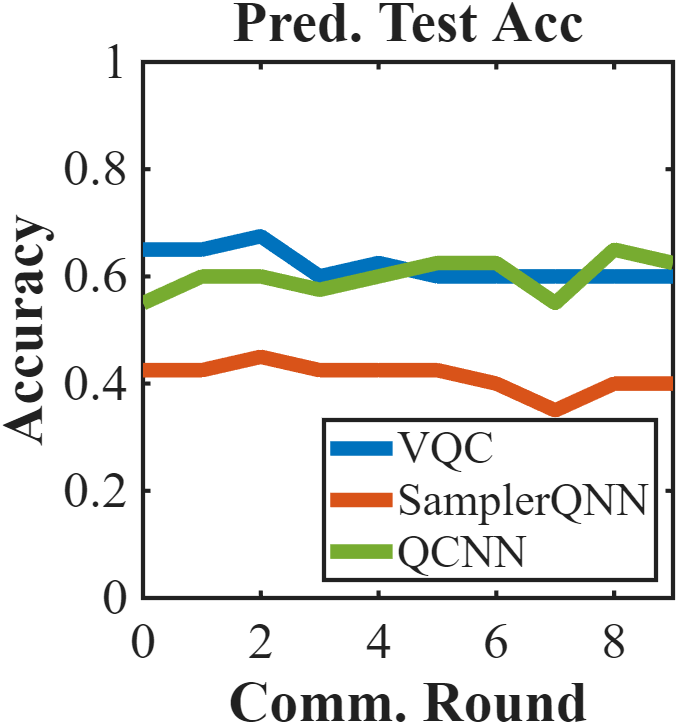}
    \caption{Pred. Test}
    \label{fig:prediction_test_acc_models_waymo}
    \end{subfigure}
    \begin{subfigure}[b]{0.24\columnwidth}
        \centering
       \includegraphics[width=\columnwidth]{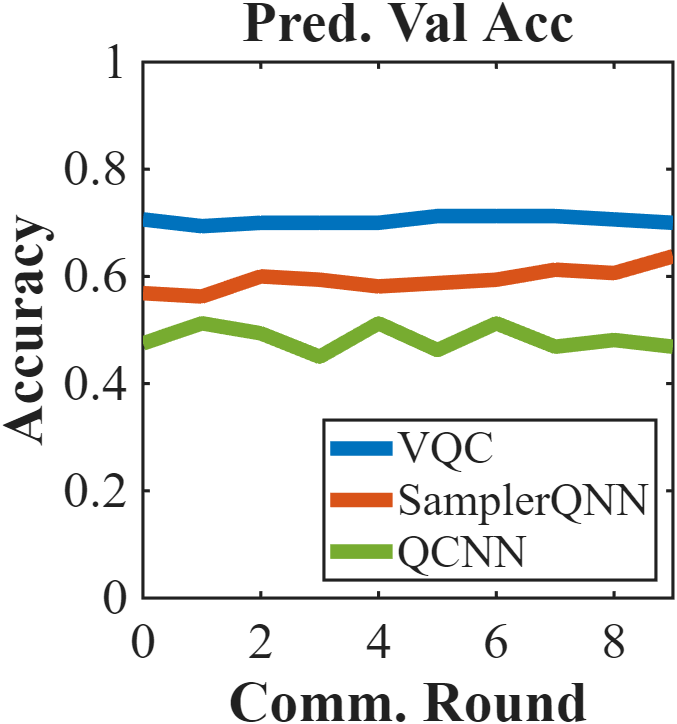}
    \caption{Pred. Val}
    \label{fig:prediction_val_acc_models_waymo}
    \end{subfigure}
  \begin{subfigure}[b]{0.24\columnwidth}
        \centering
      \includegraphics[width=\columnwidth]{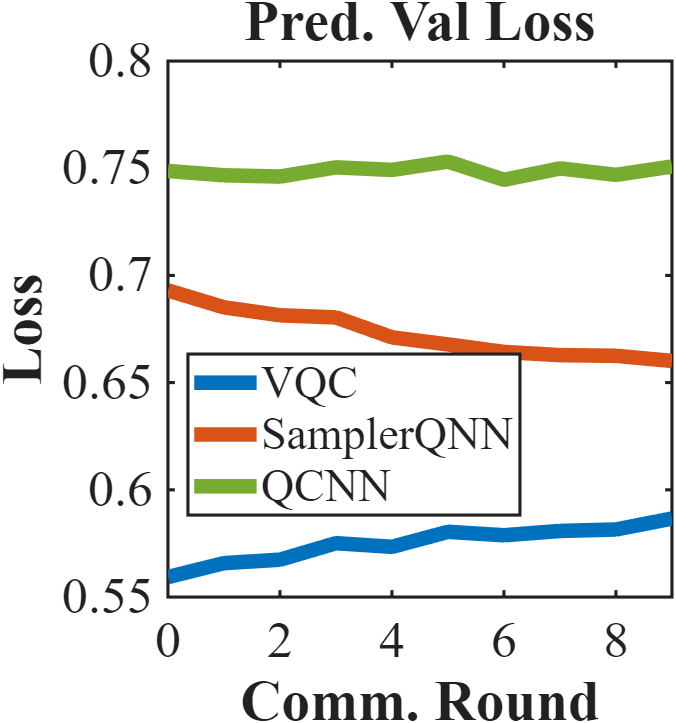}
    \caption{Pred. Loss}
    \label{fig:prediction_val_loss_models_waymo}
    \end{subfigure}
     \begin{subfigure}[b]{0.24\columnwidth}
        \centering
       \includegraphics[width=\columnwidth]{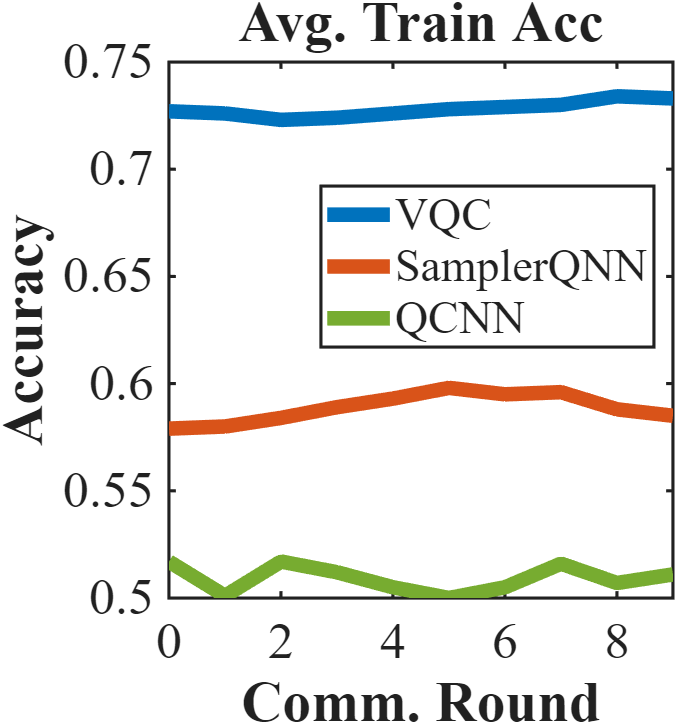}
    \caption{Avg. Train}
    \label{fig:avg_devices_train_acc_models_waymo}
    \end{subfigure}
    \begin{subfigure}[b]{0.24\columnwidth}
        \centering
       \includegraphics[width=\columnwidth]{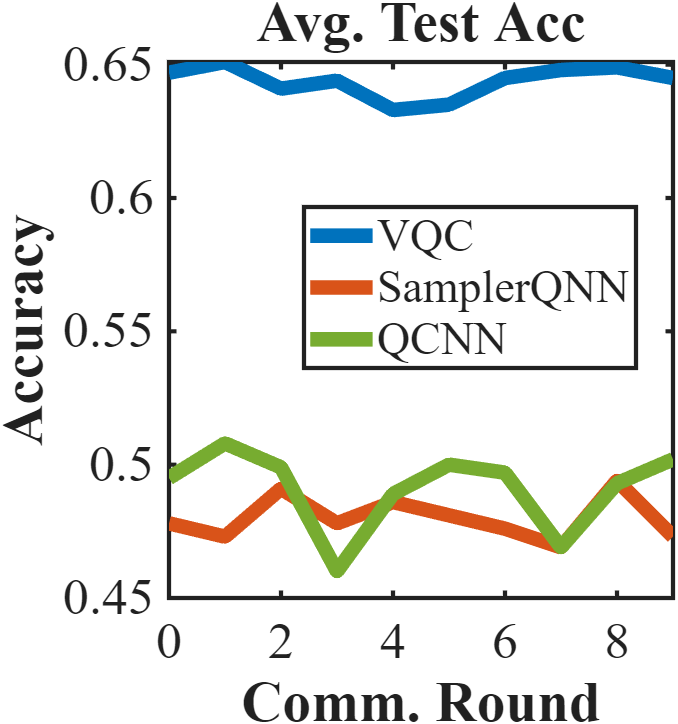}
    \caption{Avg. Test}
    \label{fig:avg_devices_test_acc_models_waymo}
    \end{subfigure}
  \begin{subfigure}[b]{0.24\columnwidth}
        \centering
      \includegraphics[width=\columnwidth]{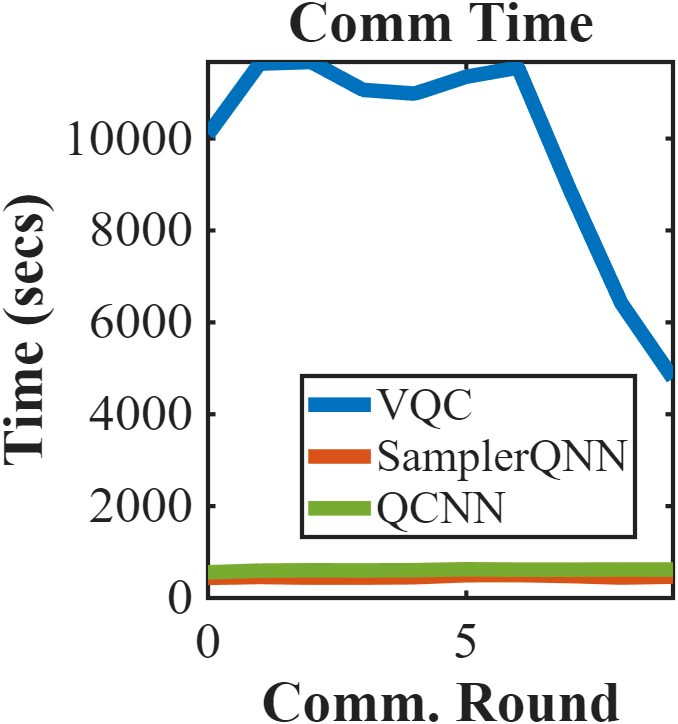}
    \caption{Time}
    \label{fig:comm_time_models_waymo}
    \end{subfigure}
     \begin{subfigure}[b]{0.24\columnwidth}
        \centering
       \includegraphics[width=\columnwidth]{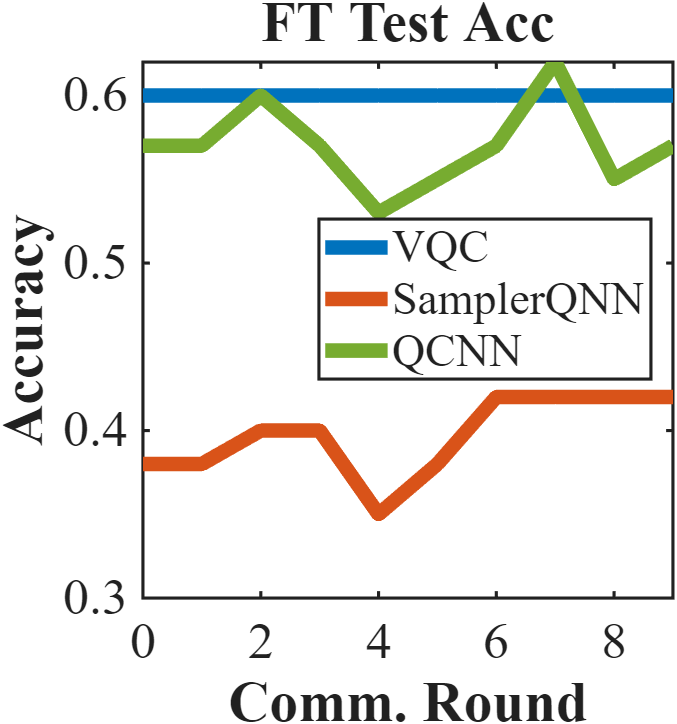}
    \caption{FT Test}
    \label{fig:global_ft_test_acc_models_waymo}
    \end{subfigure}
    \begin{subfigure}[b]{0.24\columnwidth}
        \centering
       \includegraphics[width=\columnwidth]{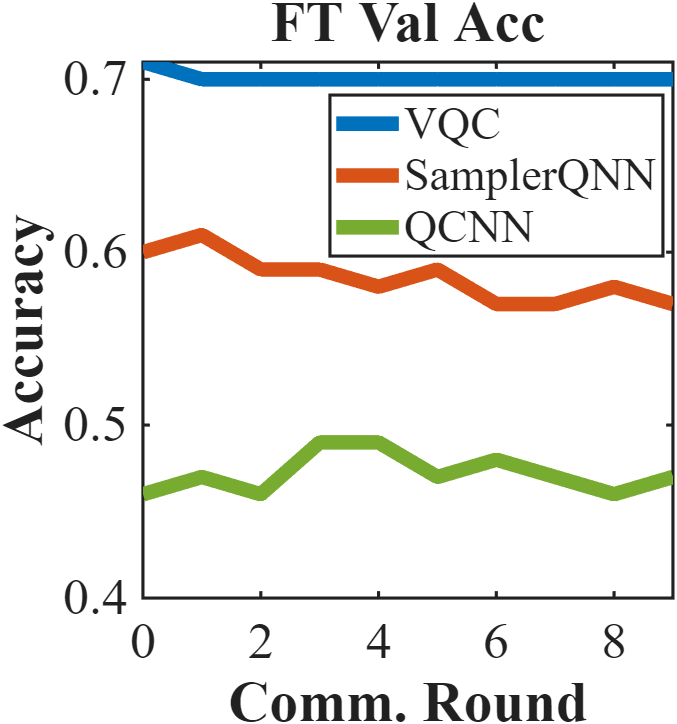}
    \caption{FT Val}
    \label{fig:global_ft_val_acc_models_waymo}
    \end{subfigure}
     \caption{Models Comparison on Waymo Dataset; Prediction Model, Fine-Tuned Model, Average Devices Performance, Communication Time}
    \label{fig:model_performance_waymo}
\end{figure}

With Waymo dataset, we only consider 2 classes, either vehicle or pedestrian.
We can observe that in Figure \ref{fig:model_performance_waymo}, QCNN performs similar to VQC and better than SamplerQNN all in terms of prediction accuracy as in Figures \ref{fig:prediction_test_acc_models_waymo} and \ref{fig:prediction_val_acc_models_waymo}, while  in Figures \ref{fig:avg_devices_train_acc_models_waymo} and \ref{fig:avg_devices_test_acc_models_waymo}, QCNN does not perform well in terms of average devices performance.
However, VQC is observed to be slower in terms of communication time as shown in Figure \ref{fig:comm_time_models_waymo}.
As waymo was experimented within wsl environment using local jupyter notebook, running multiple jupyter notebooks in parallel might have impacted the results in terms of communication time.
For fine tuned results, VQC performs overall best in both test accuracy and validation accuracy as shown in Figures \ref{fig:global_ft_test_acc_models_waymo} and \ref{fig:global_ft_val_acc_models_waymo} respectively.

\begin{figure}[!htbp]
    \centering
    \begin{subfigure}[b]{0.24\columnwidth}
        \centering
       \includegraphics[width=\columnwidth]{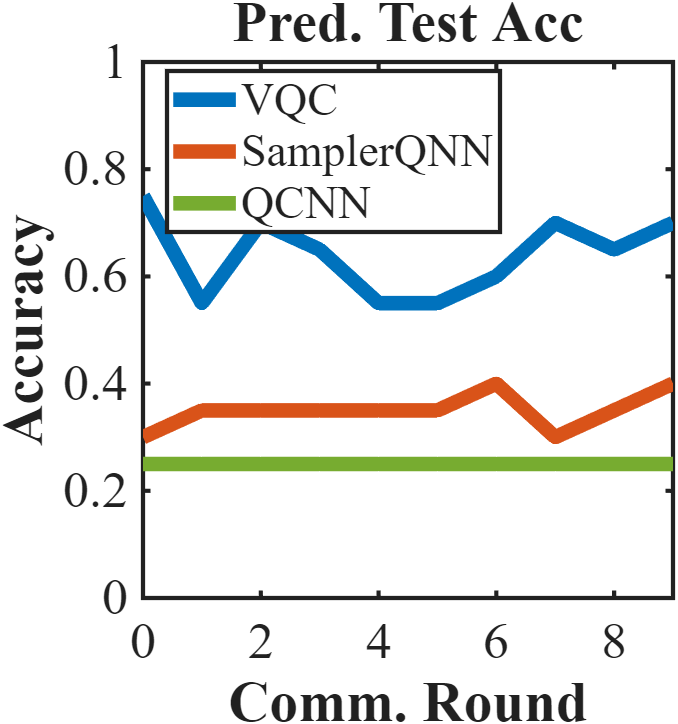}
    \caption{Pred. Test}
    \label{fig:prediction_test_acc_models_nuscenes}
    \end{subfigure}
    \begin{subfigure}[b]{0.24\columnwidth}
        \centering
       \includegraphics[width=\columnwidth]{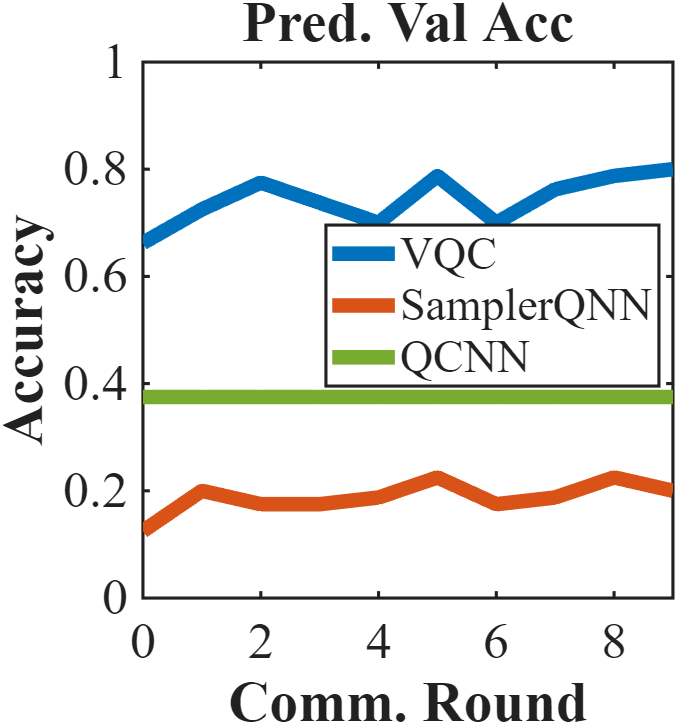}
    \caption{Pred. Val}
    \label{fig:prediction_val_acc_models_nuscenes}
    \end{subfigure}
  \begin{subfigure}[b]{0.24\columnwidth}
        \centering
      \includegraphics[width=\columnwidth]{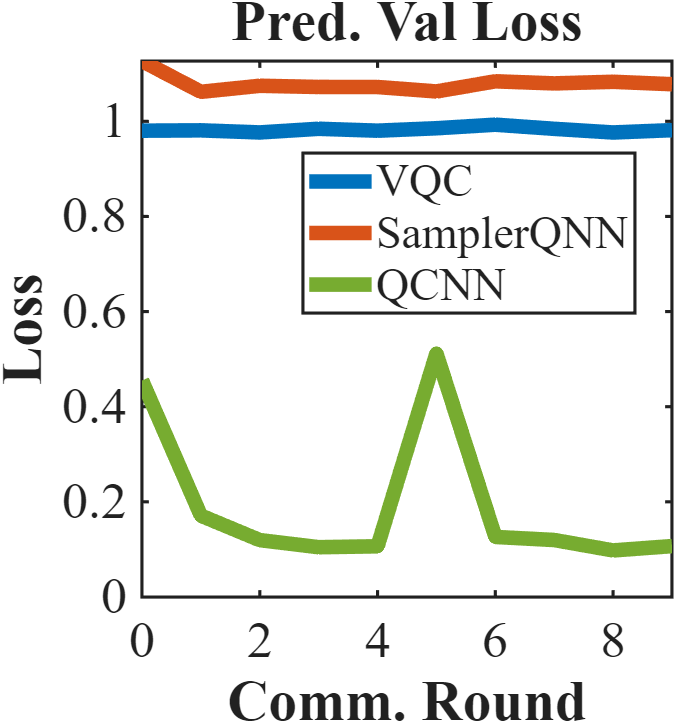}
    \caption{Pred. Loss}
    \label{fig:prediction_val_loss_models_nuscenes}
    \end{subfigure}
     \begin{subfigure}[b]{0.24\columnwidth}
        \centering
       \includegraphics[width=\columnwidth]{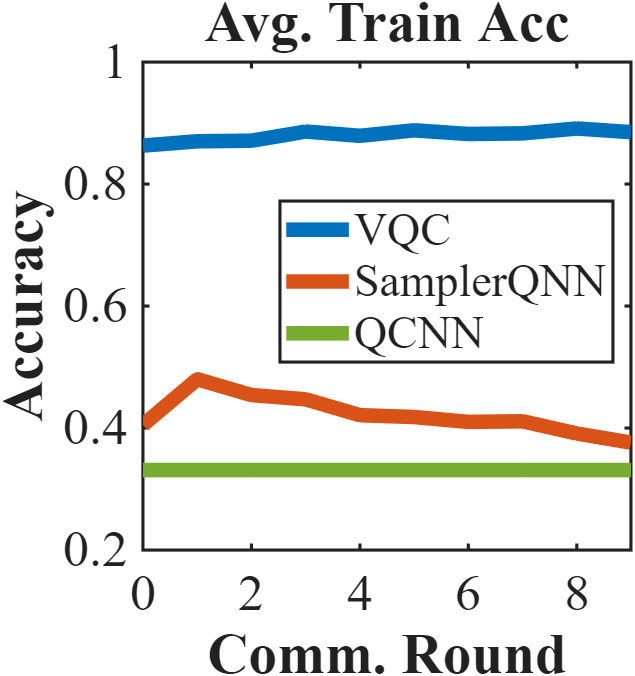}
    \caption{Avg. Train}
    \label{fig:avg_devices_train_acc_models_nuscenes}
    \end{subfigure}
    \begin{subfigure}[b]{0.24\columnwidth}
        \centering
       \includegraphics[width=\columnwidth]{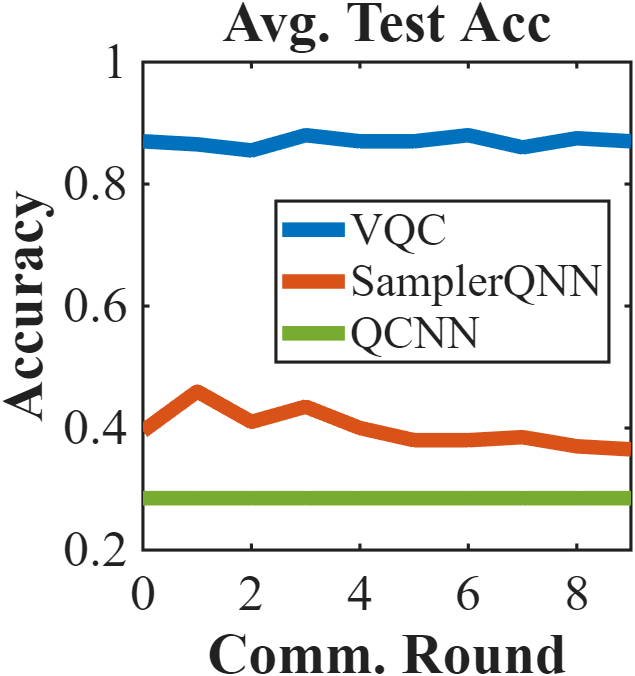}
    \caption{Avg. Test}
    \label{fig:avg_devices_test_acc_models_nuscenes}
    \end{subfigure}
  \begin{subfigure}[b]{0.24\columnwidth}
        \centering
      \includegraphics[width=\columnwidth]{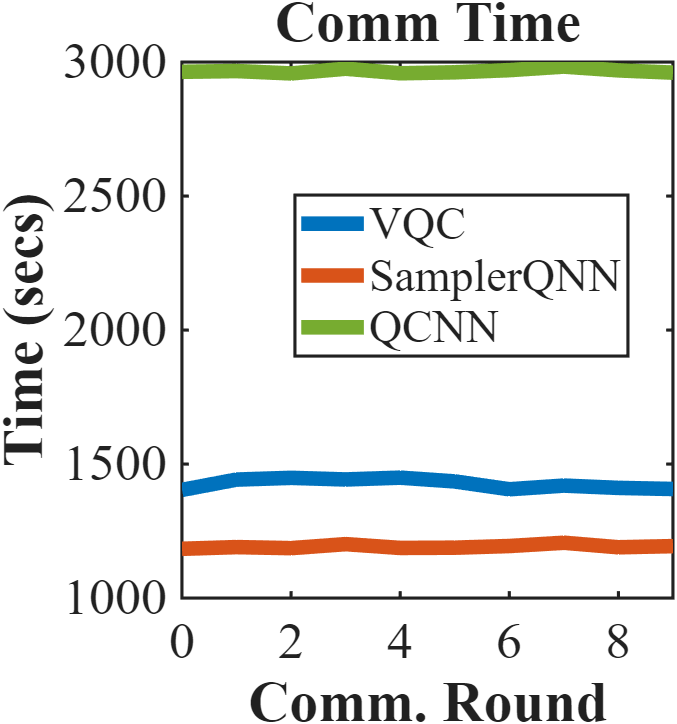}
    \caption{Time}
    \label{fig:comm_time_models_comparison_nuscenes}
    \end{subfigure}
     \begin{subfigure}[b]{0.24\columnwidth}
        \centering
       \includegraphics[width=\columnwidth]{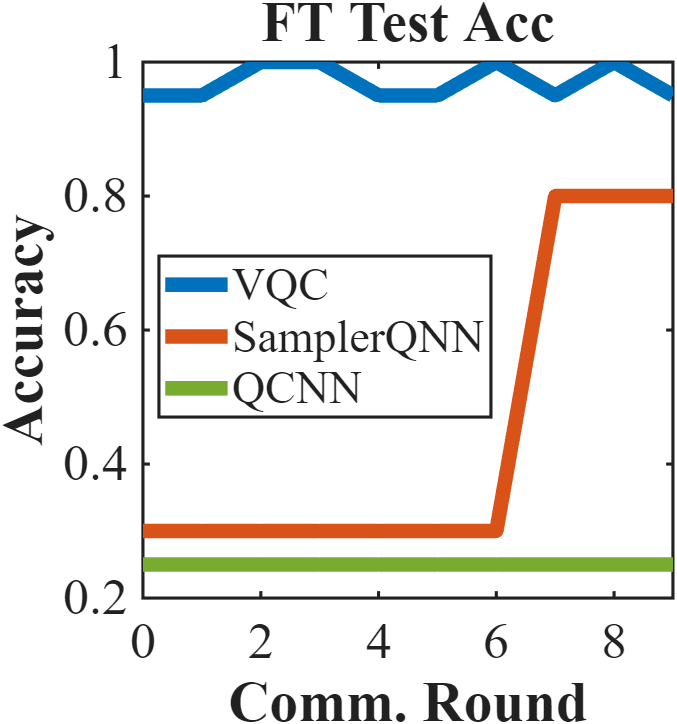}
    \caption{FT Test}
    \label{fig:global_ft_test_acc_models_nuscenes}
    \end{subfigure}
    \begin{subfigure}[b]{0.24\columnwidth}
        \centering
       \includegraphics[width=\columnwidth]{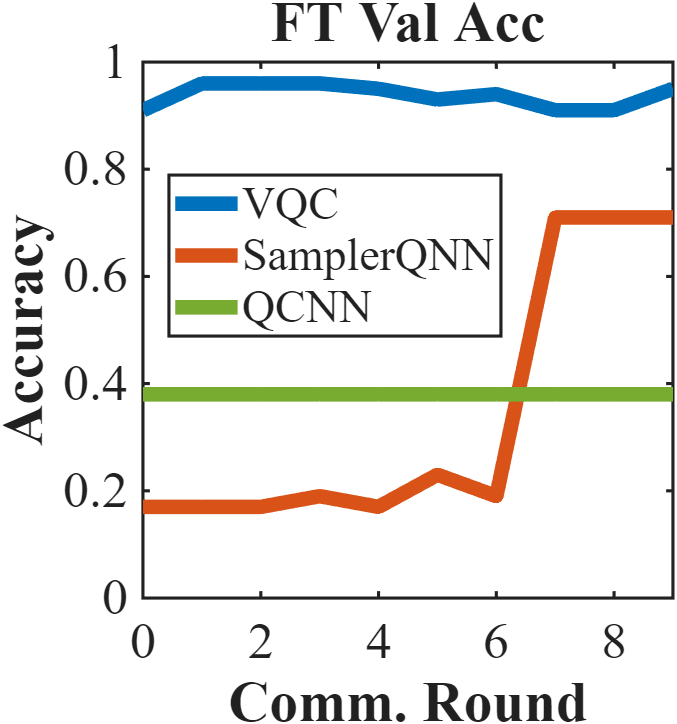}
    \caption{FT Val}
    \label{fig:global_ft_val_acc_models_nuscenes}
    \end{subfigure}
     \caption{Models Comparison on nuScenes Dataset; Prediction Model, Fine-Tuned Model, Average Devices Performance, Communication Time}
    \label{fig:model_performance_nuscenes}
\end{figure}

With nuScenes data, VQC performs better in terms of devices, prediction, and finetuned model results, as in Figure \ref{fig:model_performance_nuscenes}.
While QCNN is slowest as shown in Figure \ref{fig:comm_time_models_comparison_nuscenes}.
In summary, we observed QCNN performed better with KITTI dataset but not with nuScenes dataset and was similar to VQC somehow with Waymo dataset.
 In most cases, VQC stands out in terms of its performance in all datasets.
This highlights the performance variation and capability of different models in performing various tasks, and thus correct selection of a specific model to do the work is required.

For the KITTI dataset, the experimental setup uses 2000 training samples partitioned across client devices and 100 samples reserved for server-side testing, with a maximum of 100 training iterations and 10 participating devices.
For the Waymo dataset, we similarly employ 2000 training samples distributed among devices and 100 server test samples, again using a maxiter setting of 100 and 10 devices.
For the nuScenes dataset, the configuration consists of 1000 training samples allocated to devices and 100 samples for server testing, with a maximum of 100 iterations and 10 devices.

Table \ref{tab:model_performance_kitti_waymo_nuscenes} shows further results in detail showing other results such as average results for the global model, which is average performance in all communication rounds and final performance is the last result at the end of all communication rounds.
The table also includes the top results chosen from all communication rounds for global prediction results.
In brief, with the KITTI dataset, VQC performs the best in terms of accuracy, prediction while for training, we can see similar results between QCNN and VQC.
Overall, VQC also performs well for the Waymo and nuScenes datatset.

\begin{table*}[!htbp]
    \centering
    \caption{Global FT Model, Prediction Model and Local Model Performance}
    \label{tab:model_performance_kitti_waymo_nuscenes}
    \resizebox{\textwidth}{!}{
    \begin{tabular}{l l
        *{2}{c} *{2}{c} *{2}{c}
        *{2}{c} *{2}{c}
        c
        *{8}{c}
    }
        \toprule
        \toprule
        \multirow{4}{*}{\textbf{Dataset}} &
        \multirow{4}{*}{\textbf{Model}} &
        \multicolumn{6}{c}{\textbf{Global Fine-tuned Model}} &
        \multicolumn{4}{c}{\textbf{Local Model}} &
        \multirow{4}{*}{\textbf{Comm. Time}} &
        \multicolumn{8}{c}{\textbf{Global Prediction Model}} \\
        \cmidrule(lr){3-8} \cmidrule(lr){9-12} \cmidrule(lr){14-21}
        & &
        \multicolumn{2}{c}{\textbf{Val Acc}} &
        \multicolumn{2}{c}{\textbf{Test Acc}} &
        \multicolumn{2}{c}{\textbf{Val Loss}} &
        \multicolumn{2}{c}{\textbf{Train Acc}} &
        \multicolumn{2}{c}{\textbf{Test Acc}} &
        &
        \multicolumn{2}{c}{\textbf{Val Loss}} &
        \multicolumn{3}{c}{\textbf{Val Acc}} &
        \multicolumn{3}{c}{\textbf{Test Acc}} \\
        \cmidrule(lr){3-4} \cmidrule(lr){5-6} \cmidrule(lr){7-8}
        \cmidrule(lr){9-10} \cmidrule(lr){11-12}
        \cmidrule(lr){14-15} \cmidrule(lr){16-18} \cmidrule(lr){19-21}
        & &
        Avg & Final &
        Avg & Final &
        Avg & Final &
        Avg & Final &
        Avg & Final &
        &
        Avg & Final &
        Avg & Final & Top &
        Avg & Final & Top \\
                \midrule \midrule
                \multirow{3}{*}{KITTI}
              & VQC & \textbf{0.7680} & \textbf{0.7600} & \textbf{0.9000} & \textbf{0.9000} & \textbf{1.03} & \textbf{0.98}
                         & 0.79 & 0.79 & \textbf{0.79} & \textbf{0.80} & 9036.35
                         & 1.1147 & 1.1807 & 0.6575 & 0.6000 & 0.7000
                         & 0.8050 & 0.8000 & 0.8500 \\
              & SamplerQNN & 0.7260 & \textbf{0.7600} & 0.5600 & 0.5500 & 1.41 & 1.25
                         & 0.68 & 0.70 & 0.68 & 0.70 & 9618.97
                         & 1.1541 & 1.1053 & 0.6113 & 0.6500 & 0.6500
                         & 0.4300 & 0.4000 & 0.4500 \\
              & QCNN & 0.7000 & 0.7000 & 0.8500 & 0.8500 & 1.23 & 0.87
                         & \textbf{0.80} & \textbf{0.80} & 0.75 & 0.75 & \textbf{1287.69}
                         & \textbf{0.0000} & \textbf{0.0000} & \textbf{0.7000} & \textbf{0.7000} & \textbf{0.7000}
                         & \textbf{0.8500} & \textbf{0.8500} & \textbf{0.8500} \\
                         \midrule
        \multirow{3}{*}{Waymo}
              & VQC & \textbf{0.7010} & \textbf{0.7000} & \textbf{0.6000} & \textbf{0.6000} & 0.8035 & 0.7886
                         & \textbf{0.73} & \textbf{0.73} & \textbf{0.64} & \textbf{0.65} & 9841.01
                         & 0.5749 & 0.5866 & \textbf{0.7044} & \textbf{0.7000} & \textbf{0.7125}
                         & \textbf{0.6200} & \textbf{0.6000} & \textbf{0.6750} \\
              & SamplerQNN & 0.5850 & 0.5700 & 0.3970 & 0.4200 & \textbf{0.4433} & \textbf{0.4353}
                         & 0.59 & 0.58 & 0.48 & 0.47 & \textbf{462.12}
                         & 0.6728 & 0.6601 & 0.5944 & 0.6375 & 0.6375
                         & 0.4125 & 0.4000 & 0.4500 \\
              & QCNN & 0.4720 & 0.4700 & 0.5700 & 0.5700 & 1.0814 & 1.0786
                         & 0.51 & 0.51 & 0.49 & 0.50 & 606.16
                         & 0.7486 & 0.7508 & 0.4838 & 0.4688 & 0.5125
                         & 0.6000 & 0.6250 & 0.6500 \\
        \midrule
         \multirow{3}{*}{nuScenes}
              & VQC & \textbf{0.9380} & \textbf{0.9500} & \textbf{0.9700} & \textbf{0.9500} & 1.0819 & 1.0124
                         & \textbf{0.88} & \textbf{0.89} & \textbf{0.87} & \textbf{0.87} & 1426.26
                         & 0.9824 & 0.9815 & \textbf{0.7438} & \textbf{0.8000} & \textbf{0.8000}
                         & \textbf{0.6400} & \textbf{0.7000} & \textbf{0.7500} \\
              & SamplerQNN & 0.3420 & 0.7100 & 0.4500 & 0.8000 & 0.9944 & 0.8036
                         & 0.42 & 0.38 & 0.40 & 0.36 & \textbf{1191.79}
                         & 1.0797 & 1.0781 & 0.1875 & 0.2000 & 0.2250
                         & 0.3500 & 0.4000 & 0.4000 \\
              & QCNN & 0.3800 & 0.3800 & 0.2500 & 0.2500 & \textbf{0.4534} & \textbf{0.4142}
                         & 0.33 & 0.33 & 0.29 & 0.29 & 2966.30
                         & \textbf{0.1920} & \textbf{0.1070} & 0.3750 & 0.3750 & 0.3750
                         & 0.2500 & 0.2500 & 0.2500 \\
        \bottomrule
        \bottomrule
    \end{tabular}
    }
\end{table*}

\subsection{Number of Devices - Waymo}
\begin{figure}[!htbp]
    \centering
    \begin{subfigure}[b]{0.24\columnwidth}
        \centering
       \includegraphics[width=\columnwidth]{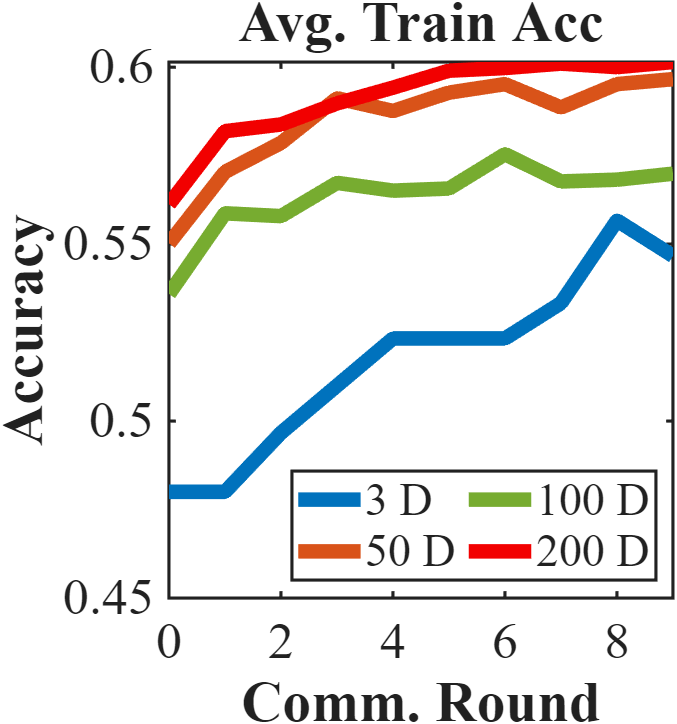}
    \caption{Avg. Train}
    \label{fig:avg_devices_train_acc_number_devices}
    \end{subfigure}
    \begin{subfigure}[b]{0.24\columnwidth}
        \centering
       \includegraphics[width=\columnwidth]{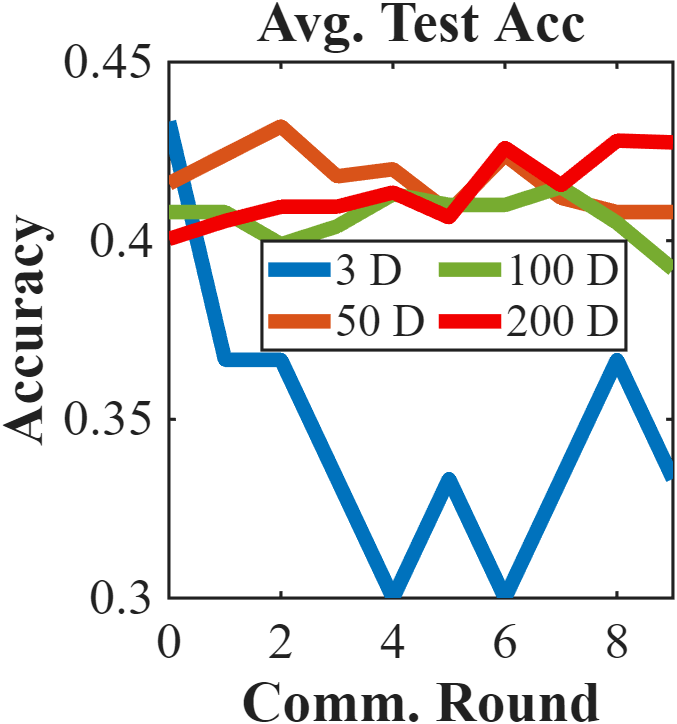}
    \caption{Avg Test}
    \label{fig:avg_devices_test_acc_number_devices}
    \end{subfigure}
  \begin{subfigure}[b]{0.24\columnwidth}
        \centering
      \includegraphics[width=\columnwidth]{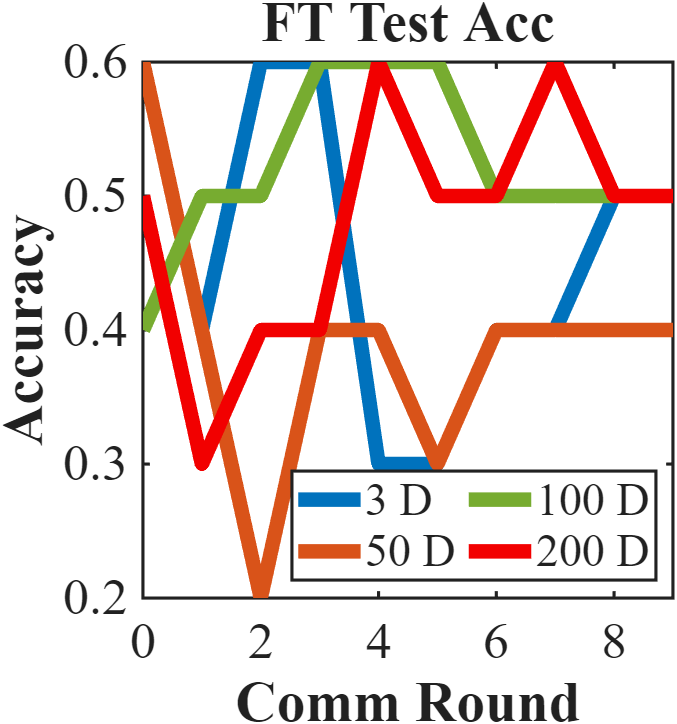}
    \caption{FT Test}
    \label{fig:ft_test_acc_number_devices}
    \end{subfigure}
    \begin{subfigure}[b]{0.24\columnwidth}
        \centering
      \includegraphics[width=\columnwidth]{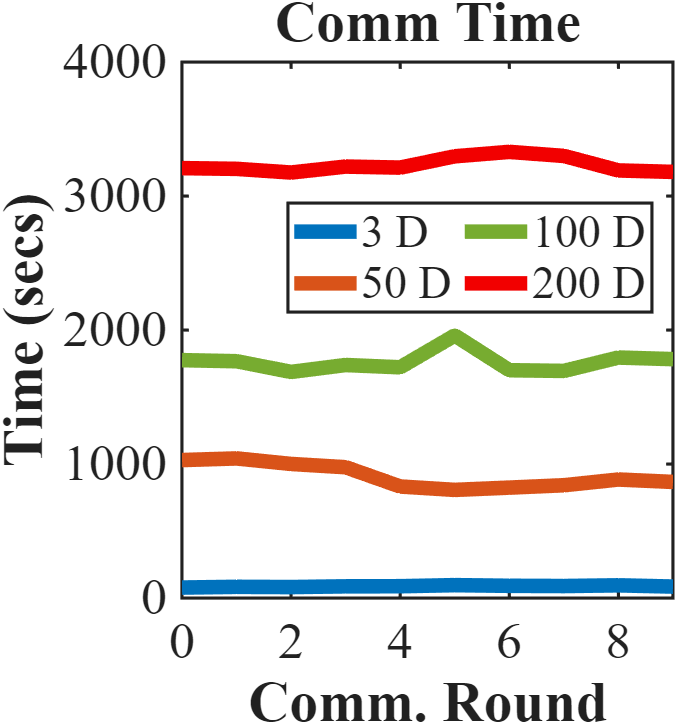}
    \caption{Time}
    \label{fig:comm_time_number_of_devices}
    \end{subfigure}
    \caption{Impact of number of devices: 3, 20, 100, 200 Devices}
    \label{fig:number_of_devices}
\end{figure}

For this experiment, we 
randomly extracted small set from the dataset so that each device has 
50 data samples for faster computation. 
Then we experiment with 3, 20, 100, 200 devices and observe the performance in terms of average devices performance (Figures \ref{fig:avg_devices_train_acc_number_devices}, \ref{fig:avg_devices_test_acc_number_devices}), fine tune model test accuracy (Figure \ref{fig:ft_test_acc_number_devices}) and communication time (Figure \ref{fig:comm_time_number_of_devices}).
We get better results with more devices (200 D) devices in terms of average device performance with only 3 devices performing worst in terms of accuracy and FT model performance but with communication overhead as expected with a greater number of devices.

\subsection{Learning Behaviour of Devices}
\begin{figure}[!htbp]
    \centering
    \begin{subfigure}[h]{0.34\columnwidth}
        \centering
       \includegraphics[width=\columnwidth]{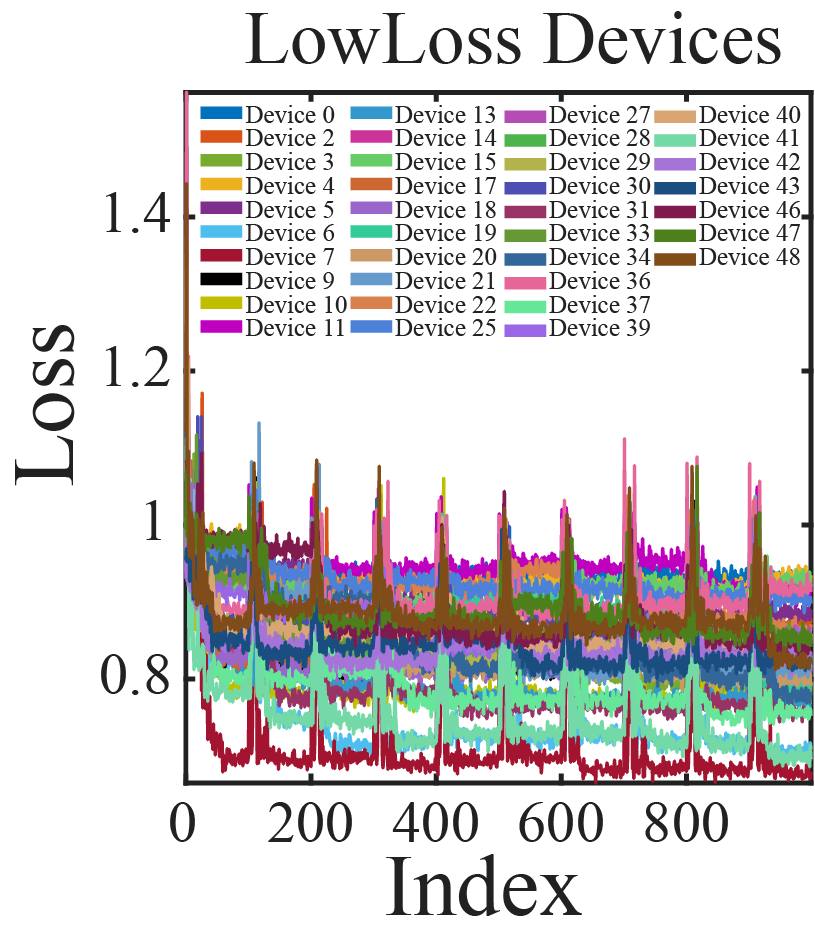}
    \caption{Low Loss Devices}
    \label{fig:low_loss_devices}
    \end{subfigure}
    \begin{subfigure}[h]{0.34\columnwidth}
        \centering
       \includegraphics[width=\columnwidth]{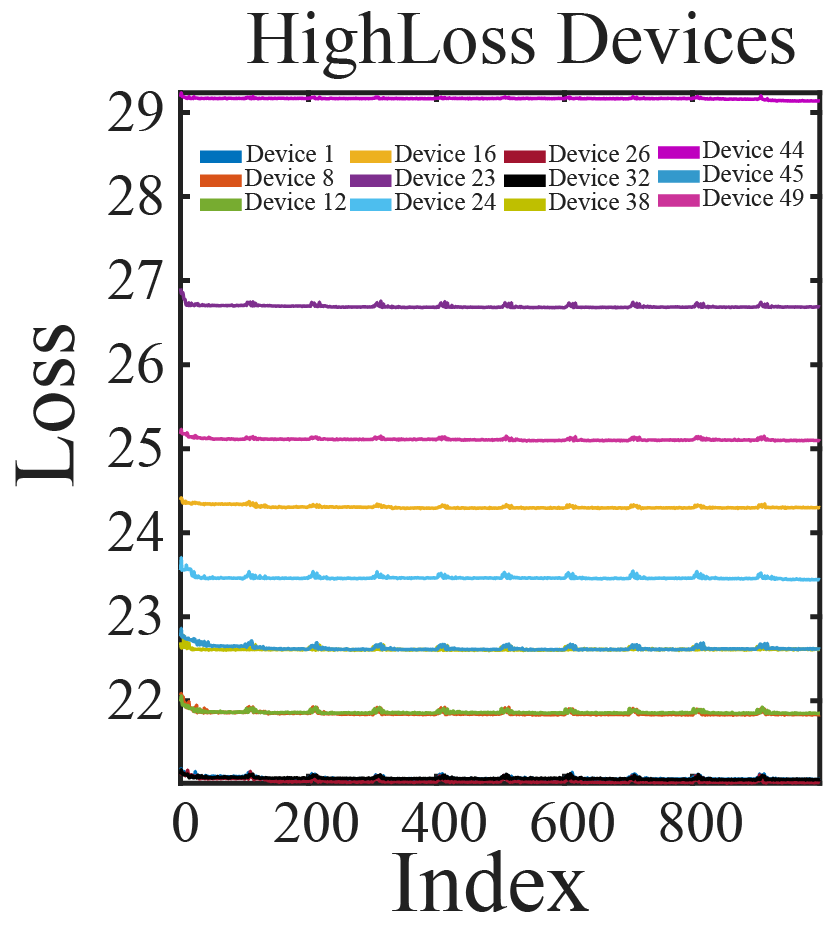}
    \caption{High Loss Devices}
    \label{fig:high_loss_devices}
    \end{subfigure}
    \caption{Devices Behavior: Two group of devices in same vQFL set up behaving different in terms of local learning with contrasting objective values.}
    \label{fig:devices_behaviors}
\end{figure}

We observe the behaviors of the devices as in Figure \ref{fig:devices_behaviors}.
We plot devices local objective values and observe that they all behave differently among 50 devices with each having 50 data samples Waymo dataset, 100 max iter for COBYLA optimizer run for 10 communication rounds. 
Figure \ref{fig:low_loss_devices} shows devices with low object values ranging from 1.2 to close to zero, while 
Figure \ref{fig:high_loss_devices} shows devices with high object values ranging from around 29 to 22 which is much poorer than the other cohort of devices.
This behavior of devices could impact the overall performance of the devices, not letting global optimization reach as desired as one group of devices performs much lower than the other group of devices.
Considering this Waymo dataset has max 3 classes and thus not completely representing non IID scenario, a proper investigation to this behavior of devices performing differently drastically is required and thus deferred to our future work.

\begin{figure}[!htbp]
    \centering
    \begin{subfigure}[b]{0.24\columnwidth}
        \centering
       \includegraphics[width=\columnwidth]{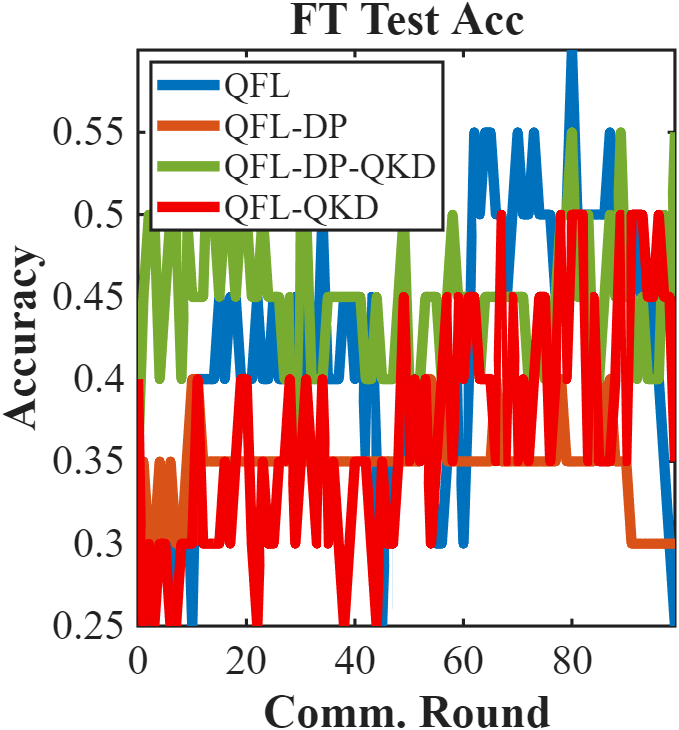}
    \caption{FT Test}
    \label{fig:server_test_acc_waymo}
    \end{subfigure}
    \begin{subfigure}[b]{0.24\columnwidth}
        \centering
       \includegraphics[width=\columnwidth]{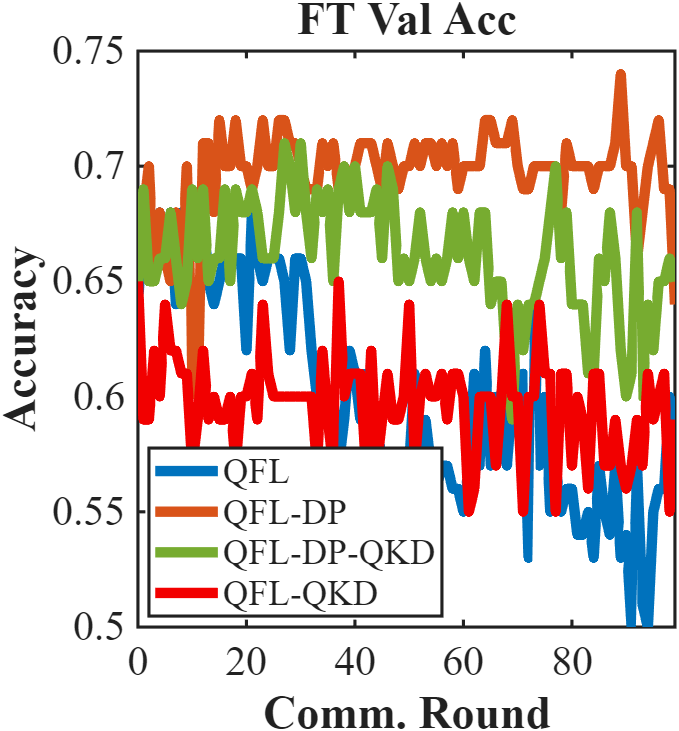}
    \caption{FT Val}
    \label{fig:server_val_acc_waymo}
    \end{subfigure}
  \begin{subfigure}[b]{0.242\columnwidth}
        \centering
      \includegraphics[width=\columnwidth]{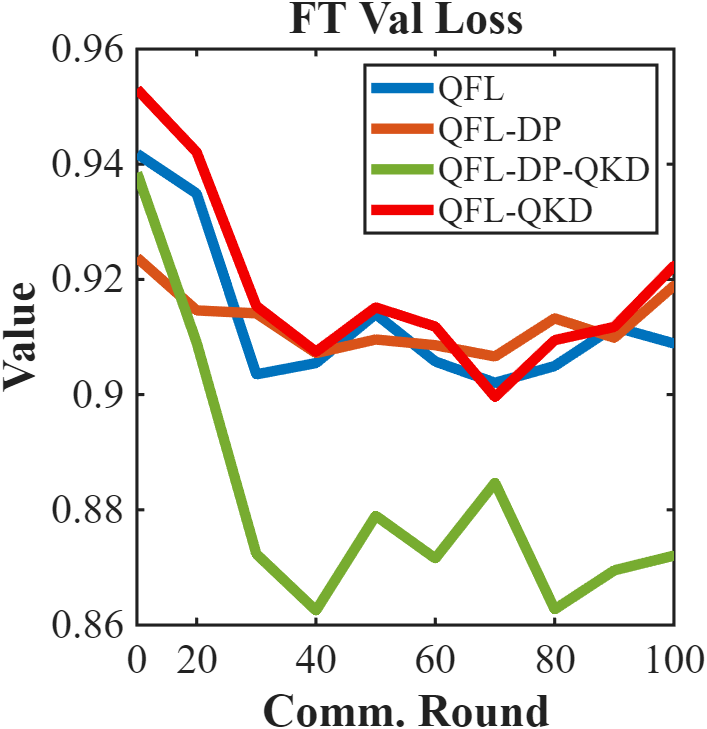}
    \caption{FT Loss}
    \label{fig:server_objective_values_waymo}
    \end{subfigure}
    \begin{subfigure}[b]{0.24\columnwidth}
        \centering
       \includegraphics[width=\linewidth]{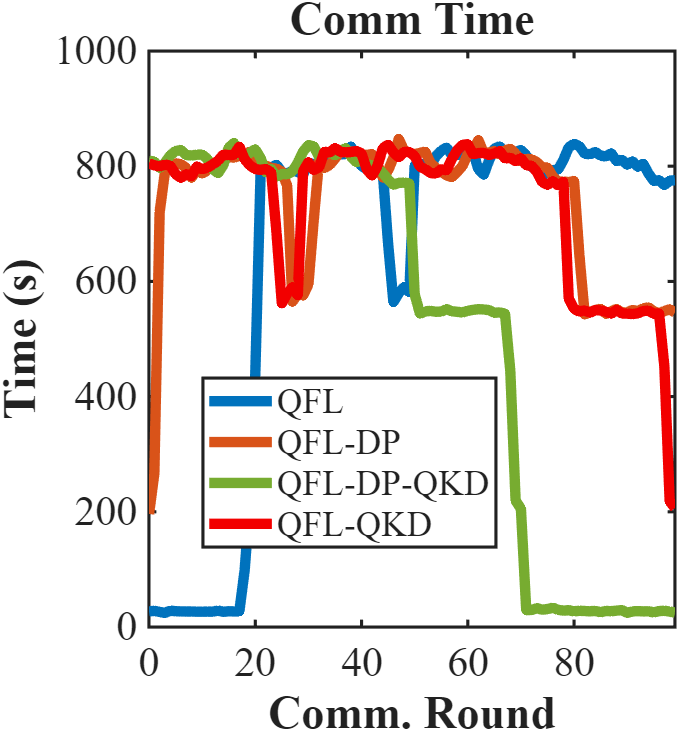}
    \caption{Comm Time}
    \label{fig:comm_time_waymo}
    \end{subfigure}
      \begin{subfigure}[b]{0.24\columnwidth}
        \centering
       \includegraphics[width=\columnwidth]{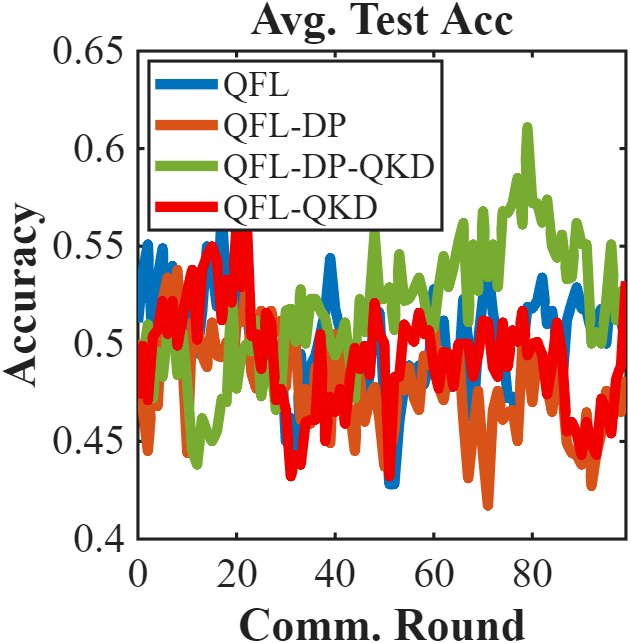}
    \caption{Avg. Test}
    \label{fig:devices_avg_test_acc_waymo}
    \end{subfigure}
    \begin{subfigure}[b]{0.24\columnwidth}
        \centering
       \includegraphics[width=\columnwidth]{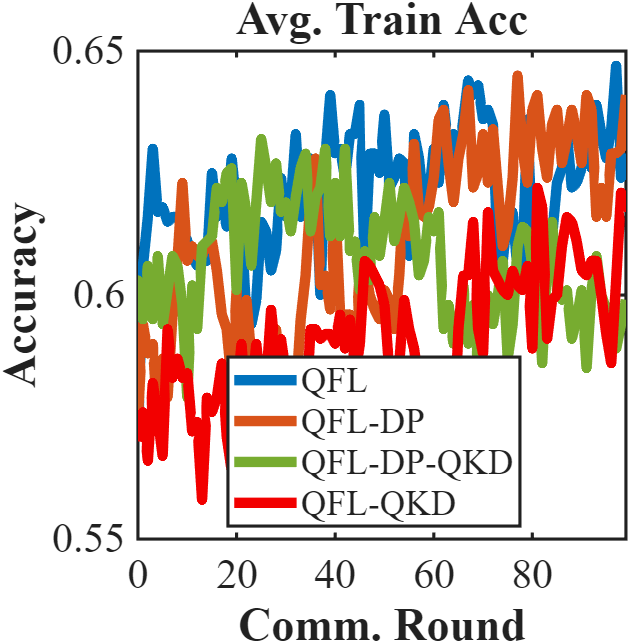}
    \caption{Avg. Train}
    \label{fig:devices_avg_train_acc_waymo}
    \end{subfigure}
  \begin{subfigure}[b]{0.226\columnwidth}
        \centering
      \includegraphics[width=\columnwidth]{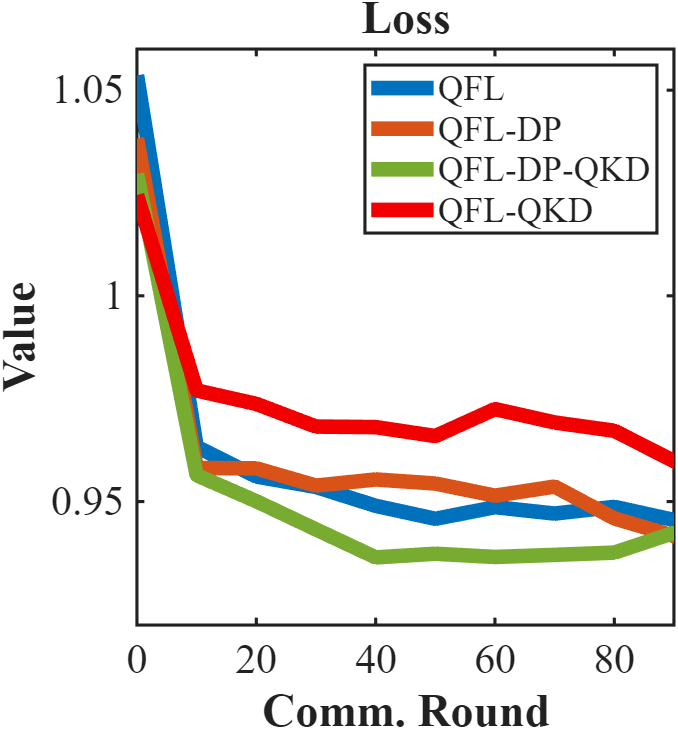}
    \caption{Avg. Loss}
    \label{fig:devices_avg_objective_values_waymo}
    \end{subfigure}
    \begin{subfigure}[b]{0.223\columnwidth}
        \centering
      \includegraphics[width=\columnwidth]{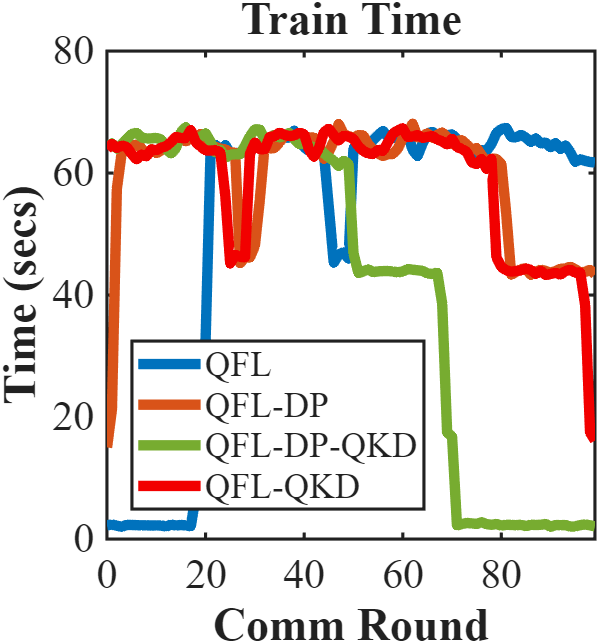}
    \caption{Train Time}
    \label{fig:devices_avg_training_time_waymo}
    \end{subfigure}
    \caption{Model Performance: Waymo Dataset; QFL, QFL-DP, QFL-QKD, QFL-DP-QKD}
    \label{fig:server_performance_waymo}
\end{figure}

\subsection{Results - DP, QKD}
These are additional results that are compared to see if there was any impact on standard QFL due to QKD, DP integration, etc. 
The experiment is performed with the methods QFL, QFL-DP, QFL-QKD and QFL-DP-QKD
representing default QFL, QFL with differential privacy, QFL with QKD implementation, and final QFL with both DP and QKD implemented.
In Figure \ref{fig:server_performance_waymo}, we observe the performance (FT model) of the server device with the Waymo dataset which was run for 100 communication rounds with 1000 train samples distributed among 10 devices with optimizer 10 maxiter.
In terms of validation accuracy and test accuracy, as in Figures \ref{fig:server_test_acc_waymo} and
\ref{fig:server_val_acc_waymo}, QFL-DP-QKD and QFL-DP perform better than other methods, respectively.
However, as seen in Figure \ref{fig:server_objective_values_waymo}, in terms of the loss values of the objective function, QFL-DP-QKD performs better.
In terms of device performance, 
devices perform best on average validation accuracy with QFL-DP-QKD as in Figure \ref{fig:devices_avg_test_acc_waymo} and 
on average train accuracy with QFL and QFL-DP as in Figure \ref{fig:devices_avg_train_acc_waymo}.
In terms of training time per device and overall communication round, with added DP and QKD implementation, there is no impact as expected.
However, real implementation of QKD can be far more complex than one implemented in this study for proof-of-concept.
Also, the pattern is similar in terms of average devices training time (Figure \ref{fig:devices_avg_training_time_waymo} and total communication round time \ref{fig:comm_time_waymo} which implies that most of the difference in communication overhead is similar other than each individual device training time.

\begin{figure}[!htbp]
    \centering
    \begin{subfigure}[b]{0.24\columnwidth}
        \centering
       \includegraphics[width=\columnwidth]{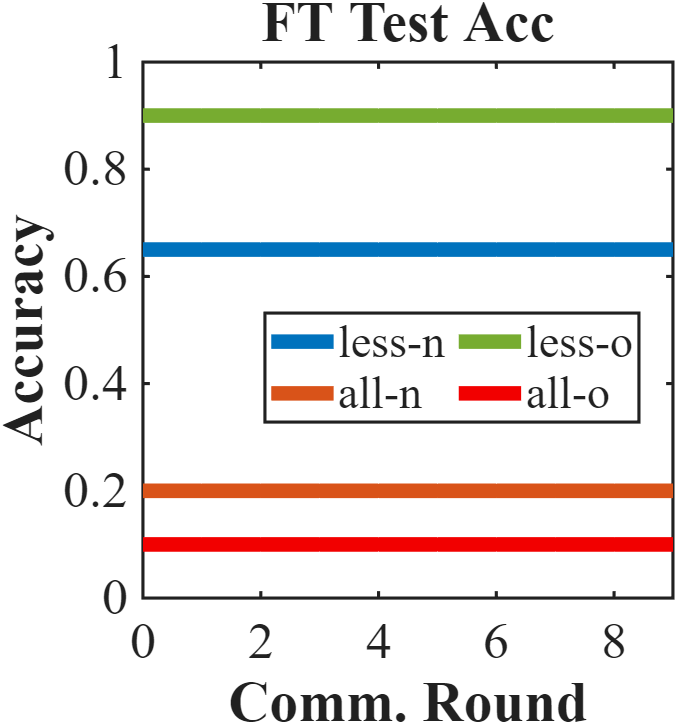}
    \caption{FT Test}
    \label{fig:less_features_server_test_acc}
    \end{subfigure}
    \begin{subfigure}[b]{0.24\columnwidth}
        \centering
       \includegraphics[width=\columnwidth]{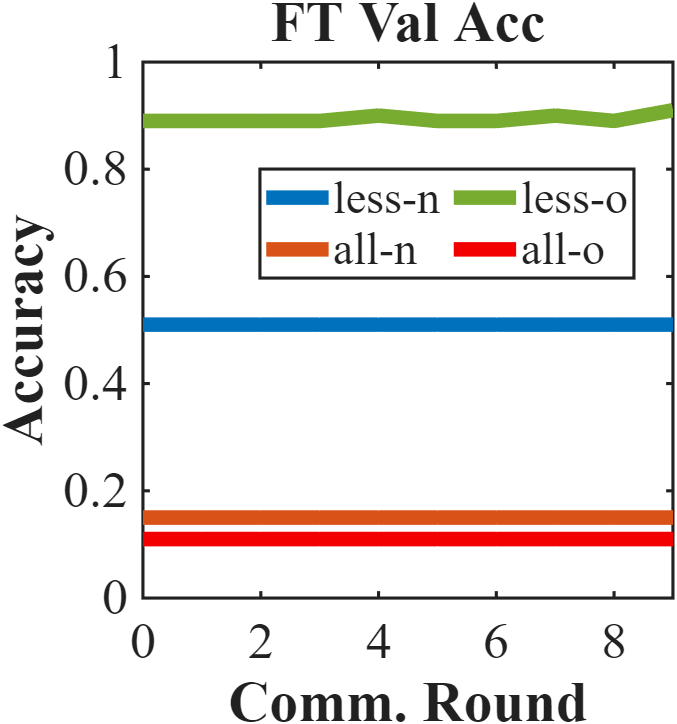}
    \caption{FT Val}
    \label{fig:less_features_server_val_acc}
    \end{subfigure}
  \begin{subfigure}[b]{0.24\columnwidth}
        \centering
      \includegraphics[width=\columnwidth]{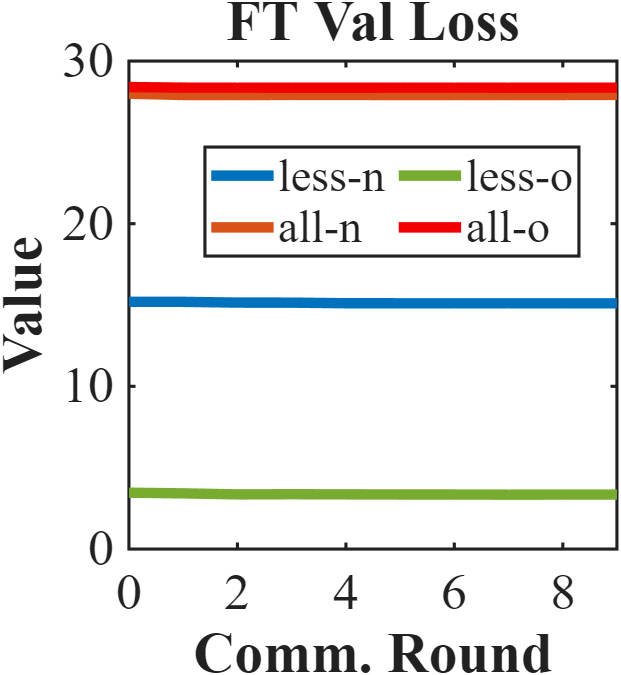}
    \caption{FT Loss}
    \label{fig:less_features_server_val_loss}
    \end{subfigure}
    \begin{subfigure}[b]{0.24\columnwidth}
        \centering
      \includegraphics[width=\columnwidth]{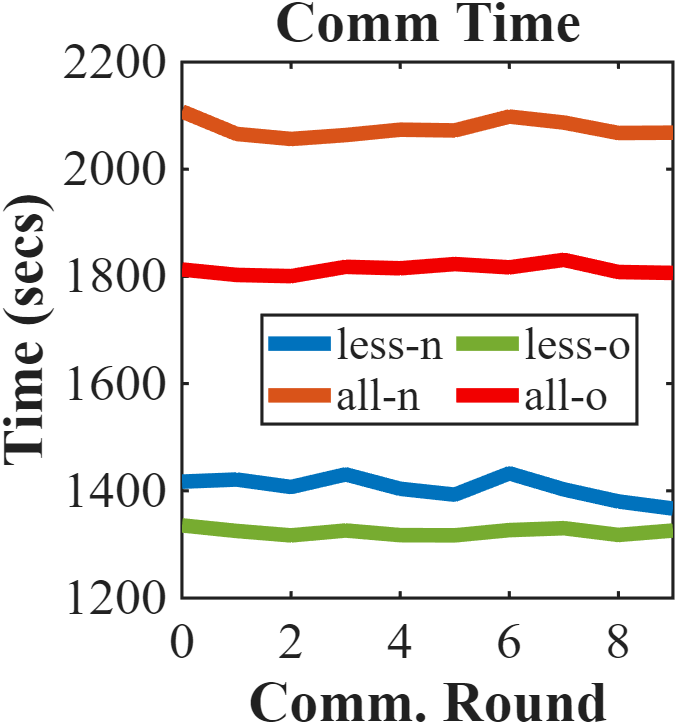}
    \caption{Comm Time}
    \label{fig:less_features_device_comm_time}
    \end{subfigure}
     \begin{subfigure}[b]{0.24\columnwidth}
        \centering
       \includegraphics[width=\columnwidth]{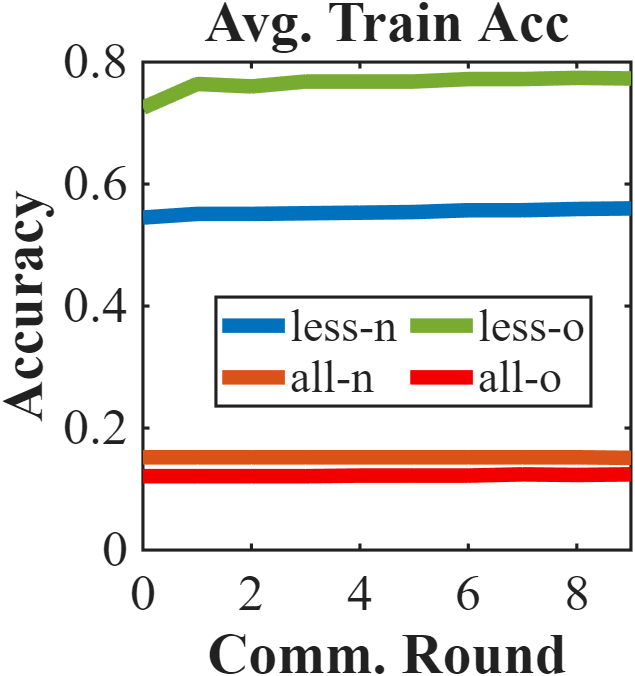}
    \caption{Avg. Train}
    \label{fig:less_features_device_train_acc}
    \end{subfigure}
    \begin{subfigure}[b]{0.24\columnwidth}
        \centering
       \includegraphics[width=\columnwidth]{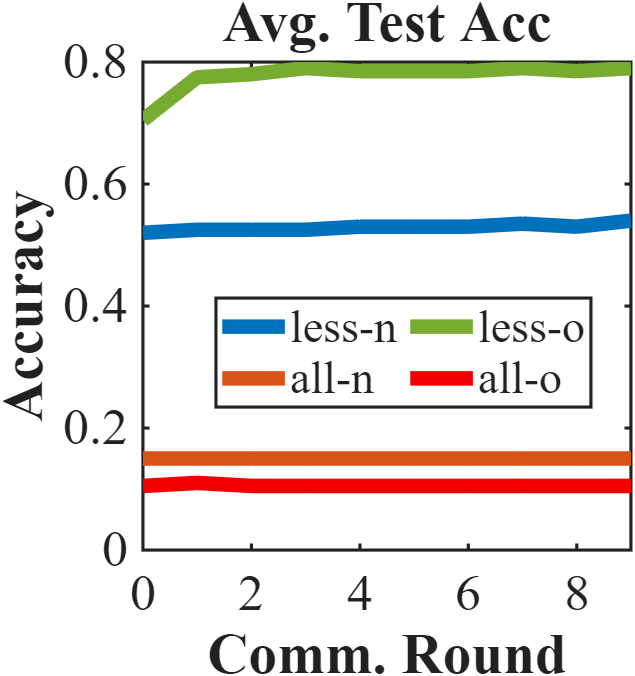}
    \caption{Avg. Val}
    \label{fig:less_features_device_val_acc}
    \end{subfigure}
  \begin{subfigure}[b]{0.24\columnwidth}
        \centering
      \includegraphics[width=\columnwidth]{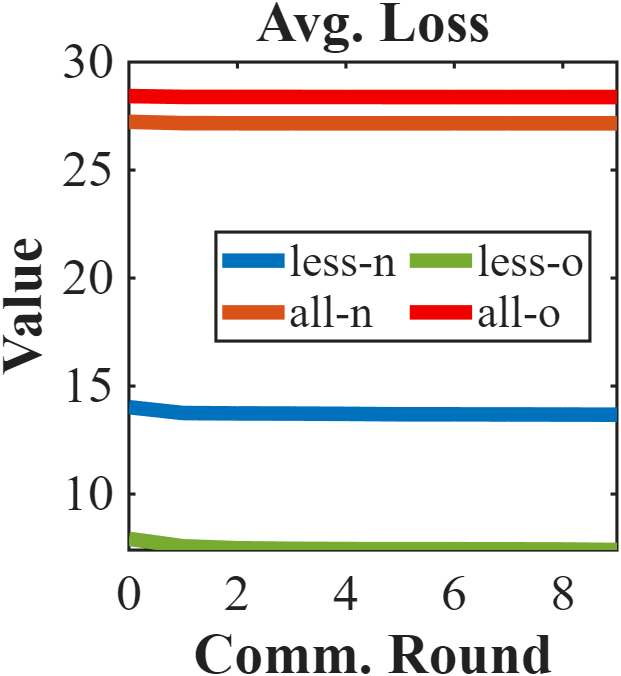}
    \caption{Avg. Loss}
    \label{fig:less_features_devices_average_loss}
    \end{subfigure}
    \begin{subfigure}[b]{0.24\columnwidth}
        \centering
      \includegraphics[width=\columnwidth]{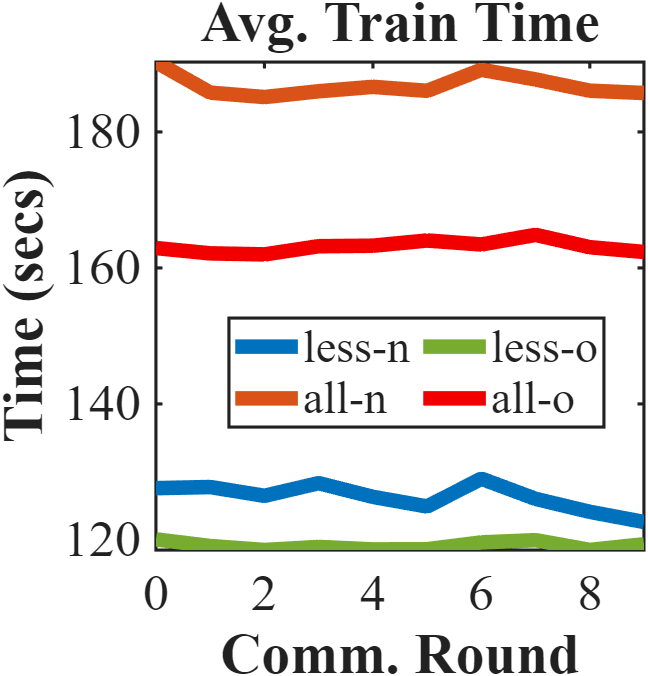}
    \caption{Train Time}
    \label{fig:less_features_training_time}
    \end{subfigure}
    \caption{Model Performance: nuScenes Less Dataset}
    \label{fig:server_performance_nuscenes_less}
\end{figure}

\begin{figure}[!htbp]
    \centering
    \begin{subfigure}[b]{0.34\columnwidth}
        \centering
       \includegraphics[width=\columnwidth]{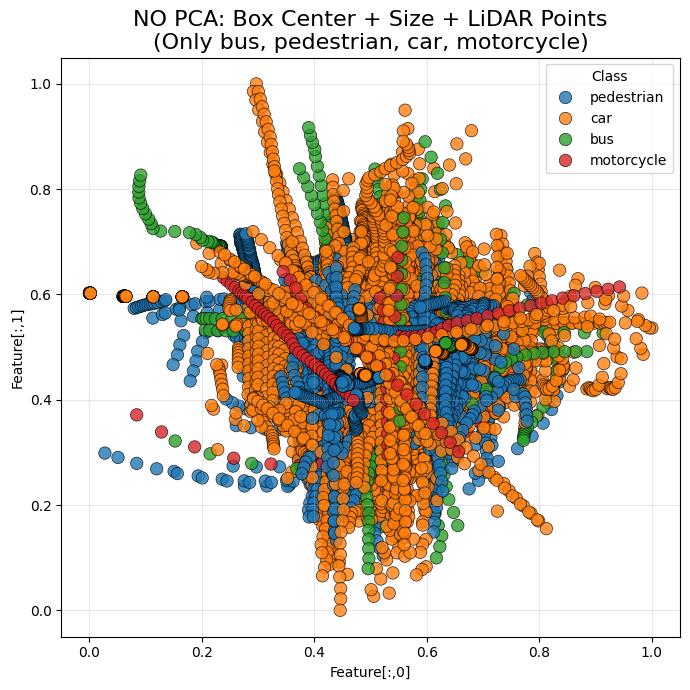}
    \caption{Feature 0 vs 1}
    \label{fig:no_pca_nuscenes_less_features_labels_features0vs1}
    \end{subfigure}
    \begin{subfigure}[b]{0.34\columnwidth}
        \centering
       \includegraphics[width=\columnwidth]{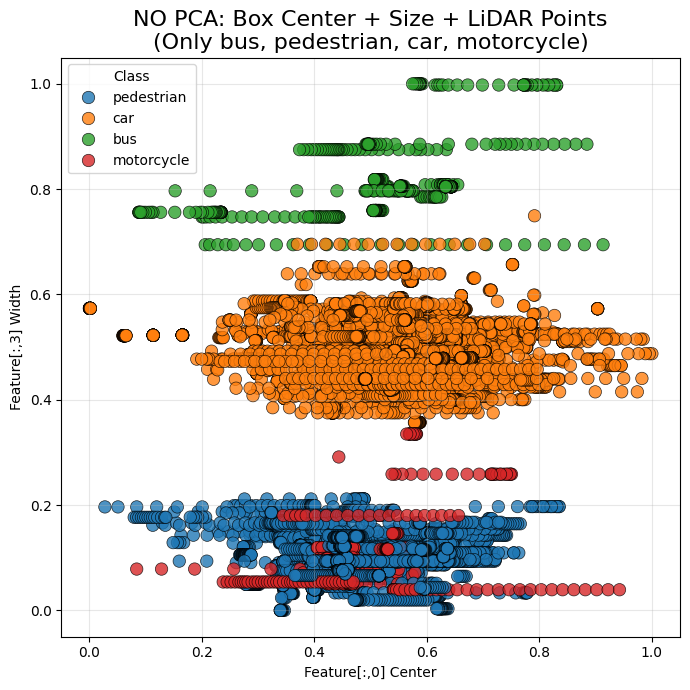}
    \caption{Feature 0 vs 3}
    \label{fig:no_pca_nuscenes_less_features_labels_features0vs3}
    \end{subfigure}
    \caption{nuScenes Less Features and No PCA}
    \label{fig:server_performance_less}
\end{figure}

\subsection{Nuscenes Dataset - Less Features.}
In nuScenes dataset, we have altogether 11 features extracted including xyz, wlh, lidar points, rotation angle, etc. with all the labels as default such as pedestrian, car, truck, etc.
But in this experiment, we have selected a few features of 7 and also reduced labels to car, truck, and pedestrian.
We did not apply PCA in these experiments and used all features, which implies an equal number of qubits (7 features) used in the circuits.
From the results in Figure \ref{fig:server_performance_nuscenes_less}, we can observe improved performance with fewer features and labels.
We compare between less-o against all-o and less-n against all-n where ``less" means fewer features (7) and ``all" means all features (11) whereas ``o" (old approach) and ``n" (new approach) refer to slightly different ways of processing data, but still with most similarity and are mentioned just so they can be compared correctly.
In all aspects, less-o performs better in terms of server test accuracy as in Figure \ref{fig:less_features_server_test_acc}, validation accuracy as in Figure \ref{fig:less_features_server_val_acc}, 
server validation loss \ref{fig:less_features_server_val_loss} and in device performance train accuracy, validation accuracy and average loss as in Figure \ref{fig:less_features_device_train_acc}, \ref{fig:less_features_device_val_acc} and \ref{fig:less_features_devices_average_loss} respectively.
This implies that, for VQC classifier, less number of classes is more favorable, as well as for nuScenes dataset, possibly not all features are relevant to the class labels which at least are observed in our experimental analysis.
The plot between features 0 and 1 and features 0 and 3 is shown in Figures \ref{fig:no_pca_nuscenes_less_features_labels_features0vs1} and \ref{fig:no_pca_nuscenes_less_features_labels_features0vs3} respectively without applying PCA analysis.

\subsection{Impact of class imbalance}
\begin{figure}[!htbp]
    \centering
      \begin{subfigure}{0.3\columnwidth}
        \centering
        \includegraphics[width=\linewidth]{images/waymo_label_distribution.png}
        \caption{Three Classes}
        \label{fig:waymo_label_distribution}
    \end{subfigure}
    \begin{subfigure}[b]{0.3\columnwidth}
        \centering
       \includegraphics[width=\columnwidth]{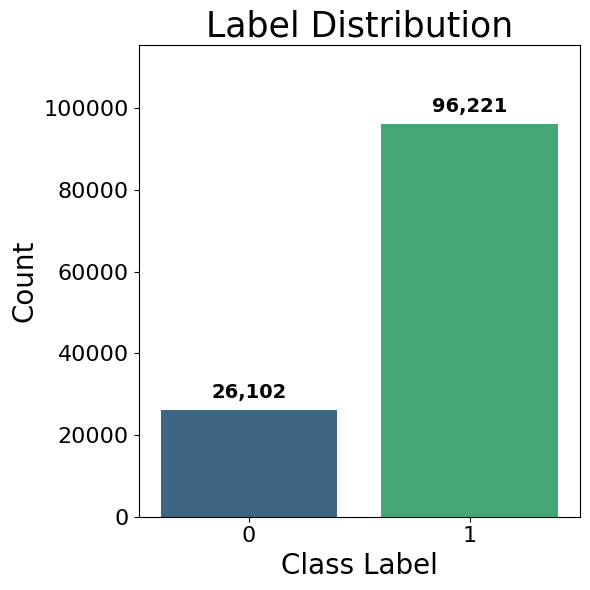}
    \caption{Unbalanced}
    \label{fig:imbalanced_class}
    \end{subfigure}
    \begin{subfigure}[b]{0.3\columnwidth}
        \centering
       \includegraphics[width=\columnwidth]{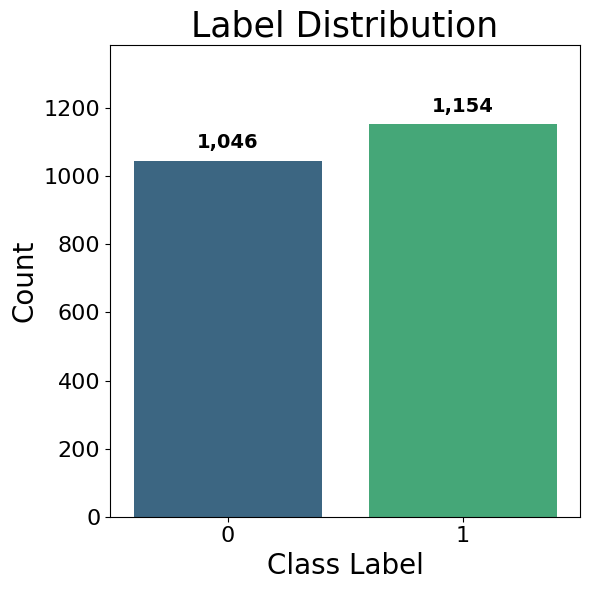}
    \caption{Balanced}
    \label{fig:balanced_class}
    \end{subfigure}
    \caption{Waymo Dataset: Class label distribution}
    \label{fig:class_label_distribution}
\end{figure}

\begin{figure}[!htbp]
    \centering
      \begin{subfigure}{0.3\columnwidth}
        \centering
        \includegraphics[width=\linewidth]{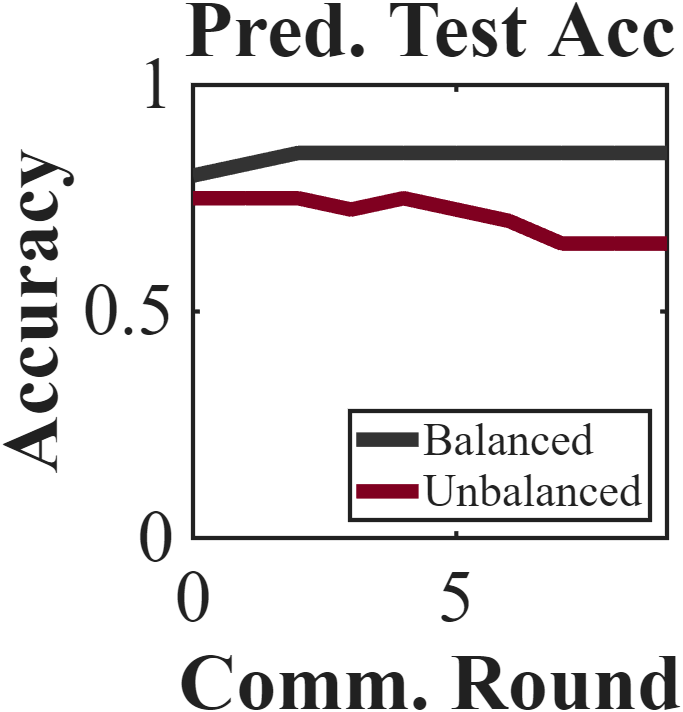}
        \caption{Pred. Test}
        \label{fig:prediction_test_waymo_balanced}
    \end{subfigure}
    \begin{subfigure}[b]{0.3\columnwidth}
        \centering
       \includegraphics[width=\columnwidth]{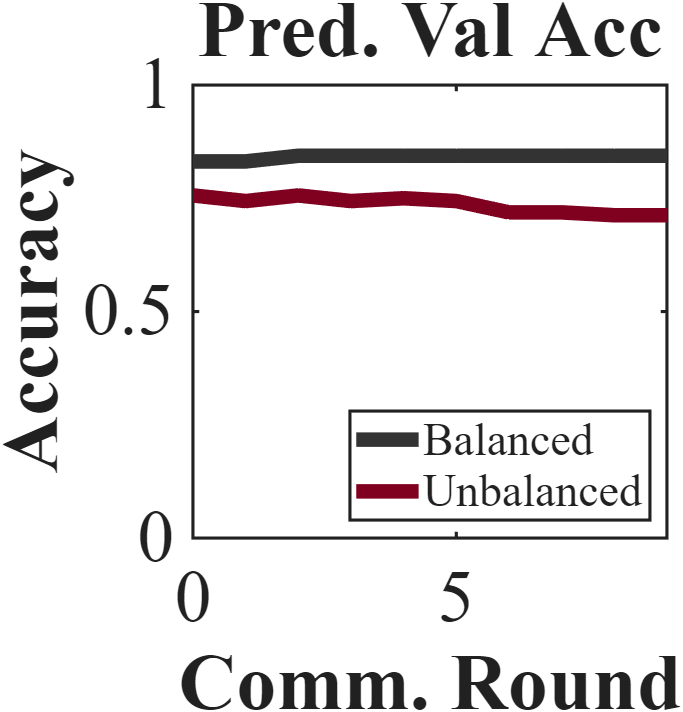}
    \caption{Pred. Val}
    \label{fig:prediction_val_waymo_balanced}
    \end{subfigure}
    \begin{subfigure}[b]{0.3\columnwidth}
        \centering
       \includegraphics[width=\columnwidth]{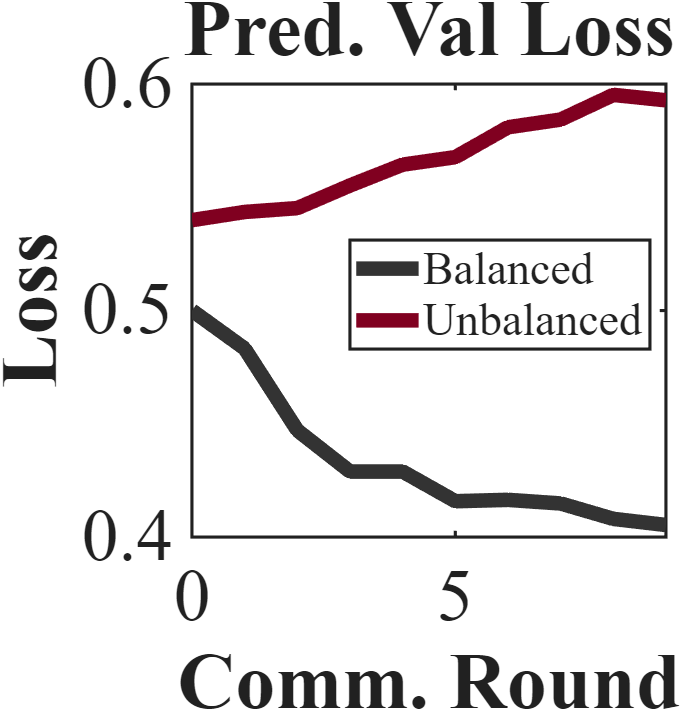}
    \caption{Pred. Loss}
    \label{fig:pediction_val_loss_waymo_balanced}
    \end{subfigure}
     \begin{subfigure}[b]{0.3\columnwidth}
        \centering
       \includegraphics[width=\columnwidth]{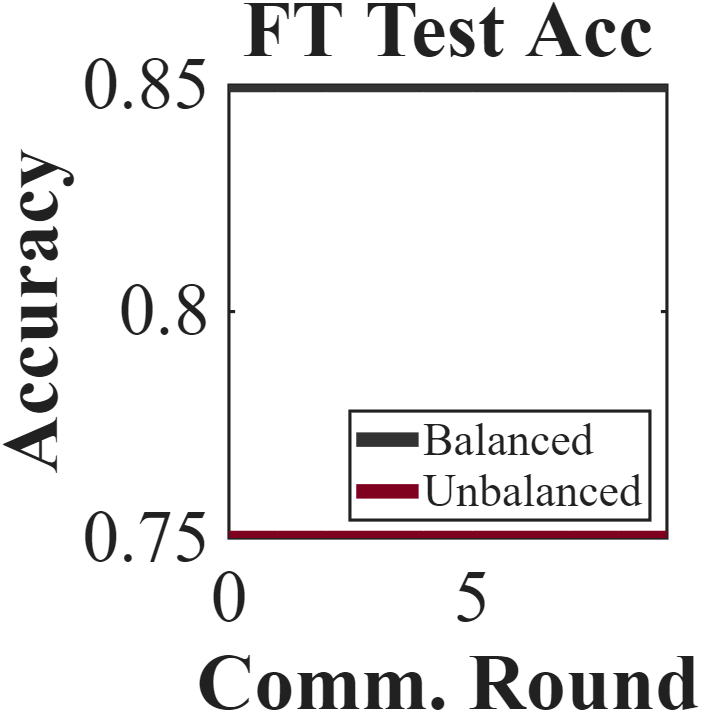}
    \caption{FT Test}
    \label{fig:server_test_acc_waymo_balanced}
    \end{subfigure}
     \begin{subfigure}[b]{0.3\columnwidth}
        \centering
       \includegraphics[width=\columnwidth]{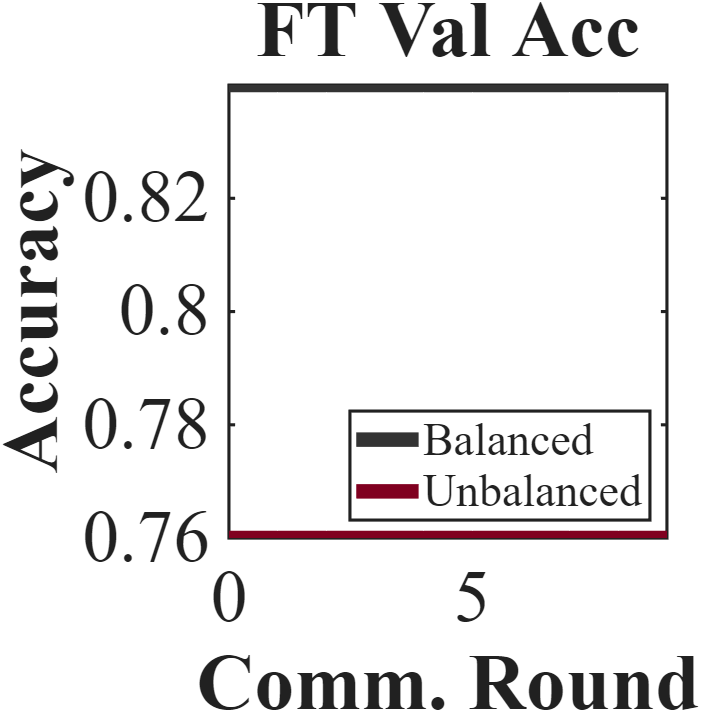}
    \caption{FT Val}
    \label{fig:server_val_acc_waymo_balanced}
    \end{subfigure}
     \begin{subfigure}[b]{0.3\columnwidth}
        \centering
       \includegraphics[width=\columnwidth]{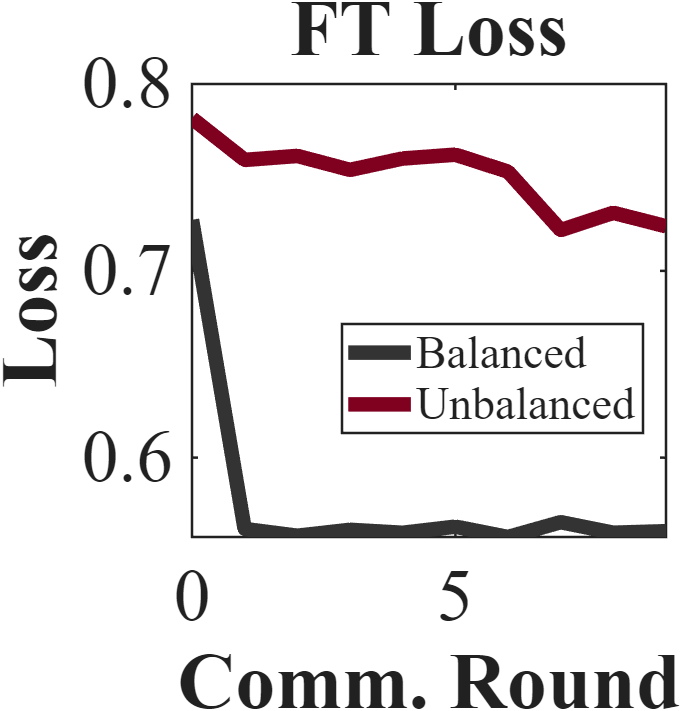}
    \caption{FT Loss}
    \label{fig:server_loss_waymo_balanced}
    \end{subfigure}
    \caption{Waymo Dataset: Global FT and Prediction Results}
    \label{fig:balanced_unbalanced_results}
\end{figure}

\begin{figure}[!htbp]
    \centering
      \begin{subfigure}{0.45\columnwidth}
        \centering
        \includegraphics[width=\linewidth]{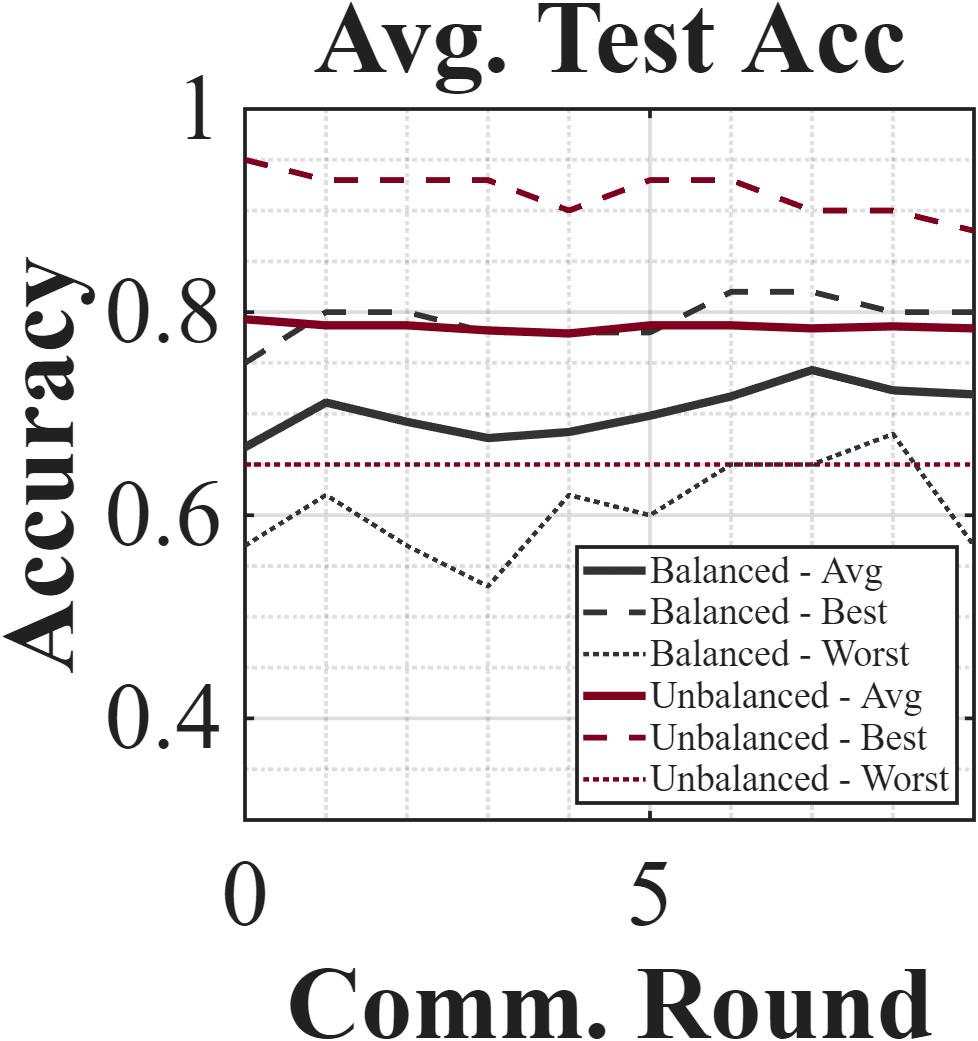}
        \caption{Average Devices Test}
        \label{fig:avg_devices_test_acc_waymo_balanced}
    \end{subfigure}
    \begin{subfigure}[b]{0.45\columnwidth}
        \centering
       \includegraphics[width=\columnwidth]{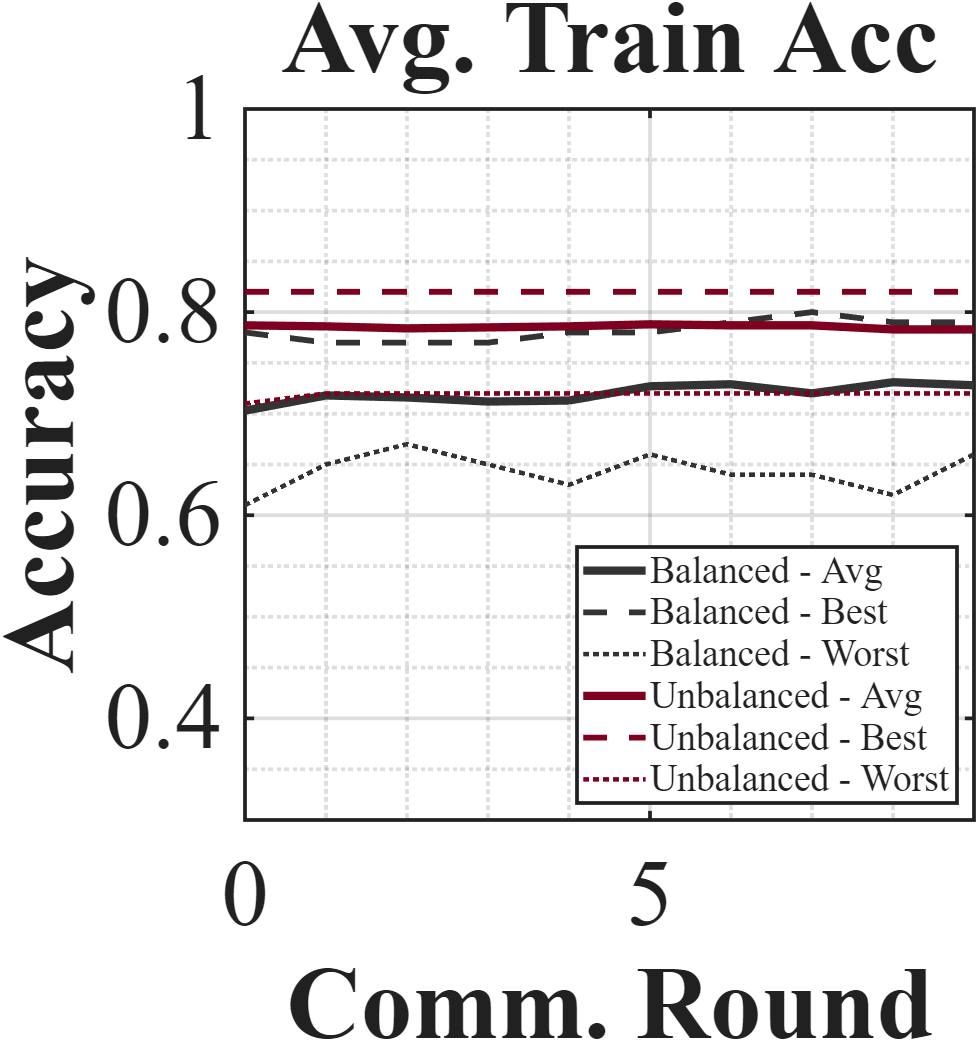}
    \caption{Average Devices Train}
    \label{fig:avg_devices_train_acc_waymo_balanced}
    \end{subfigure}
    \caption{Average Devices Performance: Best, Worst and Average Results}
    \label{fig:avg_devices_train_test_balanced}
\end{figure}

In this section, we study the impact of class balance and imbalance.
Figure \ref{fig:waymo_label_distribution} shows all three labels found in the selected .tar file we are using for this experiment.
In Figures \ref{fig:imbalanced_class} and \ref{fig:balanced_class}, we observe a balanced and unbalanced number of classes, respectively.
Imbalance labels are by default present in the dataset, while the balanced label is the one we prepared to study the impact.
During the processing of the dataset, most of the labels are imbalanced in all datasets, with the main one label having a dominance. 
We observe the impact of balanced and unbalanced in Figures \ref{fig:balanced_unbalanced_results} and \ref{fig:avg_devices_train_test_balanced}.
In terms of prediction results, results from the FT model, the results are better with balanced class labels 
as in Figures \ref{fig:prediction_test_waymo_balanced}, \ref{fig:prediction_val_waymo_balanced}, \ref{fig:pediction_val_loss_waymo_balanced}, \ref{fig:server_test_acc_waymo_balanced}, \ref{fig:server_loss_waymo_balanced}, \ref{fig:server_val_acc_waymo_balanced}.
It clearly shows the impact of an imbalanced class balance within the dataset.
However, in terms of device performance, performance is still better with an unbalanced data set as in Figures \ref{fig:avg_devices_test_acc_waymo_balanced} and \ref{fig:avg_devices_train_acc_waymo_balanced}.
This contrast result against the FT model or prediction performance shows that, in fact, within the device itself, training and testing may not have been impacted by overall distribution labels across the datasets.
However, the best results in each communication round are slightly similar to both average test accuracy and the average train accuracy.

\subsection{Server Side Optimization (FT) Results}
In this result, we compare three results.
\begin{enumerate}
    \item No server side optimization (No FT): In this case, we train local models, average a global model. This average model is used to update local devices. After that, we first use this global model to see how well it can optimize on the server validation set and then test on the test set. This optimization and testing has no effect on device local models and is not used for prediction accuracy and loss.
    \item Server side optimization (FT): In this case, after we create the model, we check how well it optimizes on the server validation set and then test on the test set. This optimization result, the optimized global model is used by local devices to update their local model and is also used for the prediction accuracy and loss.
    \item Server side optimization and averaging (FT+Avg): In addition to the second step (FT), before sending optimized model to the local devices and using it for prediction accuracy, we further average this optimized model with untouched global average model developed in that communication round.
\end{enumerate}
The results are device local model performance, FT global model performance, and prediction test results. The experiment was on balanced labels on the wAYMO dataset with only two labels (Vehicle and Pedestrian - 33,115 label 1, 26,102 label 0) as in Figure \ref{fig:balanced_class}.
However, we only randomly take 2000 for training and 200 for test samples.

\begin{figure}[!htbp]
    \centering
      \begin{subfigure}{0.3\columnwidth}
        \centering
        \includegraphics[width=\linewidth]{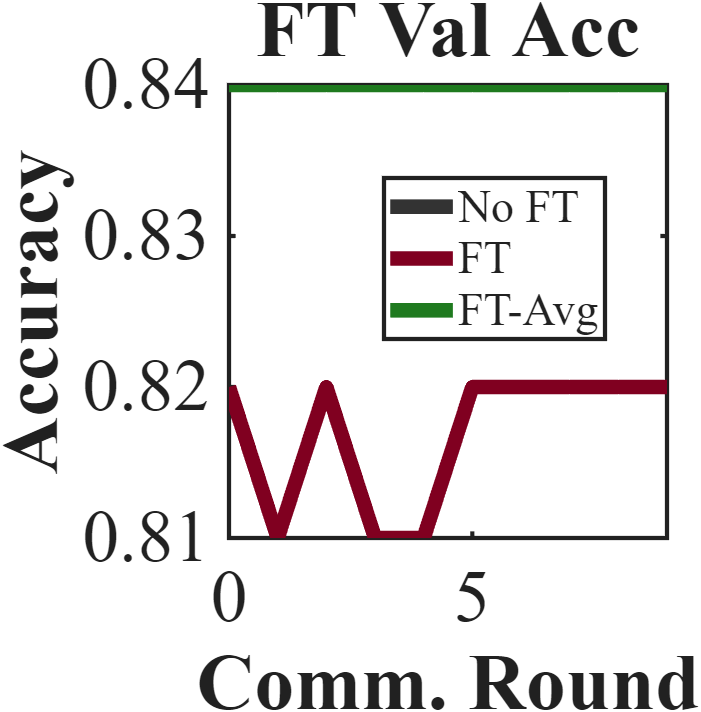}
        \caption{FT Val}
        \label{fig:server_val_accuracy_waymo_finetune}
    \end{subfigure}
    \begin{subfigure}{0.3\columnwidth}
        \centering
        \includegraphics[width=\linewidth]{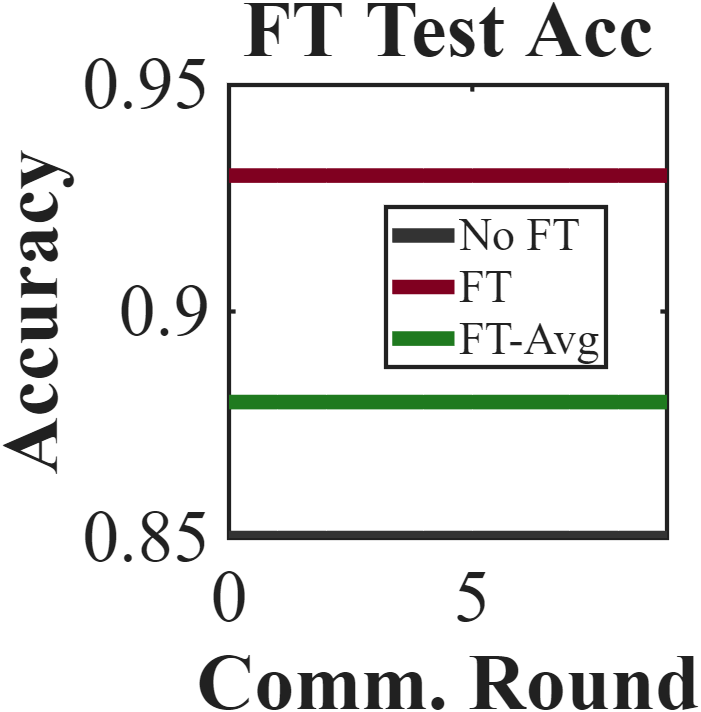}
        \caption{FT Test}
        \label{fig:server_test_accuracy_waymo_finetune}
    \end{subfigure}
    \begin{subfigure}[b]{0.3\columnwidth}
        \centering
       \includegraphics[width=\columnwidth]{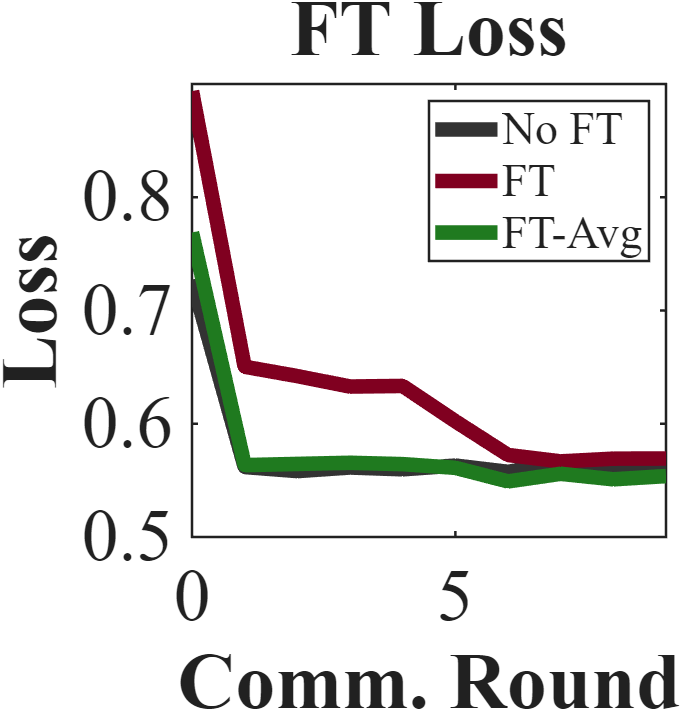}
    \caption{FT Loss}
    \label{fig:server_objective_values_waymo_finetune}
    \end{subfigure}
    \begin{subfigure}[b]{0.3\columnwidth}
        \centering
       \includegraphics[width=\columnwidth]{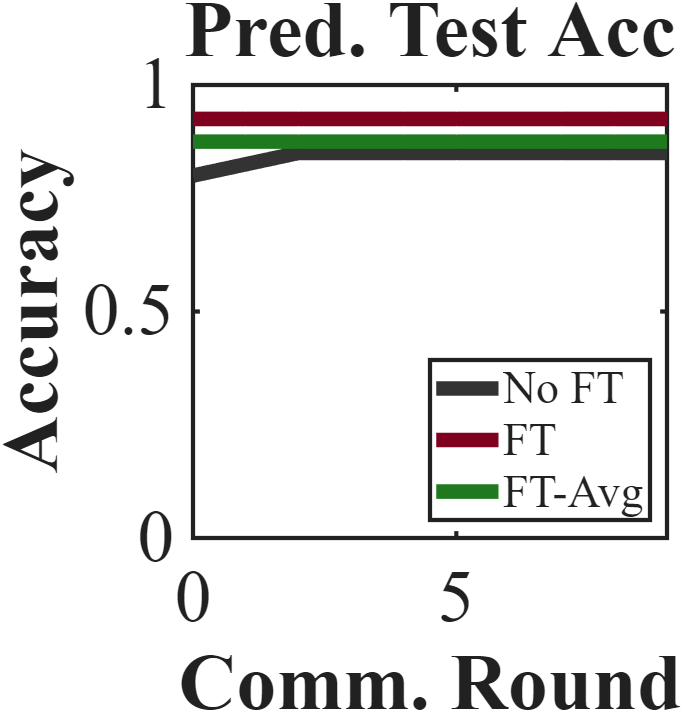}
    \caption{Pred. Test}
    \label{fig:prediction_test_accuracy_waymo_finetune}
    \end{subfigure}
    \begin{subfigure}[b]{0.3\columnwidth}
        \centering
       \includegraphics[width=\columnwidth]{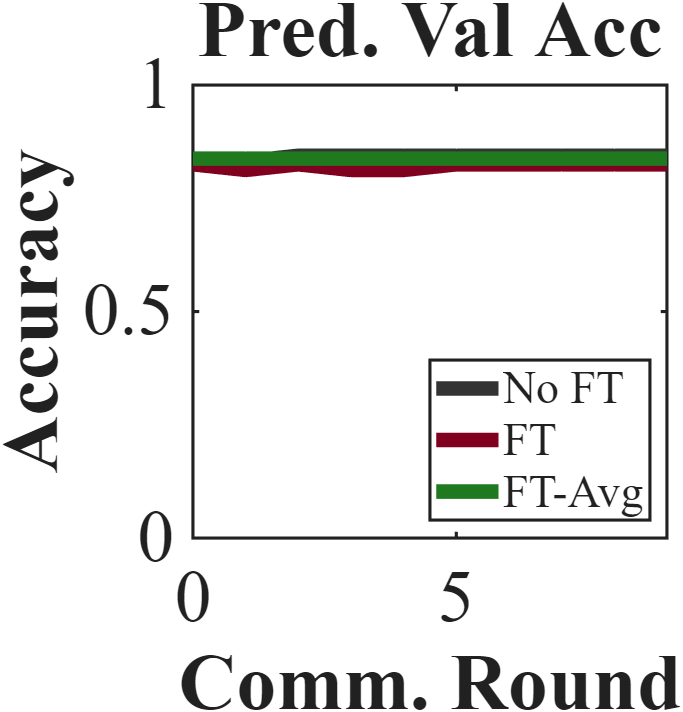}
    \caption{Prediction Val}
    \label{fig:prediction_val_accuracy_waymo_finetune}
    \end{subfigure}
    \begin{subfigure}[b]{0.3\columnwidth}
        \centering
       \includegraphics[width=\columnwidth]{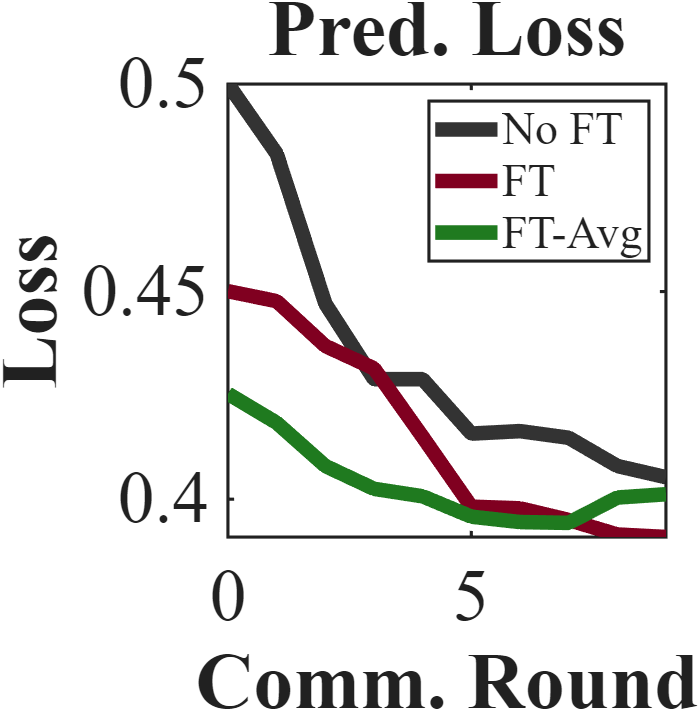}
    \caption{Pred. Loss}
    \label{fig:prediction_loss_waymo_finetune}
    \end{subfigure}
      \begin{subfigure}[b]{0.3\columnwidth}
        \centering
       \includegraphics[width=\columnwidth]{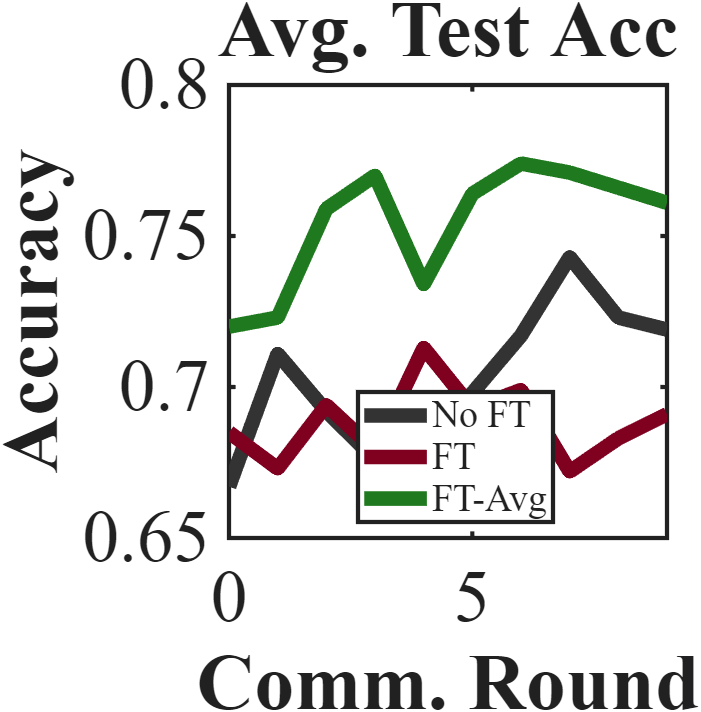}
    \caption{Avg. Test}
    \label{fig:avg_devices_test_acc_waymo_finetune}
    \end{subfigure}
      \begin{subfigure}[b]{0.3\columnwidth}
        \centering
       \includegraphics[width=\columnwidth]{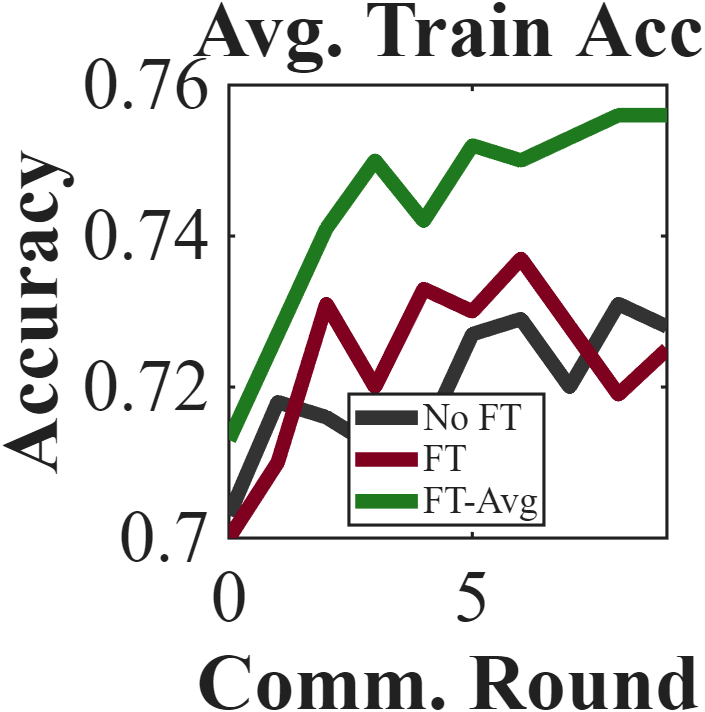}
    \caption{Avg. Train}
    \label{fig:avg_devices_train_acc_waymo_finetune}
    \end{subfigure}
       \begin{subfigure}[b]{0.3\columnwidth}
        \centering
       \includegraphics[width=\columnwidth]{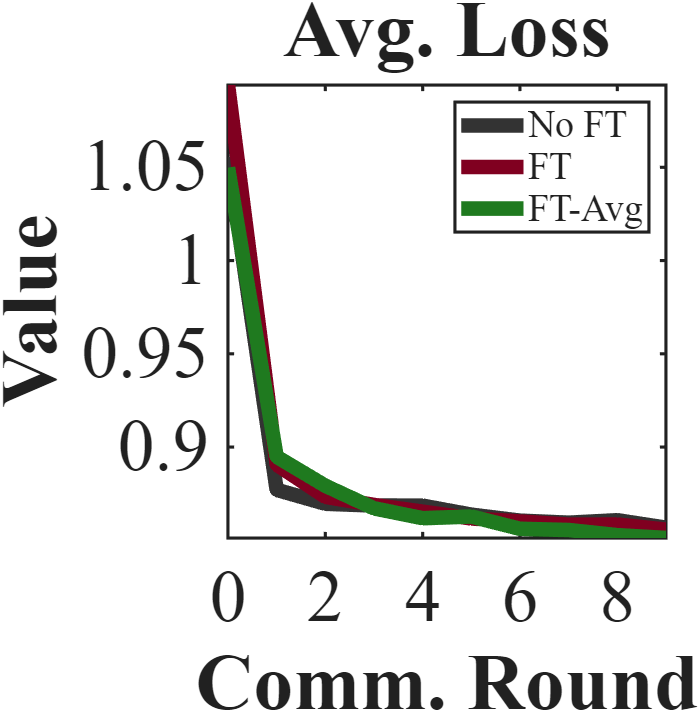}
    \caption{Avg. Loss}
    \label{fig:avg_devices_loss_waymo_finetune}
    \end{subfigure}
    
    \caption{Server Side Optimization (Fine tuning/Adaptation) results on Waymo dataset; FT Model performance, local model performance and prediction results.}
    \label{fig:fine_tune_results}
\end{figure}

In Figure \ref{fig:fine_tune_results}, we can observe that in general FT-Avg performs the best in the test accuracy of the average devices, the training accuracy, and the average training loss as in Figures \ref{fig:avg_devices_test_acc_waymo_finetune}, \ref{fig:avg_devices_train_acc_waymo_finetune} and \ref{fig:avg_devices_loss_waymo_finetune}, respectively.
This shows that fine tuning along with also considering the average model created on which the optimization or fine tuning was done, the performance is well improved in local level also.
However, in terms of server performance, FT performs better than others while default No FT performs the least among the methods, as shown in Figures \ref{fig:server_test_accuracy_waymo_finetune} and \ref{fig:server_objective_values_waymo_finetune} for server test accuracy and loss, respectively.
This highlights the need for further study on how to maintain balance between the created global model and the fine tuned global model to perform optimally on both server and device levels. However, both FT and FT-Avg perform better than No FT.
Thus, it highlights that FedAvg might benefit from some mechanism of fine tuning or optimization at the server side to improve overall performance at both the local and device level.

\section{Conclusion}
In this work, we proposed a novel QFL framework for autonomous vehicles, performed study of various datasets like Waymo, nuScenes, KITTI with various quantum models like QCNN, VQC etc. 
In addition to the default QFL framework, we proposed a private and quantum secure vehicular QFL framework which is provable DP private providing security against model inversion and QKD integration providing further quantum security with encryption and decrypted communication channel.
Further, we provide an optimized version of vQFl where we adapt the global model on the server side to improve the performance both at the local and global level.
We believe that our work presents a crucial foundation for future work towards robust, reliable, and secure autonomous vehicle systems for the field of QFL.
In the future, our aim is to expand our work further in tasks other than object detection such as route planning, lane detection, and semantic segmentation exploring other crucial decision-making aspects of autonomous driving in various traffic scenarios.

\section*{Acknowledgments}
This research was supported by an Australian Government Research Training Program (RTP) Scholarship. 



\end{document}